%% file: ejor2019-R3-ArXiV.tex
\documentclass[12pt]{article}

\usepackage{natbib}
 \bibpunct[, ]{(}{)}{,}{a}{}{,}%
 \def\newblock{\ }%

\usepackage[english]{babel}
\usepackage{amssymb,amsmath,amsfonts,graphicx}
\usepackage{float}
\usepackage{afterpage}
\usepackage{color}
\usepackage{enumitem}
\usepackage{setspace}

\usepackage[norelsize,boxed,linesnumbered,noend]{algorithm2e}
\SetArgSty{textrm} 
\SetAlCapSkip{0.7em} 

\usepackage[table]{xcolor}
\usepackage{multirow}
\usepackage{microtype}
\usepackage[hidelinks]{hyperref}
\usepackage{cleveref}
\usepackage{mathtools}
\usepackage{booktabs}
\usepackage{upgreek}
\usepackage{tikz}
\usepackage{multicol}
\usepackage{lscape}
\usepackage{glossaries}
\usepackage{longtable}
\usepackage{nicefrac}

\newcommand{\cN}{{\mathcal{N}}}
\newcommand{\cO}{{\mathcal{O}}}
\newcommand{\cP}{{\mathcal{P}}}

\newcommand{\cU}{{\mathcal{U}}}
\newcommand{\cL}{{\mathcal{L}}}

\newacronym{HGS}{HGS}{hybrid genetic search}
\newacronym{HGSADC}{HGSADC}{hybrid genetic search with advanced diversity control}
\newacronym[longplural={industrial and tramp ship routing and scheduling problems}]{ITSRSP}{ITSRSP}{industrial and tramp ship routing and scheduling problem}
\newacronym{ITSRSPSO}{ITSRSPSO}{industrial and tramp ship routing and scheduling problem with speed optimization}
\newacronym{PDPTW}{PDPTW}{pickup and delivery problem with time windows}
\newacronym{CG}{CG}{column generation}
\newacronym{SP}{SP}{set partitioning}
\newacronym{VRP}{VRP}{vehicle routing problem}
\newacronym{UHGS}{UHGS}{unified hybrid genetic search}
\newacronym{BnP}{B\&P}{branch-and-price}
\newacronym{BCP}{BCP}{branch-cut-and-price}
\newacronym{BnB}{B\&B}{branch-and-bound}
\newacronym{MDA}{MDA}{monotonic decomposition algorithm}
\newacronym{RSA}{RSA}{recursive smoothing algorithm}
\newacronym{SOP}{SOP}{speed optimization problem}
\newacronym{ALNS}{ALNS}{adaptive large neighborhood search}
\newacronym{CVRP}{CVRP}{capacitated vehicle routing problem}
\newacronym{VRPTW}{VRPTW}{vehicle routing problem with time windows}
\newacronym{ESPPRC}{ESPPRC}{elementary shortest path problem with resource constraints}
\newacronym{RAPNC}{RAP-NC}{resource allocation problem with nested constraints}
\newacronym{RAP}{RAP}{resource allocation problem}
\newacronym{DSSR}{DSSR}{decremental state-space relaxation}
\newacronym{DAT}{DAT}{discretized arrival times}
\newacronym{LS}{LS}{local search}
\newacronym{DP}{DP}{dynamic programming}
\newacronym{DTI}{DTI}{delivery triangle inequality}
\newacronym{3-SRC}{3-SRC}{3-Subset-Row Cut}
\newacronym{MIP}{MIP}{mixed-integer programming}
\newcolumntype{H}{>{\setbox0=\hbox\bgroup}c<{\egroup}@{}}

\allowdisplaybreaks

\title{Industrial and Tramp Ship Routing Problems: Closing the Gap for Real-Scale Instances}

\author{Gabriel Homsi, Rafael Martinelli, Thibaut Vidal, Kjetil Fagerholt}

\begin{document}

\vspace*{-1.8cm}

\begin{scriptsize}
\noindent
This is the post-peer-review, pre-copyedit version of the article published in the \emph{European Journal of Operational Research}. The published article is available at \url{https://doi.org/10.1016/j.ejor.2019.11.068}. This manuscript version is made available under the CC-BY-NC-ND 4.0 license \url{http://creativecommons.org/licenses/by-nc-nd/4.0/}.\par
\end{scriptsize}

\begin{center}


\begin{LARGE}
{Industrial and Tramp Ship Routing Problems:\vspace*{0.25cm}\linebreak Closing the Gap for Real-Scale Instances}
\end{LARGE}

\vspace*{0.3cm}

\textbf{Gabriel Homsi} \\
D\'epartement d'informatique et de recherche op\'erationnelle, \\ 
CIRRELT, Universit\'e de Montr\'eal,
gabriel.homsi@cirrelt.ca \\
\vspace*{0.1cm}
\textbf{Rafael Martinelli$^*$} \\
Departamento de Engenharia Industrial, \\ 
Pontif\'{\i}cia Universidade Cat\'olica do Rio de Janeiro,
martinelli@puc-rio.br \\
\vspace*{0.1cm}
\textbf{Thibaut Vidal} \\
Departamento de Inform\'atica, \\ 
Pontif\'{\i}cia Universidade Cat\'olica do Rio de Janeiro,
vidalt@inf.puc-rio.br \\
\vspace*{0.1cm}
\textbf{Kjetil Fagerholt} \\
Department of Industrial Economics and Technology Management, \\
Norwegian University of Science and Technology,
kjetil.fagerholt@ntnu.no \\

\end{center}
\noindent
\textbf{Abstract.}
Recent studies in maritime logistics have introduced a general ship routing problem and a benchmark suite based on real shipping segments, considering pickups and deliveries, cargo selection, ship-dependent starting locations, travel times and costs, time windows, and incompatibility constraints, among other features. Together, these characteristics pose considerable challenges for exact and heuristic methods, and some cases with as few as 18 cargoes remain unsolved. To face this challenge, we propose an exact \gls{BnP} algorithm and a hybrid metaheuristic. Our exact method generates elementary routes, but exploits decremental state-space relaxation to speed up column generation, heuristic strong branching, as well as advanced preprocessing and route enumeration techniques. Our metaheuristic is a sophisticated extension of the unified hybrid genetic search. It exploits a set-partitioning phase and uses problem-tailored variation operators to efficiently handle all the problem characteristics. As shown in our experimental analyses, the \gls{BnP} optimally solves 239/240 existing instances within one hour. Scalability experiments on even larger problems demonstrate that it can optimally solve problems with around 60 ships and 200 cargoes (i.e., 400 pickup and delivery services) and find optimality gaps below $1.04\%$ on the largest cases with up to 260 cargoes. The hybrid metaheuristic outperforms all previous heuristics and produces near-optimal solutions within minutes. These results are noteworthy, since these instances are comparable in size with the largest problems routinely solved by shipping companies.\vspace*{-0.2cm}\\

\noindent
\textbf{Keywords.} OR in maritime industry, ship routing, branch-and-price, hybrid genetic search

\noindent
\textbf{$^*$ Corresponding author.}

\newpage

\glsresetall

\section{Introduction}

International trade depends heavily on ship transportation, as it is the only cost-effective means for the transportation of large volumes over long distances. It is common to distinguish between three main modes of operation in maritime transportation: liner, industrial, and tramp shipping. Liner shipping, which includes container shipping, is similar to a bus service: fixed schedules and itineraries must be followed. In industrial shipping, the operator owns the cargoes and controls the fleet, trying to minimize the cargo transportation cost. Finally, a tramp shipping operator follows the availability of cargoes in the market, often transporting a mix of mandatory and optional cargoes with the goal of maximizing profit.

In this work, we focus on \glspl{ITSRSP}, typically arising in the shipping of bulk products such as crude oil; chemicals and oil products (wet bulk); and iron ore, grain, coal, bauxite, alumina, and phosphate rock (dry bulk). In $2016$, these product types constituted more than $60\%$ of the weight transported in international seaborne trade. Yet, in the wake of the financial crisis in 2008, the freight rates in the dry bulk shipping segment dropped dramatically: the Baltic Dry Index dropped more than~$80\%$, and experienced record lows in 2016. This led to a continuing situation of shipping overcapacity and downward pressure on freight rates \citep{UNCTAD2017}. In this environment, a shipping company can be profitable only if its fleet is routed effectively.

In the \gls{ITSRSP}, a shipping company has a mix of mandatory and optional cargoes for transportation.
Each cargo in the planning period must be picked up at its loading port within a specified time window, transported, and then delivered at its destination port, also within a given time window.
The shipping company controls a heterogeneous fleet of ships; each ship has a given initial position and time for when it becomes available for new transportation tasks.
Compatibility constraints may restrict which cargoes a ship can transport (for example, due to draft limits in the ports).
The shipping company may charter ships from the spot market to transport some of the cargoes.
The planning objective in the \gls{ITSRSP} is to construct routes and schedules, deciding which spot cargoes to transport and which cargoes will be transported by a spot charter, so that all mandatory cargoes are transported while maximizing profit or minimizing costs.
The \gls{ITSRSP} extends the \gls{PDPTW} with a heterogeneous fleet, compatibility constraints, different ship starting points and starting times, and service flexibility with penalties.
The interplay of these complex attributes requires the joint optimization of multiple decision sets.
Moreover, the ITSRSP is particularly relevant to decision-makers, as it closely follows the daily practice of several shipping companies that operate in the shipping segments described above. We refer the reader to \citet{Fagerholt2004} for a description of the development process of an optimization-based decision support system for the ITSRSP with several shipping companies.

In a recent work from \citet{Hemmati2014}, a set of benchmark instances based on real shipping segments with seven to 130 cargoes (pickup-and-delivery pairs) has been made available to the academic community. The authors also presented a compact mathematical formulation, and solved it with a branch-and-cut algorithm to obtain initial results. However, some instances with as few as 18~cargoes remain unsolved. Clearly, given the current scale of industrial applications, a significant methodological gap must be bridged to respond to practical needs. To compensate for the lack of exact solutions, \citet{Hemmati2014} designed an \gls{ALNS} heuristic, and subsequently investigated the impact of randomization as well as that of various search operators \citep{Hemmati2016}. However, due to the lack of available lower bounds or optimal solutions, the true performance of these methods is unknown for large problems.

This paper contributes to fill this methodological gap, from an exact and heuristic standpoint.
\begin{itemize}[leftmargin=*]
\item Firstly, we introduce an efficient \gls{BnP} algorithm for the \gls{ITSRSP}. It relies on the generation of elementary routes, but exploits \gls{DSSR} and extensive preprocessing to speed up labeling and pricing, as well as strong branching. Efficient correction strategies allow to maintain the \gls{DTI}, which can become invalid due to the dual costs but is fundamental for dominance. The \gls{BnP} is then extended using route enumeration (possible thanks to a sophisticated sequence of completion bounds), inspection pricing, and separation of subset-row cuts. This is first time these methodological building blocks are adapted, improved and combined into an efficient algorithm for ship routing.

\item Secondly, to quickly generate high-quality solutions, we introduce a \gls{HGS}. Our approach follows the same principles as the \gls{UHGS} of \citet{Vidal2014}. Yet, the \gls{UHGS} was never applied to heterogeneous fixed fleet and pickup-and-delivery problems as these problems require structurally-different local-search neighborhoods and variation operators (e.g., crossover) to be efficiently handled. Built on a completely new code base, our algorithm bridges these gaps. It uses problem-tailored crossover, \gls{LS} operators and ship-dependent neighborhood restriction strategies to efficiently optimize all aspects of the \gls{ITSRSP} and take into account its numerous constraints. It is also complemented by a set partitioning intensification procedure so as to stimulate a faster convergence~towards~good~route~combinations.

\item Finally, we report extensive experimental analyses on the industrial instances from \citet{Hemmati2014} to measure the performance of the new methods. The \gls{BnP} algorithm is able to solve $239$ out of the $240$ available instances to optimality within a time limit of one hour. The last instance is solved in $4$ hours and $25$ minutes.
Moreover, our \gls{HGS} finds very accurate solutions within a few minutes, with an average gap of $0.01\%$ relative to the optima. The quality of these solutions is far beyond that of the previously existing \gls{ALNS}.
These results are remarkable because the instances of \citet{Hemmati2014} were built to withstand the test of time. The ability to solve \emph{all of them} within five years reflects the considerable progress made by exact methods for rich routing applications. To evaluate the scalability of our methods, we also conduct experiments on larger instances. The largest mixed load instance solved to optimality has 56 ships and 195 cargoes, and therefore 390 pickup and delivery services. This size is comparable to the largest problems solved routinely by shipping companies. For example, Wilson, which is among the largest bulk shipping companies, operates 117 bulk ships between 1.500 and 8.500 deadweight tons and divides its operations into three main segments with 40 ships each \citep{Wilson2018}. We therefore reach a turning point where state-of-the-art \emph{exact} methods become sufficient for daily maritime practice.
\end{itemize}

\section{Problem Statement and Related Literature}
\label{sec:statement}

The \gls{ITSRSP} is defined on a complete digraph $G = (V,A)$, where $V$ is the union of a set of pickup nodes $P = \{1, \dots, n\}$, delivery nodes $D = \{n+1,\dots,2n\}$, and starting locations $\{0_1, \dots, 0_m\}$. 
A tramp or industrial shipping operator owns a fleet of $m$ ships $K = \{1, \dots, m\}$, and $n$ cargoes are available for transportation. Each cargo $i \in \{1, \dots, n\}$ is characterized by a load $q_i$ and must be transported from a pickup $i \in P$ to a corresponding delivery location $n + i \in D$. Therefore, $q_i \geq 0$ for $i \in P$, and $q_{n + i} = -q_i$. Every node $i \in P \cup D$ is associated with a hard time window of allowable visit times $[a_i, b_i]$. Each ship $k \in K$ becomes available at time $s^{\textsc{d}}_{0k}$, at location $0_k$. It has a capacity $Q_k$ and can traverse any arc $(i, j) \in A$ for a cost $c^k_{ij}$ (including fuel and canal costs) and duration $\delta^k_{ij}$. For every visit involving some ship $k \in K$ and some node~$i \in P \cup D$, there is an associated port service cost $s^{\textsc{c}}_{ik} \geq 0$ and duration $s^{\textsc{d}}_{ik} \geq 0$. There may be incompatibilities between ships and cargoes (e.g.\, due to draft limits in the ports). For each $i \in \{1, \dots, n\}$ and~$k \in K$, the boolean $I_{ik}$ defines whether cargo $i$ can be serviced by ship $k$. Finally, a penalty $s^{\textsc{c}}_{i0}$ is paid if cargo $i$ is not transported by the fleet. This penalty corresponds to the revenue loss (or charter cost) due to not transporting an optional cargo.

The objective of \gls{ITSRSP} is to form routes that minimize the sum of the total travel cost and the possible penalties in the case where charter ships are used or some cargoes are not transported. The routes begin at their respective starting points but have no specified endpoint, since ships operate around the clock. Every route must be feasible: ships cannot exceed their capacity, cargoes can be serviced only within their prescribed time windows, and ships cannot transport incompatible cargoes. Furthermore, the routes must respect pairing and precedence constraints. The pairing constraint states that any pair $(i \in P, n + i \in D)$ must belong to the same route, and the precedence constraint states that any pickup $i \in P$ must occur before its delivery $n + i \in D$.

\vspace{0.5em} 
\noindent
\textbf{Related literature.}
Early studies of ship routing and scheduling optimization date back to the 1970--80s.
In a seminal study, \citet{Ronen1983} discusses the differences between classical vehicle routing and ship routing and lists possible explanations for the scarcity of research at the time. The author also provides a comprehensive classification scheme for various types of ship routing and scheduling problems. Since  this article, research on ship routing has flourished, as highlighted by a general survey of maritime transportation \citep{Christiansen2007}, and reviews focusing on routing and scheduling \citep{Christiansen2013, Christiansen2014}.

Many variations of ship routing and scheduling problems have been formulated, and these problems have grown in richness, complexity and accuracy over the years. To name a few, \citet{Brown1987} introduced a \gls{SP} model to solve a full shipload routing and scheduling problem for a fleet of crude oil tankers.
\citet{Fagerholt2000} proposed a \gls{DP} algorithm to solve a traveling salesman problem with time windows and pickups and deliveries, encountered when solving ship scheduling subproblems. The same algorithm was later exploited by \citet{Fagerholt2000b} to solve subproblems for a multi-ship \gls{PDPTW}.
\citet{Sigurd2005} introduced a heuristic \gls{BnP} algorithm for a periodic ship scheduling problem with visit separation requirements.
A maritime \gls{PDPTW} with split loads and optional cargoes was studied by \citet{Andersson2011} and \cite{Stlhane2012}.
\citet{Andersson2011} proposed two path-flow models and an exact algorithm that generates single ship schedules a priori, while \citet{Stlhane2012} designed a \gls{BCP} algorithm.

Heuristics and metaheuristics have also been applied to solve several variants of ship routing problems. Some notable examples are the multi-start \gls{LS} of~\citet{Bronmo2007}, the unified tabu search of~\citet{Korsvik2009}, and the large neighborhood searches of \citet{Korsvik2011} and \citet{Hemmati2014}. \citet{Borthen2017} used a hybrid genetic search algorithm with great success to solve a multi-period supply vessel planning problem for offshore installations. Furthermore, the \gls{UHGS} methodology of \citet{Vidal2012, Vidal2014} has led to highly accurate solutions for a considerable number of \gls{VRP} variants, including the classical capacitated \gls{VRP}, the \gls{VRPTW} \citep{Vidal2013}, and several prize-collecting \glspl{VRP} with profits and service selections \citep{Vidal2014bb, Bulhoes2017}. However, this methodology has never been extended to heterogeneous fixed fleet or pickup-and-delivery problems, which require structurally different neighborhood searches and proper precedence and pairing between the pickups and deliveries in the crossover and split operators.

\section{Branch-and-Price}
\label{sec:bp}

A simple \gls{SP} formulation of the \gls{ITSRSP} is given in Equations (\ref{eq:sp-obj}) to (\ref{eq:sp-c4}). Let $\Omega_k$ be the set of all feasible routes for ship $k \in K$. This formulation uses a binary variable $\lambda^k_\sigma$ to indicate whether or not route $\sigma \in \Omega_k$ of ship $k$ is used in the current solution for a cost of $c^k_\sigma$. Moreover,~$a_{\sigma i}^k$ is a binary constant that is equal to $1$ if and only if the route $\sigma$ of ship $k$ transports cargo $i$, and $0$ otherwise. Each variable $y_i$ is equal to $1$ if and only if cargo $i$ is transported by a charter instead of being included in a route.
\begin{align}
	\qquad \textrm{Minimize} \hspace*{0.7cm} \sum_{k \in K}\sum_{\sigma \in \Omega_k}c^k_\sigma \lambda^k_\sigma & + \sum_{i \in P} s^{\textsc{c}}_{i0} y_i \label{eq:sp-obj} &                                                         \\
	\textrm{subject to} \hspace*{1.6cm} \sum_{\sigma \in \Omega_k} \lambda^k_\sigma                                & \leq 1                                                     & \forall k \in K \label{eq:sp-c1}                      \\
	\sum_{k \in K}\sum_{\sigma \in \Omega_k}a^k_{\sigma i} \lambda^k_\sigma + y_i                                & = 1                                                        & \forall i \in P \label{eq:sp-c2}                    \\
	\lambda^k_\sigma                                                                                               & \in \{0, 1\}                                               & \forall k \in K, \sigma \in \Omega_k \label{eq:sp-c3} \\
	y_i                                                                                                            & \in \{0, 1\}                                               & \forall i \in P. \label{eq:sp-c4}
\end{align}

Objective~\eqref{eq:sp-obj} minimizes the routing and charter costs. Constraints~\eqref{eq:sp-c1} ensure that each ship is used at most once, and Constraints~\eqref{eq:sp-c2} guarantee that each cargo is either transported or chartered.

\subsection{Column generation}
\label{sec:cg}

Formulation (\ref{eq:sp-obj}--\ref{eq:sp-c4}) clearly contains an exponential number of variables, and therefore we will use a \gls{CG} algorithm to solve its linear relaxation.
Moreover, each ship in the \gls{ITSRSP} has a different starting location, cargo compatibility, capacity, travel cost, and time matrix. For this reason, we must solve a collection of pricing subproblems (one for each ship) rather than a single one.

Let $\gamma_k$ and $\beta_i$ be the dual variables associated with Constraints~\eqref{eq:sp-c1} and~\eqref{eq:sp-c2}. The reduced cost of a route, defined in Equation~\eqref{eq:rc-route}, can be distributed into reduced costs for each arc, as shown in Equation~\eqref{eq:rc-arc}.
\begin{align}
 \bar{c}^k_\sigma &= c^k_\sigma - \gamma_k - \sum\limits_{i \in P} a^k_{\sigma i} \beta_i & \forall k \in K, \sigma \in \Omega_k. \label{eq:rc-route}\\
 \bar{c}^k_{ij} &= \left\{\begin{array}{l}
 c^k_{ij} - \gamma_k \\
 c^k_{ij} - \beta_i \\
 c^k_{ij}
 \end{array}\right. & \begin{array}{r}
 \forall k \in K, i = 0, j \in V, \\
 \forall k \in K, i \in P, j \in V, \\
 \forall k \in K, i \in D, j \in V.
 \end{array}
 \label{eq:rc-arc}
\end{align}
Since the dual costs are originally associated with cargoes (p--d pairs), we opted to associate all the dual costs with the out-arcs from the pickup nodes and depot, and none with those emerging from delivery nodes. This methodological choice led to better performance in our initial experiments, and it is in agreement with previous studies on the \gls{PDPTW} \citep[e.g.\,][]{Ropke2009}.

\vspace{0.5em} \noindent \textbf{Pricing. }
The pricing subproblem is an \gls{ESPPRC}, which is $\mathcal{NP}$-hard and often difficult to solve for large instances. Various studies present ways to solve it more efficiently by using route relaxation techniques, at the cost of a slightly weaker linear relaxation \citep{Christofides1981, Baldacci2011}. Again based on preliminary experiments, we decided to maintain the pairing and precedence constraints, since relaxing these constraints leads to a strong deterioration in the linear relaxation bound. In contrast, elementarity tends to be ``naturally'' satisfied in most situations without any specific measure in the labeling and dominance. This occurs because after performing a pickup and delivery, the ship is usually far from the pickup point from a spatial and temporal viewpoint, and a new visit to the pickup may be impossible because of the time-window constraints. A similar property has been exploited in \cite{Bertsimas2018} to optimize taxi-fleets services.

We take advantage of this observation by initially relaxing the elementarity and reintroducing it using \gls{DSSR}, to accelerate the solution of the pricing subproblems \citep{Righini2008}. This is done by defining a set $\Gamma \subseteq P$ of pickups that cannot be opened again. The set is initialized as $\Gamma = \varnothing$ at the start of the process, and it is augmented each time that a repeated service is identified.

The \gls{ESPPRC} is solved using a forward \gls{DP} algorithm. For each path~$\cP$, we define a label $\mathcal{L}(\cP) = (v(\cP), \bar{c}(\cP), q(\cP), t(\cP), \mathcal{O}(\cP), \mathcal{U}(\cP))$ containing, respectively, the last vertex of the path, the accumulated reduced cost, the total load, the arrival time, the set of opened p--d pairs, and the set of unreachable pairs. As in \citet{Dumas1991} and \citet{Ropke2009}, the set of opened pairs contains the visited pickup nodes for which the corresponding delivery node has not been visited. A pair is unreachable if the pickup node has already been visited. Finally, a valid route is a feasible path $\cP$ such that $\mathcal{O}(\cP) = \varnothing$.

Given $i = v(\cP)$, extending the path $\cP$ to a vertex $j \in V$ is allowed only if $q(\cP) + q_j \leq Q_k$, $t(\cP) + s^{\textsc{d}}_{ik} + \delta^k_{ij} \leq b_j$, and:
\begin{equation}
\begin{cases}
	j \notin \cO(\cP)  & \text{if } j \in P \backslash \Gamma,              \\
	j \notin \cU(\cP)  & \text{if } j \in \Gamma, \\
	j - n \in \cO(\cP) & \text{if } j \in D.
\end{cases}
\end{equation}

If an extension is allowed, it generates the new label presented in Equation~\eqref{eq:label}:
\begin{equation}
\cL(\cP^\prime) = \begin{cases}
	(j, \bar{c}(\cP) + \bar{c}^k_{ij}, q(\cP) + q_j, \max\{a_j, t(\cP) + s^{\textsc{d}}_{ik} + \delta^k_{ij}\}, \cO(\cP) \cup \{j\}, \cU(\cP)) & \text{if } j \in P \backslash \Gamma, \\
	(j, \bar{c}(\cP) + \bar{c}^k_{ij}, q(\cP) + q_j, \max\{a_j, t(\cP) + s^{\textsc{d}}_{ik} + \delta^k_{ij}\}, \mathcal{O}(\cP) \cup \{j\}, \mathcal{U}(\cP) \cup \{j\}) & \text{if } j \in \Gamma, \\
	(j, \bar{c}(\cP) + \bar{c}^k_{ij}, q(\cP) + q_j, \max\{a_j, t(\cP) + s^{\textsc{d}}_{ik} + \delta^k_{ij}\}, \mathcal{O}(\cP) \backslash \{n - j\}, \mathcal{U}(\cP)) & \text{if } j \in D.
\end{cases}
\label{eq:label}
\end{equation}

To reduce the number of labels during the \gls{DP} algorithm, we use the following dominance rule: a path $\cP_1$ dominates a path $\cP_2$ if Condition \eqref{eq:dom} holds.
\begin{equation}
\begin{aligned}
	& v(\cP_1)       = v       (\cP_2)        \text{ and }
	  \bar{c}(\cP_1) \leq      \bar{c}(\cP_2) \text{ and }
	  t(\cP_1)       \leq      t(\cP_2)       \text{ and } \\
	& \cO(\cP_1)     \subseteq \cO(\cP_2)     \text{ and } 
	  \cU(\cP_1)     \subseteq \cU(\cP_2).
\end{aligned}
\label{eq:dom}
\end{equation}

Note that $\cO(\cP_1) \subseteq \cO(\cP_2)$ implies that $q(\cP_1) \leq q(\cP_2)$. However, as discussed in \citet{Ropke2009}, a subset-based dominance between $\cO(\cP_1)$ and $\cO(\cP_2)$ is valid only if the reduced costs satisfy the \gls{DTI}: $\bar{c}^k_{ij} \leq \bar{c}^k_{i\ell} + \bar{c}^k_{\ell j}, \forall i \in V, j \in V, \ell \in D, k \in K$. When we define the reduced costs as in Equation \eqref{eq:rc-arc}, the \gls{DTI} is valid if the original distances satisfy it. However, it may become violated by branching constraints that introduce new dual costs. Correction techniques need to be applied in these situations, as discussed in~\Cref{sec:bb}.

To the best of our knowledge, the \gls{DSSR} approach has not been tested for \gls{PDPTW} problems, but it was suggested as a promising research avenue in \citet{Ropke2009}. The same authors also suggested relaxing the $\cO(\cP)$ sets. We tested this option, but observed that it led to a much slower convergence.

\vspace{0.5em} \noindent \textbf{Ship ordering.}
Since the \gls{ITSRSP} includes a heterogeneous fleet of ships, it becomes necessary to solve the pricing subproblems associated with each ship type. To reduce the number of subproblems, we tested various approaches based on ship grouping and different orderings. We opted to simply include all the ships in a circular list, and we systematically call the pricing subproblem for the last ship with which a route was last obtained. When the pricing algorithm fails to generate a negative reduced-cost route for the current ship, the procedure selects the next one in the list. The \gls{CG} terminates when a full round has been performed without generating any new routes.

\vspace{0.5em} \noindent \textbf{Initialization and heuristic pricing.}
Because of the charter variables in the SP formulation, we simply start with empty $\Omega_k$ sets. To reduce the computational effort, the \gls{CG} initially uses a fast heuristic pricing in which only the label with the minimum reduced cost for each vertex and time value is kept. The \gls{CG} starts using the exact pricing when a full round on the ship list fails to generate a new route with the heuristic pricing.

\vspace{0.5em} \noindent \textbf{Preprocessing.}
Finally, our \gls{CG} exploits various preprocessing techniques to eliminate arcs from the pricing subproblem. A simple version of these strategies was used in \citet{Dumas1991}. These procedures are extended to consider the different attributes of the \gls{ITSRSP} and quickly determine which requests cannot be closed from a given (node, time) pair, in such a way that it is possible to filter label extensions in the pricing algorithms using bitwise operations in $\cO(n/64)$.

\subsection{Branch-and-bound}
\label{sec:bb}

The \gls{CG} presented in the previous section produces strong lower bounds for the \gls{ITSRSP}. To obtain integer optimal solutions, we embed it into a branch-and-bound algorithm to form a Branch-and-Price (B\&P) method.

\vspace{0.5em} \noindent \textbf{Branching rules.}
The \gls{BnP} uses three branching rules, giving priority to the most fractional element, as explained below.
\begin{itemize}[nosep,leftmargin=0cm]
\item[B1)] Branching on charters. If any $y_i$ variable is fractional, generate two branches with $y_i = 0$ and \mbox{$y_i = 1$}.
\item[B2)] Branching on ships. If $\sum_{\sigma \in \Omega_k} \lambda_\sigma^k$ is fractional for any ship $k$, generate two branches with $\sum_{\sigma \in \Omega_k} \lambda_\sigma^k = 0$ and $\sum_{\sigma \in \Omega_k} \lambda_\sigma^k = 1$.
\item[B3)] Branching on edges for all ships. Given $b^k_{ra}$, a binary constant indicating whether or not route $\sigma$ of ship~$k$ traverses arc $a = (i, j)^k$, let the number of times any ship traverses $a_1 = (i, j)$ or~$a_2 = (j, i)$ be $x_e = \sum_{k \in K}\sum_{\sigma \in \Omega_k} (b^k_{ra_1} + b^k_{ra_2})\lambda^k_\sigma$. If $x_e$ is fractional, generate two branches~with~\mbox{$x_e = 0$}~and~\mbox{$x_e = 1$}.
\end{itemize}

\vspace{0.5em} \noindent \textbf{Delivery triangle inequality.}
Rule B1 has no impact on the pricing subproblems. In contrast, each new constraint generated by rules B2 and B3 introduces a new dual variable that is included in the reduced costs. The dual variables associated with rule B2 cannot lead to a \gls{DTI} violation, since the first node of every route must be a pickup node. In contrast, the constraints resulting from rule B3 introduce a dual variable that will be subtracted from the right-hand side of Equation~\eqref{eq:rc-arc}, and can lead to violations of the \gls{DTI}. To circumvent this issue, we use a method similar to that of~\citet{Ropke2009} to fix the \gls{DTI}. To reduce the computational complexity of this approach, we check and fix violations in an incremental manner, focusing on the newly generated dual variables.

\vspace{0.5em} \noindent \textbf{Artificial variables.}
The branching rules presented in this section may make the solution of a child node infeasible. However, it is not immediately possible to be sure about the infeasibility because the solution may simply be missing some columns. In addition, as we start with an empty route set, infeasibility may also happen at the root node. For this reason, at every node of the \gls{BnP} that results in an infeasible solution, the algorithm uses an approach like that of the two-phase simplex method. It introduces an artificial variable on each violating constraint and changes the objective function to minimize their sum, thus minimizing the infeasibility. When the solution becomes feasible again, the artificial variables are removed and the original objective function is restored. If the \gls{CG} terminates before reaching this state, then the solution is confirmed to be infeasible.

\vspace{0.5em} \noindent \textbf{Heuristic strong branching.}
We apply strong branching to predict which element will result in better solutions, thus reducing the size of the \gls{BnP} tree. After solving the \gls{CG} of each node, we build a set of branching candidates from the most fractional elements found by the branching rules, and we simulate the branching for each element by solving both child nodes. Since it is prohibitively expensive to solve the exact pricing several times, we perform heuristic strong branching by executing the \gls{CG} with the heuristic pricing. Even the non-optimal linear solution gives a good prediction for the quality of each child node and can be used to compare the candidates. The method then retains the best one, i.e.\, the branching with the best worst child node. Moreover, a branching with one infeasible child node (from the heuristic pricing viewpoint) is always considered to be better than one with no infeasible child node, and a branching with two infeasible child nodes is immediately chosen. When the branching candidate is chosen, the exact pricing is executed on both child nodes.

\subsection{Route enumeration}
\label{sec:enumeration}

The techniques discussed to this point lead to an efficient method, the results of which will be discussed in \Cref{sec:experiments}. Since a good upper bound is known from the heuristic presented in \Cref{sec:hgs}, we decided to test additional route enumeration techniques that, when applicable, can allow us to solve larger problem instances. Given the known upper bound, the algorithm attempts to enumerate all feasible routes within the integrality gap, and it aborts if more than $2|K|$ million routes are created.
The route enumeration is done by a \gls{DP} algorithm similar to that for the exact pricing procedure, using dominance rule~\eqref{eq:enum-dom}:
\begin{equation}
\begin{aligned}
& v(\cP_1) = v(\cP_2)  \text{ and } \bar{c}(\cP_1) \leq \bar{c}(\cP_2)\text{ and } t(\cP_1) \leq t(\cP_2) \text{ and } \\
& \cO(\cP_1) = \cO(\cP_2)  \text{ and } \cU(\cP_1) = \cU(\cP_2).
\end{aligned}
\label{eq:enum-dom}
\end{equation}

This rule is weaker since it cannot discard a route unless there is another with the exact same set of opened and unreachable nodes, and therefore it leads to a much larger number of labels during the \gls{DP} algorithm. To deal with this issue, we first execute a backward pricing using completion bounds (the best reduced cost of a path ending at a given node and time) calculated from the last run of the forward exact pricing. From the results of the backward pricing we then calculate backward completion bounds (the best reduced cost of a path starting at a given node and time). As the name suggests, the backward pricing is the exact pricing algorithm executed from the end of the time horizon to the start, changing dominance rule \eqref{eq:dom} into ($\cO(\cP_1) = \cO(\cP_2)$) to avoid enforcing the pickup triangle inequality (see \citealt{Gschwind2018}). At first the completion bounds from the forward pricing subproblems seem to be incompatible with the backward pricing because of the \gls{DTI} fix. However, we observe that they are an underestimate of the correct completion bounds and therefore can be used to fathom labels. 
Finally, we generate completion bounds from the backward pricing and use them to fathom labels during the route enumeration procedure.

Upon success, the enumerated routes may be fed into the \gls{SP} formulation to obtain an optimal solution. However, the number of routes is often prohibitively large to be directly tackled with a \gls{MIP} solver. Therefore, we continue the search with the same \gls{BnP} approach and rely on \glspl{3-SRC} to improve the value of the linear relaxations \citep{Jepsen2008}. For the \gls{ITSRSP}, the \glspl{3-SRC} are Chv\'atal-Gomory rank-1 cuts obtained from a subset of three constraints from Constraints \eqref{eq:sp-c2} and a $\frac{1}{2}$ multiplier, resulting in the valid inequality presented in \eqref{eq:3src}, where $\alpha^k_{\sigma}$ represents the number of times the route $\sigma$ of ship $k$ visits the nodes of the \gls{3-SRC}.
\begin{equation}
\sum\limits_{k \in K}\sum\limits_{\sigma \in \Omega_k} \left\lfloor\frac{\alpha^k_{\sigma}}{2}\right\rfloor \lambda^k_{\sigma} \leq 1
\label{eq:3src}
\end{equation}

This leads to a \gls{BnP} algorithm where the separation and pricing is done by simple route inspection, as in \cite{Contardo2014}. In addition, note that the \glspl{3-SRC} can be first separated at the root node to improve the linear relaxation and reduce the number of routes resulting from the enumeration. Moreover, the \glspl{3-SRC} are not separated when evaluating candidates during strong branching; they are instead used on the two chosen branches.

\section{Hybrid Genetic Search}
\label{sec:hgs}

As demonstrated in \Cref{sec:experiments}, the proposed exact approaches lead to remarkable results for industrial-size ship routing instances, but the variance of CPU times can be a drawback for industrial applications. To provide heuristic solutions in a more consistent manner, as well as initial upper bounds for the exact method, we now introduce an \gls{HGS}, a sophisticated extension of the \gls{UHGS} of~\citet{Vidal2012,Vidal2014} which includes problem-tailored search operators and an \gls{SP}-based intensification procedure.

As \Cref{alg:hgs} indicates, our method follows the same general scheme as the \gls{UHGS} with the addition of the \gls{SP} procedure. It jointly evolves a feasible and an infeasible subpopulation of individuals representing solutions. At each iteration, two parents are selected from the union of the subpopulations. A crossover operator is applied to generate an offspring, which is improved by \gls{LS} and inserted into the adequate subpopulation according to its feasibility (Lines~3--6).
If the offspring is infeasible, a repair procedure is called in an attempt to generate a feasible solution (Lines~7--10).

\begin{algorithm}[htbp]
\setlength{\algomargin}{2em}
\linespread{1.1}\selectfont
 \SetAlCapSkip{0.5em}
 Initialize population\;
 \While{number of iterations without improvement $< It_\textsc{ni}$ and $time < T_\textsc{max}$}{
 Select parent solutions $P_1$ and $P_2$\;
 Apply the crossover operator on $P_1$ and $P_2$ to generate an offspring $C$\;
 Educate offspring $C$ by local search\;
 Insert $C$ into respective subpopulation\;
 \If{$C$ is infeasible}{
 With probability $p_\textsc{rep}$, repair $C$ (local search) and \\
 insert it into respective subpopulation\;
 }
 \If{maximum subpopulation size reached}{
 Select survivors\;
 }
 \If{best solution not improved for $It_\textsc{div}$ iterations}{
 Diversify population\;
 }
 \If{best solution not improved for $It_\textsc{sp}$ iterations}{
 Run set partitioning\;
 }
 Adjust penalty coefficients for infeasibility\;
 }
 Return best feasible solution\;
\caption{\textsc{Hybrid Genetic Search (HGS)}}\label{alg:hgs}
\end{algorithm}

Whenever a subpopulation reaches a maximum size, a survivor selection procedure is applied to remove individuals based on their quality and contribution to the population diversity (Lines~11--13). If no improving solution is found after $It_\textsc{div}$ successive iterations, new individuals are added to the population in order to diversify it (Lines~14--16).
Similarly, if no improving solution is found after $It_\textsc{sp}$ successive iterations, an intensification procedure is triggered, in the form of an \gls{SP} model that aims to construct better solutions from high-quality routes identified in the search history.
Finally, the penalty coefficients are periodically adjusted to control the proportion of feasible individuals in the search.

Hybrid genetic searches with a similar structure have been used with great success for a variety of \glspl{VRP} \citep[see, e.g.,][]{Vidal2014,Vidal2014bb,Borthen2017}. Still, the \gls{ITSRSP} includes such a diversity of constraints that most components of the method had to be carefully tailored in order to obtain an effective algorithm. The following subsections describe each component in more detail.

\subsection{Search space}

Previous studies have demonstrated that the controlled use of penalized infeasible solutions can help converging towards high-quality feasible solutions \citep{Glover2009, Vidal2015}. This is especially relevant for the \gls{ITSRSP}, since this problem includes time windows, capacity constraints and incompatibility constraints, as well as precedence and pairing restrictions for p--d pairs.
In the proposed \gls{HGS}, we allow the exploration of infeasible solutions in which:
\begin{itemize}[nosep]
\item ship capacity constraints may be exceeded;
\item some cargoes may not be picked up or delivered within their respective time windows;
\item each ship may transport incompatible cargoes; but
\item no component of the \gls{HGS} creates solutions that violate precedence or pairing constraints.
\end{itemize}
The load infeasibility is proportional to the difference between the peak load (largest load over the trip) and the ship capacity. To relax the time-window constraints, we use the ``time-warp'' approach of~\citet{Nagata2010b} and \citet{Vidal2013}, which allows penalized ``returns in time'' upon a late arrival to a node. Finally, the penalty associated with ship-cargo incompatibilities is proportional to the number of incompatible cargoes carried by each ship.

Let $\sigma = (\sigma_0, \dots, \sigma_{n(\sigma)})$ be a route for ship $k$, starting from the initial position ($\sigma_0 = 0_k$) and servicing a sequence of (pickup or delivery) nodes $(\sigma_1, \dots, \sigma_{n(\sigma)})$.
The start-of-service time $t^k_{\sigma_i}$ at the $i$th node can be defined as:
\begin{equation}
t^k_{\sigma_i} =
\begin{cases}
	s^{\textsc{d}}_{0k} \text{ if $i = 0$,}                                                                                \\
	\min \{\max\{a_{\sigma_i}, t_{\sigma_{i - 1}} + s^{\textsc{d}}_{\sigma_{i - 1}} + \delta^k_{\sigma_{i - 1}, \sigma_i}\}, b_{\sigma_i}\} \text{ otherwise.}
\end{cases}
\end{equation}
Route $\sigma$ can be characterized by the following quantities:
\begin{align}
	 & \text{Travel cost: } \hspace*{-2.5cm}       &  & C_k(\sigma) =       \sum_{i=1}^{n(\sigma) - 1} (c^k_{\sigma_i, \sigma_{i + 1}} + s^{\textsc{c}}_{\sigma_i}) \label{eq:tc_linear}                                          \\
	 & \text{Peak load: }  \hspace*{-2.5cm}        &  & Q^\textsc{max}_k(\sigma) =  \underset{1 \leq i \leq j \leq n(\sigma)}{\max} \sum_{l=i}^{j} q_{\sigma_l} \label{eq:pk_linear}                                              \\
	 & \text{Time warp use: } \hspace*{-2.5cm}     &  & TW_k(\sigma) =    \sum_{i=1}^{n(\sigma)} \max\{ t_{\sigma_{i - 1}} + s^{\textsc{d}}_{\sigma_{i - 1}} + \delta^k_{\sigma_{i - 1}, \sigma_i} - b_{\sigma_i}, 0\} \label{eq:tw_linear} \\
	 & \text{Incompatibilities: } \hspace*{-2.5cm} &  & I_k(\sigma) = \sum_{i=1}^{n(\sigma)} I_{\sigma_i k}. \label{eq:in_linear}
\end{align}
Finally, we define the penalized cost of route $\sigma$ for ship $k$ as:
\begin{equation}
\phi_k(\sigma) = C_k(\sigma) + \omega^{Q} \max\{0, Q^\textsc{max}_k(\sigma) - Q_k\} + \omega^\textsc{tw} TW_k(\sigma) + \omega^\textsc{I} I_k(\sigma),
\end{equation}
where $\omega^\textsc{q}$, $\omega^\textsc{tw}$, and $\omega^\textsc{I}$ are the respective penalty coefficients for peak-load, time-window, and incompatibility-constraint violations. The penalty coefficients will be adjusted during the search as described in \Cref{subsec:population_management}. The penalized cost of a solution $S$ is the sum of the penalized costs of all its routes, that is, $\phi(S) = \sum_{(r,k) \in S} \phi_k(\sigma)$. 

\subsection{Solution representation and evaluation}

A solution $S$ is represented in \gls{HGS} as a giant tour $\pi^S$ that holds a permutation of nodes in $P \cup D$ and satisfies the precedence constraints between the p--d pairs. Such a representation greatly facilitates the design of an effective crossover operator. Moreover, segmenting this giant tour into different routes can be done efficiently using a variant of the \textsc{Split} algorithm \citep{Prins2004,Vidal2016} to form a complete solution after crossover.

\textsc{Split} is a \gls{DP} algorithm that was originally designed for the capacitated VRP but is flexible enough to be adapted to a variety of constraints and objectives \citep[see, e.g.,][]{Prins2009a,Velasco2009}. When dealing with \glspl{VRP} with heterogeneous fleets, previous authors have assumed that \textsc{Split} should jointly optimize the giant tour segmentation \emph{and} the choice of ship for each route \citep{Duhamel2011a}. This extension, unfortunately, leads to a special case of the shortest path problem with resource constraints, for which only pseudo-polynomial algorithms are currently available. To avoid this issue, we opted to fix the sequence of ships and restrict the \textsc{Split} algorithm to the segmentation of the tours, so that the ships are considered one by one in their order of appearance. To avoid any possible bias from the instance representation, we shuffle the order of the ships when reading the data and keep this order fixed during the solution process. The possible use of spot charters for optional cargoes is modeled via a dummy ship of index $m+1$ with zero distance cost and a service cost equal to the charter price for each cargo.

The \textsc{Split} graph is defined as follows. Let $G^S$ be a directed acyclic graph with nodes $V^S = \{v^0_0, \dots, v^0_{2n}, v^1_0, \dots, v^1_{2n}, \dots, v^{m+1}_0, \dots, v^{m+1}_{2n}\}$ and arcs $A^S = \{(v^{k-1}_i, v^{k}_j) : 0 \leq i \leq j \leq 2n, 1 \leq k \leq m + 1\}$. 
Each arc $(v^{k-1}_i, v^{k}_j) \in A^S$ represents a route
\begin{equation}
\sigma^{k}_{i j} = (0_{k}, \pi^S_{i + 1}, \pi^S_{i + 2}, \dots, \pi^S_j)
\end{equation}
associated with ship $k$.
If $i = j$, then $\sigma^{k}_{i j} = (0_{k})$, representing an empty route.
The cost of arc $(v^{k-1}_i, v^{k}_j)$ is set to $\phi_k(\sigma^{k}_{i j})$ when the route satisfies the pairing constraints (no open pickup or delivery), and infinity otherwise. With these definitions, an optimal segmentation of the giant tour $\pi^S$ into routes assigned to the ships $1$ to $m+1$ corresponds to a shortest path between nodes $v^0_0$ and~$v^{m+1}_{2n}$~in~$G^S$. This shortest path can be obtained via Bellman's algorithm in topological order, with a time complexity of $\cO(m n^2)$ and space complexity of $\cO(m n)$.
Moreover, note that there always exists at least one feasible path without necessarily relying on spot charters: a single route for ship $1$ containing all visits naturally satisfies the pairing and precedence restrictions, despite its high cost related to the penalized violation of all the other (load, time, and incompatibility) constraints.

\vspace{0.5em} \noindent \textbf{Individual evaluation. }
As in the \gls{UHGS}, the quality of an individual $S$ is not based solely on its cost but also on its contribution to the subpopulation diversity. The combination of these two metrics is referred to as the \textit{biased fitness} of $S$ in its subpopulation $\cP$, and it is defined as
\begin{equation}
f_\cP(S) = f_\cP^\phi(S) + \left(1 - \frac{\mu^\textsc{elite}}{|\cP|}\right) f_\cP^\textsc{div}(S),
\end{equation}
where $f_\cP^\phi(S)$ is the penalized cost rank of $S$ in $\cP$, and $f_\cP^\textsc{div}(S)$ is the diversity contribution rank of $S$ in $\cP$. Both ranks are relative to the subpopulation size, and parameter $\mu^\textsc{elite}$ balances the weight of each rank. The diversity contribution of $S$ in $\cP$ is defined as the average distance to its $\mu^\textsc{close}$ closest individuals. We use the \textit{broken pairs} distance, measuring the proportion of different edges between two solutions.

\subsection{Parent selection and crossover}

At each iteration, the algorithm selects two parents $P_1$ and $P_2$ by binary tournament based on their biased fitness. To produce a child $C$, these parents are submitted to a specialized one-point crossover operator \citep{Velasco2009}, designed to enforce the pairing and precedence constraints between the pickups and deliveries:
\begin{itemize}[nosep,leftmargin=0cm]
\item[] Step 1) A cutting point $s \in \{1, \dots, 2n\}$ is randomly selected with uniform probability in the giant tour $\pi^{P_1}$ of the first parent, and the sequence of visits $\sigma = (\pi^{P_1}_1, \dots, \pi^{P_1}_s)$ is copied into~$\pi^C$.
\item[] Step 2) The second parent is swept from beginning to end, and any pending delivery ($n+i \notin \sigma$ such that $i \in \sigma$) is inserted at the end of $\pi^C$.
\item[] Step 3) The second parent is swept a second time, and any missing node is inserted at~the~end~of~$\pi^C$.
\end{itemize}

\subsection{Education and repair}
\label{sec:education_and_repair}

Each individual resulting from the crossover operator is decoded with the \textsc{Split} algorithm to obtain a complete solution, and then improved (i.e., educated) by an efficient \gls{LS} based on a variety of neighborhoods tailored for pickup-and-delivery problems.
The moves are evaluated in random order, and any improving move is directly applied (first-improvement strategy). The \gls{LS} terminates as soon as no improving move exists.
After thorough computational analyses, we selected five neighborhoods:
\begin{itemize}[nosep]
\item[$\cN_1$] -- \textsc{Relocate Pickup}: Relocate a pickup $i \in P$ in the same route after a node $j \in V$ (located before the corresponding delivery $n+i \in D$).
\item[$\cN_2$] -- \textsc{Relocate Delivery}: Relocate a delivery $n+i \in D$ in the same route after a node $j \in V$ (located at or after the corresponding pickup $i \in P$).
\item[$\cN_3$] -- \textsc{Relocate Pair}: Relocate a p--d pair $(i, n + i)$, placing $i$ after a node $j \in V$ and placing $n + i$ no more than $\Delta$ nodes after $i$.
\item[$\cN_4$] -- \textsc{Swap Pair}: Given two pairs $(i, n + i)$ and $(j, n + j)$, swap $i$ with $j$ and $n + i$ with $n + j$.
\item[$\cN_5$] -- \textsc{Swap Ships}: Exchange the ships assigned to two different routes.
\end{itemize}

All these neighborhoods preserve the p--d pairing and precedence constraints.
Neighborhoods $\cN_1$ and $\cN_2$ specialize in intra-route modifications, while $\cN_3$ and $\cN_4$ may be used for both intra- and inter-route modifications, and they can therefore change cargo-ship allocations. Finally, to maintain a low complexity, neighborhood $\cN_3$ is limited to $\Delta \in \{0, 1, 2\}$.

\vspace{0.5em} \noindent \textbf{Efficient move evaluations.}
All the move evaluations are performed in $\cO(1)$ amortized time thanks to \emph{concatenation} strategies \citep{Vidal2014,Vidal2015b}. These strategies are based on the fact that all moves in  $\cN_1$ to  $\cN_5$ create routes that correspond to the concatenation of a \emph{constant} number of route subsequences of the current solution. Therefore, preprocessing meaningful information on subsequences of consecutive visits prior to move evaluations (as well as after each route update) can facilitate the evaluation of complex constraints and objectives.

\begin{table}[htbp]
\centering
\renewcommand{\arraystretch}{1.2}
\caption{Preprocessing and move evaluations by concatenation.}\label{tab:resources}
\setlength{\tabcolsep}{0.33cm}
\scalebox{0.87}{
\begin{tabular}{@{\hspace{0.00cm}}lll@{\hspace{0.00cm}}}
	\toprule
	Name                              & Base case                             & Induction step -- Concatenation                                                                                                             \\ \midrule
	Travel cost                       & $C_k(\sigma^0) = s^{\textsc{c}}_{ik}$ & $C_k(\sigma \oplus \sigma^{\prime}) = C_k(\sigma) + C_k(\sigma^{\prime}) + c^k_{\sigma_{n(\sigma)} \sigma^{\prime}_1}$                      \\
	Load                              & $Q_k(\sigma^0) = q_i$                 & $Q_k(\sigma \oplus \sigma^{\prime}) = Q_k(\sigma) + Q_k(\sigma^{\prime}) $                                                                  \\
	Peak load                         & $Q^\textsc{max}_k(\sigma^0) = q_i$    & $Q^\textsc{max}_k(\sigma \oplus \sigma^{\prime}) = \max \{Q^\textsc{max}_k(\sigma), Q_k(\sigma) + Q^\textsc{max}_k(\sigma^{\prime})\}$      \\
	Time warp use                     & $TW_k(\sigma^0) = 0$                  & $TW_k(\sigma \oplus \sigma^{\prime}) = TW_k(\sigma) + TW_k(\sigma^{\prime})+ \Delta^k_{TW}$                                                 \\
	Earliest possible completion time & $E_k(\sigma^0) = a_i$                 & $E_k(\sigma \oplus \sigma^{\prime}) = \max\{E_k(\sigma^{\prime}) - \Delta^k, E_k(\sigma)\} - \Delta^k_{WT}$                                 \\
	Latest feasible starting time     & $L_k(\sigma^0) = b_i$                 & $L_k(\sigma \oplus \sigma^{\prime}) = \min\{L_k(\sigma^{\prime}) - \Delta^k, L_k(\sigma)\} + \Delta^k_{TW}$                                 \\
	Duration                          & $D_k(\sigma^0) = s^{\textsc{d}}_{ik}$ & $D_k(\sigma \oplus \sigma^{\prime}) = D_k(\sigma) + D_k(\sigma^{\prime}) + \smash{\delta^k_{\sigma_{n(\sigma)} \sigma^{\prime}_1}} + \Delta^k_{WT}$ \\
	Incompatibilities                 & $I_k(\sigma^0) = I_{ik}$              & $I_k(\sigma \oplus \sigma^{\prime}) = I_k(\sigma) + I_k(\sigma^{\prime})$                                                                   \\ \midrule
	Auxiliary computations            &                                       & $\Delta^k = D_k(\sigma) - TW_k(\sigma) + \delta^k_{\sigma_{n(\sigma)} \sigma^{\prime}_1}$                                                   \\
	                                  &                                       & $\Delta^k_{WT} = \max\{E_k(\sigma^{\prime}) - \Delta^k - L_k(\sigma), 0\}$                                                                  \\
	                                  &                                       & $\Delta^k_{TW} = \max\{E_k(\sigma) + \Delta^k - L_k(\sigma^{\prime}), 0\}$                                                                  \\ \bottomrule
\end{tabular}
 }
\end{table}

The information preprocessed on subsequences for the \gls{ITSRSP} is listed in \Cref{tab:resources}. This is done by induction on the operation of concatenation $\oplus$ of two visit sequences, starting from the base case of a single node $\sigma^0 = (i)$ in the sequence. 
Note that, in contrast with other problems, this information depends on the ship type $k$, requiring $\cO(n^2 m)$ preprocessing time and space if a brute force approach is employed, since a solution contains $\cO(n^2)$ subsequences of consecutive visits and~$m$ ships. Fortunately, this complexity can be reduced to $\cO(n^2 + n m)$ time and $\cO(n^2 + m)$ space by observing that the following information is sufficient to evaluate all the moves:
\begin{itemize}[nosep,leftmargin=*]
\item 
The information on all $\cO(n^2)$ subsequences of consecutive nodes in the incumbent solution, for their current ship type (for neighborhoods $\cN_1$ to $\cN_4$) ; 
\item 
The information on each single node for all ship types, in $\cO(n m)$ (for $\cN_3$ and $\cN_4$);
\item 
The information on each sequence representing a complete route for all ship types, which can be computed in $\cO(n m)$ time and stored in $\cO(m)$  (for $\cN_5$).
\end{itemize}

Finally, to avoid redundant move evaluations, the \gls{HGS} uses a simple memory scheme that registers the \emph{last-modified} time of a route and the \emph{last-evaluated} time for each move. By comparing these values, one can decide whether or not to re-evaluate a move. This strategy is as efficient as and much simpler than the ``static move descriptors'' discussed in \cite{Zachariadis2010b}. Moreover, \emph{time} stands for any non decreasing counter, e.g., the number of moves applied or tested in the method.

\vspace{0.5em} \noindent \textbf{Ship-dependent neighborhood restrictions.}
Our \gls{LS} uses static neighborhood restrictions similar to those of \citet{Vidal2013}, by limiting move evaluations to those that create at least one directed arc $(i, j)$ with ship $k$ such that $\{i \in \{0_1, \dots, 0_k\}$ and $j \in P\}$ or $\{i \in P \cup D$ and $j \in \Gamma_k(i)\}$. 
The set $\Gamma_k(i)$ contains the $|\Gamma|$ \emph{most promising successors} of vertex $i$ for ship $k$, ranked according to a metric $\gamma_k(i, j)$ of spatial and temporal proximity between nodes $i$ and $j$:
\begin{equation}
\begin{aligned}
\gamma_k(i, j) = \gamma^\textsc{unit} c^k_{ij} &+ \gamma^\textsc{wt} \max\{a_j - s^\textsc{d}_{ik} - \delta^k_{ij} - b_i, 0\} \\ &+ \gamma^\textsc{tw} \max\{a_i + s^\textsc{d}_{ik} + \delta^k_{ij} - b_j, 0\}.
\end{aligned}
\end{equation}
The first term of the equation represents the spatial proximity (distance), scaled by the ratio between travel time and distance
$
\gamma^\textsc{unit} = \nicefrac{\delta^k_{ij}}{c^k_{ij}} 
$
to ensure that all terms have the same unit.
The next two terms measure the temporal proximity, based on the \emph{unavoidable} amount of  waiting time and time warp when servicing $i$ and $j$ consecutively, with weights $\gamma^\textsc{wt}$ and $\gamma^\textsc{tw}$.

\vspace{0.5em} \noindent \textbf{Repair. }
The routes and solutions explored in the \gls{LS} can include penalized violations of time-window, capacity, and incompatibility constraints. Therefore, this procedure may lead to an infeasible solution that will be stored in the infeasible subpopulation. In this event, a \textsc{Repair} phase is additionally called on this solution with probability $p_\textsc{rep}$. Repair temporarily multiplies all the penalty coefficients by $10$ and runs the \gls{LS}. If the resulting solution remains infeasible, then the coefficients are again multiplied, this time by $100$, and the \gls{LS} is run again. If it is successful, the resulting feasible solution will be added to the feasible subpopulation.

\subsection{Population management}
\label{subsec:population_management}

As in the \gls{UHGS}, we rely on survivor selection, population diversification, and adaptive penalty mechanisms to find a good balance between population diversity and elitism. We also incorporate an additional intensification phase, in the form of an \gls{SP} procedure that aims to build a better solution from existing routes from the search history.

\vspace{0.5em} \noindent \textbf{Initialization.}
To initialize the population, the \gls{HGS} generates $4\mu^\textsc{min}$ random individuals. Random individuals are generated as giant tours where a sequence of pickups is shuffled and then deliveries are placed immediately after the respective pickups. These individuals are educated, possibly repaired, and inserted into their respective subpopulations.

\vspace{0.5em} \noindent \textbf{Survivor selection and diversification.}
A survivor selection mechanism occurs whenever a subpopulation reaches the maximum size of $\mu^\textsc{min} + \mu^\textsc{gen}$ individuals. As in \cite{Vidal2012}, the $\mu^\textsc{gen}$ individuals with maximum biased fitness are discarded, prioritizing individuals that have a clone. This selection procedure preserves the best $\mu^\textsc{elite}$ individuals with respect to the penalized cost and finds a good balance between elitism and diversity.
To explore an even wider diversity of solution characteristics, a diversification procedure is called whenever no improving solution was found during the last $It_\textsc{div} = 0.4 \cdot It_\textsc{ni}$ iterations. It discards all but the $\mu^\textsc{min} / 3$ individuals with the smallest biased fitness in each subpopulation, and then generates $4\mu^\textsc{min}$ new random individuals that are educated and possibly repaired.

\vspace{0.5em} \noindent \textbf{Set partitioning.}
Because of the many constraints of the \gls{ITSRSP}, even generating promising feasible routes can be a challenging task. In this context, it is natural to exploit as far as possible high-quality routes from the search history. Thus, in a similar way to \cite{Muter2010} and \cite{Subramanian2013}, HGS triggers an intensification procedure whenever no improving solution has been found over $It_\textsc{sp} = 0.2 It_\textsc{ni}$ consecutive iterations. This procedure formulates an \gls{SP} model (Equations \ref{eq:sp-obj}--\ref{eq:sp-c4}) that considers all feasible routes from local minima in the past search, and solves it with a \gls{MIP} solver subject to a time limit of $T^\textsc{sp}_\textsc{max}$. Any improved solution obtained is inserted into the population. This intensification procedure complements the other operators well: instead of performing local improvements, it seeks good combinations of previously found routes and can be viewed as a form of large neighborhood search.

\vspace{0.5em} \noindent \textbf{Adaptive penalties.}
Finally, the penalty coefficients are adjusted during the search to maintain a target proportion $\xi^\textsc{ref}$ of feasible individuals with respect to the relaxed constraints (load, time windows, and incompatibilities). For each constraint $X$, HGS records the proportion~$\xi^X$ of feasible solutions w.r.t. constraint $X$ over the last $100$ iterations. This calculation is performed after the \gls{LS}, before any possible repair. The penalty coefficients are then adjusted every $It_\textsc{ni} / 100$ iterations: if $\xi^{X} \leq \xi^\textsc{ref} - 5\%$, $\xi^{X}$ is multiplied by $1.2$, and if $\xi^{X} \geq \xi^\textsc{ref} + 5\%$, $\xi^{X}$~is~multiplied~by~$0.85$.

\section{Experimental Analysis}
\label{sec:experiments}

This section reports our computational experiments with the two \gls{BnP} algorithms (with and without enumeration) and the \gls{HGS}. Our aim is fourfold:
\begin{itemize}[nosep,leftmargin=*]
\item We compare the proposed exact algorithms, and evaluate their ability to solve practical-size ship routing problems to optimality in limited time; 
\item In situations where a faster response is sought, we evaluate the quality of the solutions produced by the HGS metaheuristic;
\item We evaluate the impact of some of our most important methodological choices and new components: e.g., ship-dependent neighborhood restrictions and set-partitioning problem parameters for the HGS; advanced preprocessing techniques, DSSR, and completion bounds for the \gls{BnP} algorithms;
\item Finally, we evaluate the scalability of the proposed approaches on new larger problem instances.
\end{itemize}

We implemented all algorithms in C++ with double precision numbers, using CPLEX 12.7 to solve the \gls{BnP} master problem and the integer \gls{SP} inside the \gls{HGS}. We conducted all experiments on a computer with an i7-3960X CPU and 64 GB of RAM.

We rely on the benchmark instances suite for the \gls{ITSRSP} based on real-life scenarios, presented in \citet{Hemmati2014} and currently available at \url{http://home.himolde.no/~hvattum/benchmarks/}.
These instances are divided into four groups of $60$, according to problem topology and cargo type: short sea mixed load (SS\_MUN), short sea full load (SS\_FUN), deep sea mixed load (DS\_MUN), and deep sea full load (DS\_FUN). Each group contains five instances for 12 different problem size values. The mixed load instances have up to $130$ cargoes and $40$ ships, whereas the full load instances have up to $100$ cargoes and $50$ ships. So far, $123/240$ instances remain open. Finally, we produced $32$ additional larger instances with up to $260$ cargoes and $74$ ships for our scalability experiment.

In the full load instances, each delivery should be visited immediately after its associated pickup due to the absence of residual capacity for other loads. This property is no longer valid for the mixed load instances. 
Furthermore, short and deep sea instances consider different geographical regions. The short sea instances represent shipments among European ports, whereas the deep sea instances involve long-distance shipments between different continents.

\subsection{Exact solutions}
\label{sec:bp_results}

To our knowledge, the study of \citet{Hemmati2014} is the only one to report lower bounds and optimal solutions for these benchmark instances, obtained by solving a \gls{MIP} formulation. We establish a comparison with our branch-and-price without route enumeration (B\&P\textsubscript{1}), as well as the same algorithm with route enumeration and \glspl{3-SRC} using inspection pricing (B\&P\textsubscript{2}). Both configurations use heuristic strong branching with 50 candidates at each iteration. As an initial upper bound, we set $Z_\textsc{ub} = Z_\textsc{hgs}+0.1$, where $Z_\textsc{hgs}$ is the best objective value found by our HGS metaheuristic. One hour of computation was allowed for each instance.

\Cref{tbl:exact} reports the experimental results. Each line gives the average results for five instances with the same characteristics. The first group of columns presents the results of \mbox{\citet{Hemmati2014}}: the time in minutes, the gap between the integer solution and the best bound found in the \gls{MIP} formulation, and the number of instances solved to optimality. The second group presents the results for B\&P\textsubscript{1}: ``Gap\textsubscript{0}'' and ``T\textsubscript{0}'' represent the percentage gap and the time in minutes for the root node. ``Gap\textsubscript{F}'' and ``T\textsubscript{F}'' are the final percentage gap and time, ``N\textsubscript{F}'' is the number of nodes in the search tree, and ``Opt'' is the number of instances solved to optimality. The last group of columns presents the results for B\&P\textsubscript{2} (with route enumeration and inspection pricing): the columns ``T\textsubscript{E}'' and ``R\textsubscript{E}'' are the time in minutes for the route enumeration and the number of routes found. ``Gap\textsubscript{0}'', ``T\textsubscript{0}'', and ``Cuts\textsubscript{0}'' are the percentage gap, the overall time and the number of \glspl{3-SRC} separated at the root node. ``R\textsubscript{F}'', ``Gap\textsubscript{F}'', ``T\textsubscript{F}'', ``Cuts\textsubscript{F}'', and ``N\textsubscript{F}'' are the final number of routes, the percentage gap, the overall time, the \glspl{3-SRC} separated, and the number of nodes in the search tree. Finally, ``Opt'' is the number of instances solved to optimality in the group.

\begin{table}[htbp]
	\renewcommand{\arraystretch}{1.05}
        \setlength\tabcolsep{4pt}
	\caption{Performance comparison -- Exact approaches}\label{tbl:exact}
  \hspace*{-0.7cm}
	\scalebox{0.65}{
		\begin{tabular}{c@{\hspace*{0.6cm}}l@{\hspace*{0.6cm}}ccccccccccccccccccccccc}\toprule
& & \multicolumn{3}{c}{\cite{Hemmati2014}} && \multicolumn{6}{c}{Branch-and-Price (B\&P\textsubscript{1})} && \multicolumn{11}{c}{Branch-and-Price + Enumeration + SRCs (B\&P\textsubscript{2})} \\\cline{3-5}\cline{7-12}\cline{14-24}
& Instances & T & Gap & Opt && Gap\textsubscript{0} & T\textsubscript{0} & Gap\textsubscript{F} & T\textsubscript{F} & N\textsubscript{F} & Opt && T\textsubscript{E} & R\textsubscript{E} & Gap\textsubscript{0} & T\textsubscript{0} &Cuts\textsubscript{0} & R\textsubscript{F} & Gap\textsubscript{F} & T\textsubscript{F} &  Cuts\textsubscript{F} & N\textsubscript{F} & Opt \\\midrule
\parbox[t]{2mm}{\multirow{12}{*}{\rotatebox[origin=c]{90}{{\large SS-MUN}}}} & C7-V3 & 0.0 & 0.00 & 5 && 0.00 & 0.00 & 0.00 & 0.00 & 1.0 & 5 && 0.00 & 9 & 0.00 & 0.00 & 0.0 & 9 & 0.00 & 0.00 & 0.0 & 1.0 & 5 \\
& C10-V3 & 0.0 & 0.00 & 5 && 0.00 & 0.00 & 0.00 & 0.00 & 1.0 & 5 && 0.00 & 13 & 0.00 & 0.00 & 0.0 & 13 & 0.00 & 0.00 & 0.0 & 1.0 & 5 \\
& C15-V4 & 1.4 & 0.00 & 5 && 1.07 & 0.00 & 0.00 & 0.00 & 3.4 & 5 && 0.00 & 75 & 0.80 & 0.00 & 3.2 & 64 & 0.00 & 0.00 & 4.6 & 1.8 & 5 \\
& C18-V5 & 42.5 & 3.12 & 4 && 0.56 & 0.00 & 0.00 & 0.00 & 5.4 & 5 && 0.00 & 83 & 0.16 & 0.00 & 2.2 & 40 & 0.00 & 0.00 & 2.2 & 1.8 & 5 \\
& C22-V6 & 50.6 & 15.47 & 1 && 1.40 & 0.00 & 0.00 & 0.01 & 6.6 & 5 && 0.00 & 468 & 0.61 & 0.00 & 13.8 & 242 & 0.00 & 0.00 & 17.0 & 4.2 & 5 \\
& C23-V13 & 60.0 & 26.88 & 0 && 0.41 & 0.01 & 0.00 & 0.05 & 7.8 & 5 && 0.00 & 180 & 0.17 & 0.01 & 5.8 & 74 & 0.00 & 0.01 & 5.8 & 3.0 & 5 \\
& C30-V6 & 60.0 & 79.46 & 0 && 0.73 & 0.01 & 0.00 & 0.11 & 15.0 & 5 && 0.00 & 913 & 0.36 & 0.01 & 25.8 & 290 & 0.00 & 0.01 & 31.4 & 2.6 & 5 \\
& C35-V7 & 60.0 & 82.66 & 0 && 0.66 & 0.03 & 0.00 & 0.36 & 23.0 & 5 && 0.00 & 2K & 0.38 & 0.03 & 42.4 & 850 & 0.00 & 0.07 & 78.6 & 7.0 & 5 \\
& C60-V13 & 60.0 & 85.48 & 0 && 0.43 & 0.49 & 0.00 & 11.82 & 96.6 & 5 && 0.07 & 28K & 0.29 & 0.48 & 57.2 & 13K & 0.00 & 0.91 & 176.0 & 20.6 & 5 \\
& C80-V20 & 60.0 & 86.63 & 0 && 0.19 & 1.20 & 0.00 & 9.06 & 17.4 & 5 && 0.19 & 35K & 0.08 & 1.15 & 51.2 & 12K & 0.00 & 1.23 & 79.0 & 5.4 & 5 \\
& C100-V30 & 60.2 & 97.59 & 0 && 0.15 & 2.88 & 0.02 & 36.02 & 40.0 & 3 && 0.47 & 23K & 0.10 & 2.81 & 34.4 & 9K & 0.00 & 3.27 & 70.0 & 19.0 & 5 \\
& C130-V40 & 61.5 & 99.95 & 0 && 0.13 & 12.68 & 0.04 & 60.00 & 18.4 & 0 && 2.82 & 111K & 0.07 & 13.44 & 41.8 & 36K & 0.00 & 15.08 & 115.6 & 23.8 & 5 \\\midrule
& Overall & 43.0 & 48.10 & 20 && 0.48 & 1.44 & 0.01 & 9.79 & 19.6 & 53 && 0.30 & 17K & 0.25 & 1.49 & 23.2 & 6K & 0.00 & 1.71 & 48.4 & 7.6 & 60 \\\midrule
\parbox[t]{2mm}{\multirow{12}{*}{\rotatebox[origin=c]{90}{{\large SS-FUN}}}} & C8-V3 & 0.0 & 0.00 & 5 && 0.00 & 0.00 & 0.00 & 0.00 & 1.0 & 5 && 0.00 & 10 & 0.00 & 0.00 & 0.0 & 10 & 0.00 & 0.00 & 0.0 & 1.0 & 5 \\
& C11-V4 & 0.0 & 0.00 & 5 && 0.00 & 0.00 & 0.00 & 0.00 & 1.0 & 5 && 0.00 & 19 & 0.00 & 0.00 & 0.0 & 19 & 0.00 & 0.00 & 0.0 & 1.0 & 5 \\
& C13-V5 & 0.0 & 0.00 & 5 && 0.00 & 0.00 & 0.00 & 0.00 & 1.0 & 5 && 0.00 & 21 & 0.00 & 0.00 & 0.0 & 21 & 0.00 & 0.00 & 0.0 & 1.0 & 5 \\
& C16-V6 & 0.0 & 0.00 & 5 && 0.00 & 0.00 & 0.00 & 0.00 & 1.0 & 5 && 0.00 & 34 & 0.00 & 0.00 & 0.0 & 34 & 0.00 & 0.00 & 0.0 & 1.0 & 5 \\
& C17-V13 & 0.0 & 0.00 & 5 && 0.01 & 0.00 & 0.00 & 0.00 & 1.8 & 5 && 0.00 & 49 & 0.01 & 0.00 & 0.0 & 49 & 0.00 & 0.00 & 0.0 & 1.8 & 5 \\
& C20-V6 & 0.8 & 0.00 & 5 && 0.03 & 0.00 & 0.00 & 0.00 & 1.8 & 5 && 0.00 & 81 & 0.00 & 0.00 & 2.6 & 76 & 0.00 & 0.00 & 5.0 & 1.4 & 5 \\
& C25-V7 & 40.5 & 0.83 & 3 && 0.00 & 0.00 & 0.00 & 0.00 & 1.4 & 5 && 0.00 & 67 & 0.00 & 0.00 & 0.8 & 62 & 0.00 & 0.00 & 0.8 & 1.0 & 5 \\
& C35-V13 & 60.0 & 8.85 & 0 && 0.00 & 0.00 & 0.00 & 0.00 & 1.4 & 5 && 0.00 & 131 & 0.00 & 0.00 & 1.4 & 130 & 0.00 & 0.00 & 1.4 & 1.4 & 5 \\
& C50-V20 & 60.0 & 13.99 & 0 && 0.03 & 0.00 & 0.00 & 0.01 & 1.8 & 5 && 0.00 & 671 & 0.03 & 0.00 & 0.0 & 671 & 0.00 & 0.01 & 0.0 & 1.8 & 5 \\
& C70-V30 & 60.1 & 60.04 & 0 && 0.12 & 0.01 & 0.00 & 0.08 & 4.6 & 5 && 0.00 & 7K & 0.12 & 0.01 & 0.0 & 7K & 0.00 & 0.04 & 3.0 & 4.6 & 5 \\
& C90-V40 & 60.3 & 78.32 & 0 && 0.00 & 0.03 & 0.00 & 0.11 & 2.2 & 5 && 0.00 & 1K & 0.00 & 0.03 & 1.0 & 1K & 0.00 & 0.04 & 1.0 & 1.8 & 5 \\
& C100-V50 & 60.9 & 79.11 & 0 && 0.01 & 0.04 & 0.00 & 0.24 & 3.8 & 5 && 0.00 & 3K & 0.01 & 0.04 & 2.0 & 2K & 0.00 & 0.10 & 2.0 & 3.4 & 5 \\\midrule
& Overall & 28.5 & 20.10 & 33 && 0.02 & 0.01 & 0.00 & 0.04 & 1.9 & 60 && 0.00 & 1K & 0.02 & 0.01 & 0.7 & 978 & 0.00 & 0.02 & 1.1 & 1.8 & 60 \\\midrule
\parbox[t]{2mm}{\multirow{12}{*}{\rotatebox[origin=c]{90}{{\large DS-MUN}}}} & C7-V3 & 0.0 & 0.00 & 5 && 1.15 & 0.00 & 0.00 & 0.00 & 1.4 & 5 && 0.00 & 12 & 0.00 & 0.00 & 0.6 & 8 & 0.00 & 0.00 & 0.6 & 1.0 & 5 \\
& C10-V3 & 0.0 & 0.00 & 5 && 1.92 & 0.00 & 0.00 & 0.00 & 3.4 & 5 && 0.00 & 30 & 1.81 & 0.00 & 1.0 & 28 & 0.00 & 0.00 & 1.0 & 1.8 & 5 \\
& C15-V4 & 0.4 & 0.00 & 5 && 1.02 & 0.00 & 0.00 & 0.00 & 1.8 & 5 && 0.00 & 80 & 0.00 & 0.00 & 1.4 & 22 & 0.00 & 0.00 & 1.4 & 1.0 & 5 \\
& C18-V5 & 16.3 & 2.18 & 4 && 0.51 & 0.00 & 0.00 & 0.00 & 2.6 & 5 && 0.00 & 97 & 0.25 & 0.00 & 5.0 & 78 & 0.00 & 0.00 & 5.0 & 1.4 & 5 \\
& C22-V6 & 26.2 & 3.28 & 4 && 1.45 & 0.00 & 0.00 & 0.00 & 5.8 & 5 && 0.00 & 154 & 1.01 & 0.00 & 2.8 & 108 & 0.00 & 0.00 & 3.4 & 2.6 & 5 \\
& C23-V13 & 26.4 & 4.62 & 3 && 0.27 & 0.00 & 0.00 & 0.00 & 2.6 & 5 && 0.00 & 79 & 0.13 & 0.00 & 0.4 & 55 & 0.00 & 0.00 & 0.4 & 1.8 & 5 \\
& C30-V6 & 60.0 & 52.25 & 0 && 1.56 & 0.00 & 0.00 & 0.04 & 15.0 & 5 && 0.00 & 4K & 0.84 & 0.00 & 16.0 & 1K & 0.00 & 0.01 & 30.6 & 3.0 & 5 \\
& C35-V7 & 60.0 & 54.25 & 0 && 1.13 & 0.01 & 0.00 & 0.06 & 14.6 & 5 && 0.00 & 8K & 0.74 & 0.01 & 20.6 & 5K & 0.00 & 0.01 & 25.6 & 3.4 & 5 \\
& C60-V13 & 60.0 & 89.22 & 0 && 0.58 & 0.10 & 0.00 & 3.62 & 45.4 & 5 && 0.02 & 33K & 0.14 & 0.10 & 32.8 & 5K & 0.00 & 0.70 & 88.2 & 9.4 & 5 \\
& C80-V20 & 60.1 & 91.56 & 0 && 0.43 & 0.37 & 0.00 & 12.57 & 49.4 & 5 && 0.10 & 231K & 0.11 & 0.42 & 43.4 & 30K & 0.00 & 0.59 & 64.0 & 7.8 & 5 \\
& C100-V30 & 60.3 & 99.03 & 0 && 0.52 & 0.57 & 0.09 & 20.72 & 63.4 & 4 && 0.21 & 1M & 0.24 & 0.70 & 32.0 & 395K & 0.00 & 2.13 & 99.4 & 22.2 & 5 \\
& C130-V40 & 61.3 & 100.00 & 0 && 0.41 & 3.81 & 0.16 & 60.00 & 30.0 & 0 && 4.86 & 13M & 0.28 & 9.07 & 52.6 & 5M & 0.01 & 17.64 & 168.0 & 19.2 & 4 \\\midrule
& Overall & 35.9 & 41.37 & 26 && 0.91 & 0.40 & 0.02 & 8.08 & 19.6 & 54 && 0.43 & 1M & 0.46 & 0.86 & 17.4 & 478K & 0.00 & 1.76 & 40.6 & 6.2 & 59 \\\midrule
\parbox[t]{2mm}{\multirow{12}{*}{\rotatebox[origin=c]{90}{{\large DS-FUN}}}} & C8-V3 & 0.0 & 0.00 & 5 && 0.00 & 0.00 & 0.00 & 0.00 & 1.0 & 5 && 0.00 & 10 & 0.00 & 0.00 & 0.0 & 10 & 0.00 & 0.00 & 0.0 & 1.0 & 5 \\
& C11-V4 & 0.0 & 0.00 & 5 && 0.00 & 0.00 & 0.00 & 0.00 & 1.0 & 5 && 0.00 & 14 & 0.00 & 0.00 & 0.0 & 14 & 0.00 & 0.00 & 0.0 & 1.0 & 5 \\
& C13-V5 & 0.0 & 0.00 & 5 && 0.00 & 0.00 & 0.00 & 0.00 & 1.0 & 5 && 0.00 & 18 & 0.00 & 0.00 & 0.0 & 18 & 0.00 & 0.00 & 0.0 & 1.0 & 5 \\
& C16-V6 & 0.0 & 0.00 & 5 && 0.01 & 0.00 & 0.00 & 0.00 & 1.4 & 5 && 0.00 & 28 & 0.01 & 0.00 & 0.0 & 28 & 0.00 & 0.00 & 0.0 & 1.4 & 5 \\
& C17-V13 & 0.0 & 0.00 & 5 && 0.02 & 0.00 & 0.00 & 0.00 & 1.4 & 5 && 0.00 & 34 & 0.02 & 0.00 & 0.0 & 34 & 0.00 & 0.00 & 0.0 & 1.4 & 5 \\
& C20-V6 & 0.0 & 0.00 & 5 && 0.11 & 0.00 & 0.00 & 0.00 & 1.4 & 5 && 0.00 & 67 & 0.11 & 0.00 & 0.0 & 67 & 0.00 & 0.00 & 0.0 & 1.4 & 5 \\
& C25-V7 & 0.1 & 0.00 & 5 && 0.00 & 0.00 & 0.00 & 0.00 & 1.0 & 5 && 0.00 & 80 & 0.00 & 0.00 & 0.0 & 80 & 0.00 & 0.00 & 0.0 & 1.0 & 5 \\
& C35-V13 & 45.6 & 4.37 & 2 && 0.03 & 0.00 & 0.00 & 0.00 & 1.8 & 5 && 0.00 & 145 & 0.03 & 0.00 & 0.0 & 145 & 0.00 & 0.00 & 0.0 & 1.8 & 5 \\
& C50-V20 & 56.3 & 8.81 & 1 && 0.00 & 0.00 & 0.00 & 0.00 & 1.0 & 5 && 0.00 & 182 & 0.00 & 0.00 & 0.0 & 182 & 0.00 & 0.00 & 0.0 & 1.0 & 5 \\
& C70-V30 & 60.1 & 11.37 & 0 && 0.00 & 0.01 & 0.00 & 0.01 & 1.4 & 5 && 0.00 & 682 & 0.00 & 0.01 & 0.4 & 682 & 0.00 & 0.01 & 0.4 & 1.4 & 5 \\
& C90-V40 & 60.2 & 48.44 & 0 && 0.00 & 0.02 & 0.00 & 0.04 & 1.8 & 5 && 0.01 & 1K & 0.00 & 0.02 & 1.4 & 1K & 0.00 & 0.03 & 5.2 & 1.8 & 5 \\
& C100-V50 & 60.5 & 52.36 & 0 && 0.00 & 0.02 & 0.00 & 0.07 & 1.8 & 5 && 0.01 & 682 & 0.00 & 0.02 & 0.6 & 651 & 0.00 & 0.04 & 1.0 & 1.4 & 5 \\\midrule
& Overall & 23.6 & 10.45 & 38 && 0.01 & 0.00 & 0.00 & 0.01 & 1.3 & 60 && 0.00 & 267 & 0.01 & 0.00 & 0.2 & 265 & 0.00 & 0.01 & 0.6 & 1.3 & 60 \\\bottomrule
		\end{tabular}
	}
\end{table}

As observed in \Cref{tbl:exact}, the \gls{MIP} formulation already fails to produce optimal solutions on some small instances with 20 to 30 cargoes.
In contrast, B\&P\textsubscript{1} solves all the full load instances in less than one minute, as well as 107 out of the 120 mixed load instances.
The column generation produces good lower bounds, with an average gap of $0.36\%$ at the root node.
However, for the mixed load instances, the average time needed to solve each pricing subproblem (ratio T\textsubscript{F}/N\textsubscript{F}) increases with the number of cargoes, whereas the lower bounds tend to deteriorate, leading to larger branch-and-bound trees. To go further, one can concentrate on improving the pricing problem solution or the lower bounds. 
For the largest open instances, the root-node gap seems small enough to allow a complete route enumeration via sophisticated DP algorithms (\Cref{sec:enumeration}). We therefore derived B\&P\textsubscript{2} from this premise: the enumeration of the routes allows subsequent pricing by inspection and permits to introduce SRCs without significant consequences on CPU time.

The remaining columns of \Cref{tbl:exact} report the performance of B\&P\textsubscript{2}. This approach outperforms the standard B\&P\textsubscript{1} in terms of number of instances solved and average solution time.
Route enumeration at the root node takes a maximum of 17.8 minutes. The number of routes may, however, rise up to 60 millions (on instance DS-MUN-C130-V40-HE-2).
Having enumerated the routes allows to efficiently separate the SRCs and decrease the root-node gap (from 0.36\% to 0.19\%). As a consequence, the route set can be reduced further (from 311K to 121K on average) as well as the size of the branch-and-bound tree (from 10.6 to 4.2 on average). Using this method, all available instances but one (DS-MUN-C130-V40-HE-2) are solved within a time limit of one hour. A larger time limit allows to solve the last remaining instance in 4 hours and 23 minutes. 

\begin{figure}[htbp]
\centering
\caption{Number of instances solved over time by B\&P\textsubscript{1} and B\&P\textsubscript{2}}\label{fig:exact-figure}
\includegraphics[width=0.55\linewidth]{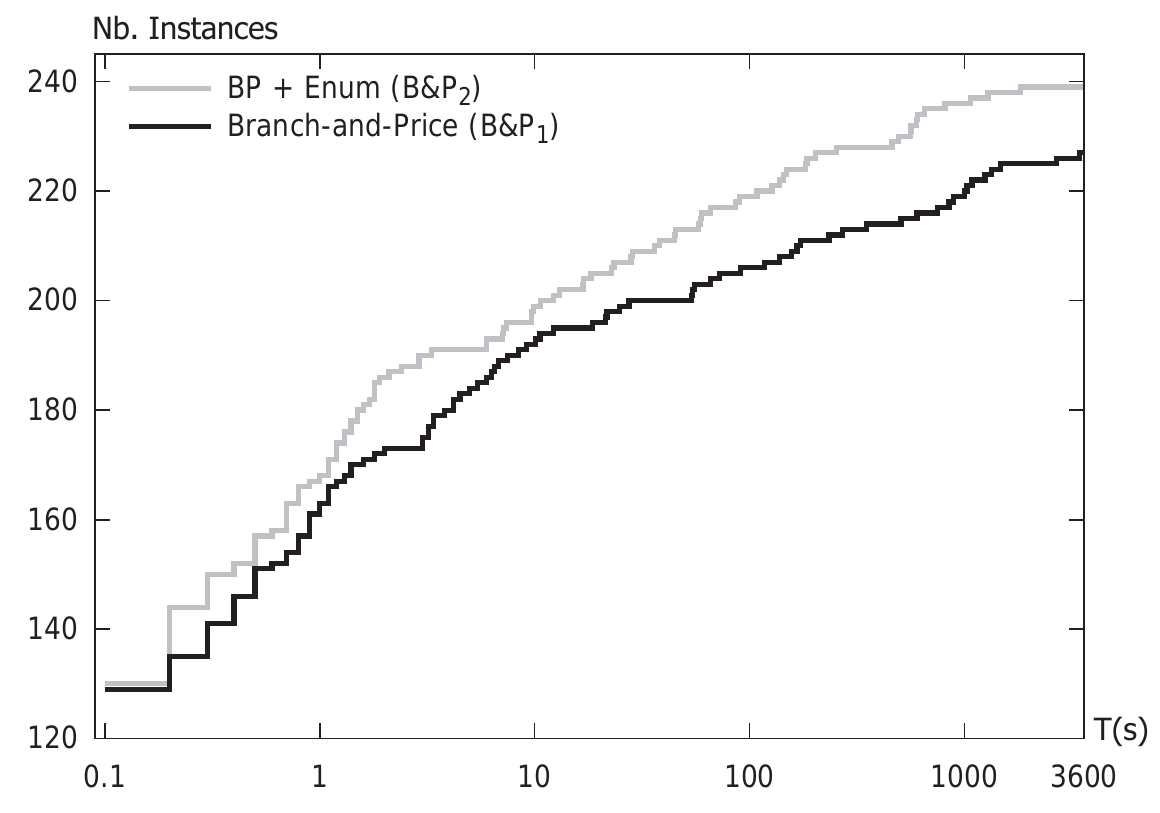}
\end{figure}

\Cref{fig:exact-figure} displays the number of instances solved by B\&P\textsubscript{1} and B\&P\textsubscript{2} as a function of the CPU time limit. B\&P\textsubscript{2} visibly produces superior results, but this method is also less flexible: its performance depends on the ability to do a complete route enumeration at the root node within the optimality gap. This process may possibly fail (due to time or memory limitations, or an unusually large gap), such that we recommend to first consider B\&P\textsubscript{1} in exploratory analyses without any prior knowledge of the structure of the \gls{ITSRSP} instances.

\subsection{Heuristic solutions}
\label{sec:metaheuristic-perf}

As seen in the previous section, our exact methods can solve the majority of the instances, but their CPU time can widely vary, even for instances with similar characteristics.
Therefore, fast heuristic solutions remain essential for applications requiring a response in a guaranteed short time. This section compares the performance of our \gls{HGS} with that of the existing \gls{ITSRSP} heuristics from \citet{Hemmati2014} and \citet{Hemmati2016}. To highlight the impact of the \gls{SP} component, we evaluated two versions of our algorithm: without \gls{SP} (HGS\textsubscript{1}) and with \gls{SP} (HGS\textsubscript{2}).

We used the same parameters as \citet{Vidal2013} to avoid any problem-specific overfit. Therefore, $(\mu^\textsc{min}, \mu^\textsc{gen}) = (25, 40)$, $\mu^\textsc{elite} = 10$, $\mu^\textsc{close} = 5$, $p_\textsc{rep} = 0.5$, $\xi^\textsc{ref} = 0.2$, $|\Gamma| = 30$, and $(\gamma^\textsc{tw}, \gamma^\textsc{wt}) = (1.0, 0.2)$. The penalty coefficients take $(\omega^\textsc{q}, \omega^\textsc{tw}, \omega^{I}) = (\bar{c} / \bar{q}, 100, \bar{c})$ as a starting value, where $\bar{c} = \frac{\sum_{k \in K} \sum_{(i, j) \in A} c^k_{ij}}{|K| |A|}$ and $\bar{q} = \frac{\sum_{i \in V} |q_i|}{n}$. To compare with previous literature in similar CPU time, we set $(It_\textsc{ni}, T_\textsc{max}) = (2.5 \cdot 10^3, 15\,\text{min})$, and $T^\textsc{sp}_\textsc{max} = 30\,\text{s}$. Finally, the full load instances (FUN) require directs visits from pickup to delivery points, such that neighborhoods $\cN_1$ and $\cN_2$ can be disabled and neighborhood $\cN_3$ is restricted to $\Delta = 0$. In the neighborhood restrictions, $i \in D$ is the only possible candidate successor of $i - n$, and only a pickup node $j \in P$ can follow a delivery~$i \in D$.

\Cref{tab:heuristic} compares the results of the ALNS of \citet{Hemmati2014} with those of HGS\textsubscript{1} and HGS\textsubscript{2}. Each line gives average results over 10 runs for five instances with the same characteristics. Each column ``Gap'' reports the percentage gap of one method relative to the \emph{optimal} solutions found in \Cref{sec:bp_results}, column ``T'' gives the total CPU time in minutes, and column ``T*'' gives the \emph{attainment} time (in minutes) needed to reach the final solution. For each instance class, the best gap is highlighted in boldface. The detailed results of \citet{Hemmati2014} were kindly provided by the authors, and our complete detailed results (per instance) are also available in the electronic companion of this paper, located at \url{https://w1.cirrelt.ca/~vidalt/en/VRP-resources.html}.

\begin{table}[htbp]
	\vspace*{-0.5cm}
	\renewcommand{\arraystretch}{1.05}
	\setlength\tabcolsep{6pt}
	\centering
	\caption{Performance comparison -- Metaheuristics}\label{tab:heuristic}
	\scalebox{0.7}{
		\begin{tabular}{cc@{\hspace*{1cm}}ccc@{\hspace*{1cm}}cccc@{\hspace*{1cm}}ccc}
			\hline
			                                                                             &           & \multicolumn{2}{c}{\hspace*{-0.8cm}\cite{Hemmati2014}} &  & \multicolumn{3}{c}{HGS\textsubscript{1}} &  & \multicolumn{3}{c}{HGS\textsubscript{2}} \\
			                                                                             & Instances &      Gap      &                   T                    &  &      Gap      &   T   &        T*        &  &      Gap      &   T   &        T*        \\ \midrule
			\parbox[t]{2mm}{\multirow{12}{*}{\rotatebox[origin=c]{90}{{\large SS-MUN}}}} &   C7-V3   & \textbf{0.00} &                  0.03                  &  & \textbf{0.00} & 0.02  &       0.00       &  & \textbf{0.00} & 0.02  &       0.00       \\
			                                                                             &  C10-V3   & \textbf{0.00} &                  0.04                  &  & \textbf{0.00} & 0.04  &       0.00       &  & \textbf{0.00} & 0.04  &       0.00       \\
			                                                                             &  C15-V4   &     0.58      &                  0.09                  &  & \textbf{0.00} & 0.08  &       0.00       &  & \textbf{0.00} & 0.08  &       0.00       \\
			                                                                             &  C18-V5   &     0.51      &                  0.13                  &  &     0.07      & 0.12  &       0.02       &  & \textbf{0.02} & 0.13  &       0.03       \\
			                                                                             &  C22-V6   &     1.82      &                  0.19                  &  &     0.05      & 0.19  &       0.05       &  & \textbf{0.00} & 0.18  &       0.04       \\
			                                                                             &  C23-V13  &     0.58      &                  0.25                  &  &     0.13      & 0.25  &       0.09       &  & \textbf{0.00} & 0.23  &       0.07       \\
			                                                                             &  C30-V6   &     1.60      &                  0.38                  &  &     0.17      & 0.36  &       0.17       &  & \textbf{0.00} & 0.33  &       0.12       \\
			                                                                             &  C35-V7   &     1.92      &                  0.54                  &  &     0.81      & 0.51  &       0.26       &  & \textbf{0.03} & 0.54  &       0.27       \\
			                                                                             &  C60-V13  &     1.69      &                  2.01                  &  &     1.03      & 2.44  &       1.81       &  & \textbf{0.14} & 2.25  &       1.37       \\
			                                                                             &  C80-V20  &     2.45      &                  4.13                  &  &     1.79      & 5.21  &       4.09       &  & \textbf{0.03} & 3.84  &       2.44       \\
			                                                                             & C100-V30  &     2.68      &                  7.77                  &  &     1.67      & 9.97  &       8.01       &  & \textbf{0.01} & 5.31  &       3.02       \\
			                                                                             & C130-V40  &     2.57      &                 16.95                  &  &     2.28      & 13.84 &      11.87       &  & \textbf{0.02} & 12.43 &       8.39       \\ \midrule
			                                                                             &  Overall  &     1.37      &                  2.71                  &  &     0.67      & 2.75  &       2.20       &  & \textbf{0.02} & 2.12  &       1.31       \\ \midrule
			\parbox[t]{2mm}{\multirow{12}{*}{\rotatebox[origin=c]{90}{{\large SS-FUN}}}} &   C8-V3   & \textbf{0.00} &                  0.03                  &  & \textbf{0.00} & 0.01  &       0.00       &  & \textbf{0.00} & 0.01  &       0.00       \\
			                                                                             &  C11-V4   &     0.13      &                  0.05                  &  & \textbf{0.00} & 0.02  &       0.00       &  & \textbf{0.00} & 0.02  &       0.00       \\
			                                                                             &  C13-V5   &     0.07      &                  0.07                  &  & \textbf{0.00} & 0.02  &       0.00       &  & \textbf{0.00} & 0.02  &       0.00       \\
			                                                                             &  C16-V6   &     0.05      &                  0.10                  &  & \textbf{0.00} & 0.03  &       0.00       &  & \textbf{0.00} & 0.03  &       0.00       \\
			                                                                             &  C17-V13  &     0.01      &                  0.14                  &  & \textbf{0.00} & 0.04  &       0.00       &  & \textbf{0.00} & 0.05  &       0.00       \\
			                                                                             &  C20-V6   &     0.14      &                  0.18                  &  & \textbf{0.00} & 0.04  &       0.00       &  & \textbf{0.00} & 0.05  &       0.00       \\
			                                                                             &  C25-V7   &     0.22      &                  0.27                  &  & \textbf{0.00} & 0.07  &       0.02       &  & \textbf{0.00} & 0.07  &       0.01       \\
			                                                                             &  C35-V13  &     0.29      &                  0.60                  &  &     0.01      & 0.19  &       0.08       &  & \textbf{0.00} & 0.17  &       0.05       \\
			                                                                             &  C50-V20  &     0.52      &                  1.38                  &  &     0.22      & 0.67  &       0.43       &  & \textbf{0.00} & 0.43  &       0.15       \\
			                                                                             &  C70-V30  &     1.29      &                  3.51                  &  &     0.68      & 1.86  &       1.28       &  & \textbf{0.00} & 1.09  &       0.42       \\
			                                                                             &  C90-V40  &     1.45      &                  6.98                  &  &     0.70      & 3.90  &       2.83       &  & \textbf{0.00} & 1.93  &       0.76       \\
			                                                                             & C100-V50  &     0.96      &                  9.79                  &  &     0.50      & 5.98  &       4.46       &  & \textbf{0.00} & 2.83  &       1.20       \\ \midrule
			                                                                             &  Overall  &     0.43      &                  1.93                  &  &     0.18      & 1.07  &       0.76       &  & \textbf{0.00} & 0.56  &       0.22       \\ \midrule
			\parbox[t]{2mm}{\multirow{12}{*}{\rotatebox[origin=c]{90}{{\large DS-MUN}}}} &   C7-V3   & \textbf{0.00} &                  0.03                  &  & \textbf{0.00} & 0.02  &       0.00       &  & \textbf{0.00} & 0.02  &       0.00       \\
			                                                                             &  C10-V3   &     0.01      &                  0.04                  &  & \textbf{0.00} & 0.04  &       0.00       &  & \textbf{0.00} & 0.04  &       0.00       \\
			                                                                             &  C15-V4   &     1.26      &                  0.08                  &  & \textbf{0.00} & 0.08  &       0.01       &  & \textbf{0.00} & 0.08  &       0.01       \\
			                                                                             &  C18-V5   &     0.47      &                  0.13                  &  & \textbf{0.00} & 0.12  &       0.01       &  & \textbf{0.00} & 0.12  &       0.01       \\
			                                                                             &  C22-V6   &     2.18      &                  0.19                  &  &     0.01      & 0.18  &       0.05       &  & \textbf{0.00} & 0.18  &       0.05       \\
			                                                                             &  C23-V13  &     0.12      &                  0.24                  &  &     0.02      & 0.19  &       0.05       &  & \textbf{0.00} & 0.20  &       0.04       \\
			                                                                             &  C30-V6   &     1.04      &                  0.37                  &  &     0.22      & 0.36  &       0.17       &  & \textbf{0.03} & 0.32  &       0.12       \\
			                                                                             &  C35-V7   &     1.08      &                  0.51                  &  &     0.14      & 0.53  &       0.28       &  & \textbf{0.00} & 0.44  &       0.17       \\
			                                                                             &  C60-V13  &     3.74      &                  1.92                  &  &     1.41      & 2.58  &       1.96       &  & \textbf{0.03} & 1.81  &       1.06       \\
			                                                                             &  C80-V20  &     3.10      &                  4.26                  &  &     1.61      & 5.69  &       4.58       &  & \textbf{0.04} & 3.56  &       2.20       \\
			                                                                             & C100-V30  &     3.69      &                  8.00                  &  &     3.27      & 9.22  &       7.27       &  & \textbf{0.01} & 9.87  &       6.84       \\
			                                                                             & C130-V40  &     5.18      &                 17.47                  &  &     5.08      & 13.66 &      11.64       &  & \textbf{0.09} & 14.81 &      12.88       \\ \midrule
			                                                                             &  Overall  &     1.82      &                  2.77                  &  &     0.98      & 2.72  &       2.17       &  & \textbf{0.02} & 2.62  &       1.95       \\ \midrule
			\parbox[t]{2mm}{\multirow{12}{*}{\rotatebox[origin=c]{90}{{\large DS-FUN}}}} &   C8-V3   & \textbf{0.00} &                  0.03                  &  & \textbf{0.00} & 0.01  &       0.00       &  & \textbf{0.00} & 0.01  &       0.00       \\
			                                                                             &  C11-V4   & \textbf{0.00} &                  0.05                  &  & \textbf{0.00} & 0.02  &       0.00       &  & \textbf{0.00} & 0.02  &       0.00       \\
			                                                                             &  C13-V5   & \textbf{0.00} &                  0.06                  &  & \textbf{0.00} & 0.02  &       0.00       &  & \textbf{0.00} & 0.04  &       0.00       \\
			                                                                             &  C16-V6   &     0.03      &                  0.10                  &  & \textbf{0.00} & 0.03  &       0.00       &  & \textbf{0.00} & 0.04  &       0.00       \\
			                                                                             &  C17-V13  & \textbf{0.00} &                  0.13                  &  & \textbf{0.00} & 0.05  &       0.00       &  & \textbf{0.00} & 0.05  &       0.00       \\
			                                                                             &  C20-V6   &     0.01      &                  0.16                  &  & \textbf{0.00} & 0.04  &       0.00       &  & \textbf{0.00} & 0.05  &       0.00       \\
			                                                                             &  C25-V7   &     0.41      &                  0.26                  &  & \textbf{0.00} & 0.07  &       0.01       &  & \textbf{0.00} & 0.07  &       0.01       \\
			                                                                             &  C35-V13  &     1.03      &                  0.59                  &  &     0.01      & 0.26  &       0.14       &  & \textbf{0.00} & 0.21  &       0.08       \\
			                                                                             &  C50-V20  &     0.61      &                  1.41                  &  &     0.23      & 0.71  &       0.46       &  & \textbf{0.00} & 0.45  &       0.17       \\
			                                                                             &  C70-V30  &     0.59      &                  3.55                  &  &     0.31      & 2.02  &       1.43       &  & \textbf{0.00} & 1.02  &       0.38       \\
			                                                                             &  C90-V40  &     1.10      &                  7.01                  &  &     0.44      & 4.27  &       3.16       &  & \textbf{0.00} & 2.20  &       1.00       \\
			                                                                             & C100-V50  &     1.07      &                  9.85                  &  &     0.41      & 6.37  &       4.80       &  & \textbf{0.00} & 3.13  &       1.44       \\ \midrule
			                                                                             &  Overall  &     0.41      &                  1.93                  &  &     0.12      & 1.16  &       0.83       &  & \textbf{0.00} & 0.61  &       0.26       \\ \midrule
		\end{tabular} 
	}
\end{table}

As visible in the results of \Cref{tab:heuristic}, both versions of the HGS largely outperform previous algorithms. HGS\textsubscript{2} obtains near-optimal solutions within an average gap of $0.01\%$, compared to $1.01\%$ for the \gls{ALNS}. The full load instances are generally easier to solve than the mixed load instances since the problem simplifies. Moreover, the \gls{SP}-based procedure largely contributes to the performance of the algorithm: contrary to intuition, it does not increase the overall CPU time, but even decreases it from $1.93$\,min (HGS\textsubscript{1}) to $1.48$\,min (HGS\textsubscript{2}) on average. Indeed, the \gls{SP} helps to reach optimal solutions more quickly, such that the method only performs $It_\textsc{ni}$ additional iterations before reaching the termination criteria instead of gradually improving over a longer time. The effect is manifest when comparing the average attainment time of HGS\textsubscript{1} (T* $= 1.49\,\text{min}$) with that of HGS\textsubscript{2} ($0.94\,\text{min}$).
In a follow-up work, \citet{Hemmati2016} investigated the impact of some design decisions and parameters of their ALNS and reported the results of six variants of their algorithm on a smaller subset of instances. Drawing a comparison with these methods leads to similar conclusions: HGS\textsubscript{2} largely outperforms these methods. For the sake of brevity, this comparison is presented in the electronic companion.

\subsection{Sensitivity analyses}
\label{sec:sensitivity}

To highlight the role of each main strategy and parameter, we have conducted extensive sensitivity analyses with the \gls{BnP} and \gls{HGS} algorithms. We started with the standard configurations of each method (B\&P\textsubscript{1}, B\&P\textsubscript{2} and HGS\textsubscript{2}) and generated a number of alternative configurations by deactivating a component or modifying a single parameter to study its impact. The results of these analyses are presented in Tables \ref{tab:analysisBP}--\ref{tab:analysisHGA}. 
In \Cref{tab:analysisBP}, Column ``Gap'' represent the average gap, Column ``T'' reports the average CPU time in minutes, Column ``Root'' counts the number of instances for which the root node solution was completed, and Column ``Opt'' counts the number of optimal solutions found. By convention, failing to solve the root node gives a Gap of 100\%.
In \Cref{tab:analysisHGA}, Column ``Gap'' refers to the average gap, and Columns ``T'' and ``T*'' represent the average CPU time and attainment time.

\begin{table}[!htbp]
\centering
\renewcommand{\arraystretch}{1.25}
\setlength{\tabcolsep}{0.2cm}
\caption{Impact of some of the key components of the B\&P}\label{tab:analysisBP}
\scalebox{0.82}
{
\begin{tabular}{lcccccccccc} 
\toprule
&& \multicolumn{4}{c}{\textbf{FUN}} && \multicolumn{4}{c}{\textbf{MUN}} \\\cline{3-6}\cline{8-11}
&& Gap & T & Root & Opt &&  Gap & T & Root & Opt  \\\midrule
Standard (B\&P\textsubscript{1}) && 0.00 & 0.02 & 120 & \textbf{120} && 0.01 & 8.94 & 120 & 107 \\
A. No Heuristic Pricing && 0.00 & 0.02 & 120 & \textbf{120} && 0.02 & 11.20 & 120 & 103 \\
B. No Strong Branching && 0.00 & 1.09 & 120 & 118 && 0.06 & 17.54 & 120 & 86 \\
C. No Preprocessing && 0.00 & 0.03 & 120 & \textbf{120} && 6.76 & 15.10 & 112 & 95 \\
D. No DSSR && 0.00 & 0.04 & 120 & \textbf{120} && 25.02 & 19.27 & 90 & 84 \\\midrule
Standard (B\&P\textsubscript{2}) && 0.00 & 0.01 & 120 & \textbf{120} && 0.00 & 1.74 & 120 & \textbf{119} \\
E. No Completion Bounds && 0.00 & 0.01 & 120 & \textbf{120} && 13.33 & 9.97 & 104 & 104 \\
F. No Subset-Row Cuts && 0.00 & 0.01 & 120 & \textbf{120} && 0.00 & 2.47 & 120 & 118 \\
\bottomrule
\end{tabular}
}
\end{table}

\begin{table}[!htbp]
\centering
\renewcommand{\arraystretch}{1.15}
\setlength{\tabcolsep}{0.2cm}
\caption{Impact of some of the key components and parameters of the HGS}\label{tab:analysisHGA}
\scalebox{0.82}
{
\begin{tabular}{l@{\hspace*{0.8cm}}cccHccccH}
	\toprule
	                                                                            & \multicolumn{4}{c}{\textbf{FUN}} &  & \multicolumn{4}{c}{\textbf{MUN}} \\ \cline{2-5}\cline{7-10}
	                                                                            &  Gap   &  T   &  T*  &    Opt    &  &  Gap   &  T   &  T*  &    Opt    \\ \midrule
	Standard (HGS\textsubscript{2})                                             & \textbf{0.00} & 0.58 & 0.24 &  0.9450   &  & \textbf{0.02} & 2.37 & 1.63 &  0.7908   \\
	G. No SP Intensification (HGS\textsubscript{1})                             & 0.15 & 1.11 & 0.80 &  0.6308   &  & 0.82 & 2.74 & 2.18 &  0.5467   \\
	H. Shorter SP Intensification ($T^\textsc{sp}_\textsc{max} = 10\,\text{s}$) & \textbf{0.00} & 0.58 & 0.24 &  0.9450   &  &\textbf{0.02} & 2.29 & 1.54 &  0.7867   \\
	I. Longer SP Intensification ($T^\textsc{sp}_\textsc{max} = 120\,\text{s}$) & \textbf{0.00} & 0.58 & 0.24 &  0.9450   &  &\textbf{0.02} & 2.37 & 1.59 &  0.7883   \\
	J. No neighborhood restrictions ($|\Gamma| = +\infty$)                      & \textbf{0.00} & 0.72 & 0.29 &  0.9675   &  & 0.05 & 3.49 & 2.35 &  0.8400   \\
	K. No diversity management ($f_\cP^\textsc{div}(\cdot) = 0$)                & \textbf{0.00} & 0.58 & 0.25 &  0.9517   &  & 0.03 & 2.39 & 1.71 &  0.7708   \\ \bottomrule
\end{tabular}
}
\end{table}

In these experiments, again, the instances of class \textbf{FUN} are solved more easily. Since most methods achieve the same solution quality for this class, we will primarily rely on the harder \textbf{MUN} instances in our analyses.

\Cref{tab:analysisBP} analyzes the components of the B\&P algorithms.
As illustrated by the results of Configuration A, heuristic pricing significantly reduces the overall pricing time. Without this component, four additional instances remain open and the CPU time increases by 25\%, though this effect is generally less marked than in other VRP variants \citep[see, e.g.][]{Desaulniers2008,Martinelli2014}.

Deactivating strong branching (Configuration B) has a larger impact. Strong branching helps predicting good branching decisions and reducing the search tree. Without it, the number of solved instances decreases from 107 to 86. The method can still nearly close the optimality gap (0.06\%~on~average), but it often fails to complete the optimality proof.

Configurations C and D evaluate the impact of our advanced preprocessing strategies and DSSR (\Cref{sec:cg}). Both components focus on enhancing the speed of the DP pricing algorithm. These components play a decisive role. Deactivating just one of these components makes it impossible to compute the root-node relaxation in a reasonable time for many instances. As an immediate consequence, the number of optimal solutions dramatically decreases (down to 84/120) and the optimality gap soars (up to 25.02\%) due to the number of incomplete root node calculations.

The remaining analyses of \Cref{tab:analysisBP} concern the B\&P\textsubscript{2} algorithm, based on route enumeration. In addition to the preprocessing strategies and DSSR, the DP algorithm used for route enumeration exploits a sophisticated succession of completion bounds (\Cref{sec:enumeration}). Deactivating these bounds, as in Configuration E, hinders the performance of the route enumeration algorithm, which fails on 16/120 largest instances. Remark that any instance which is successfully enumerated is subsequently solved. Finally, deactivating the SRC separation is moderately detrimental: it leads to an 42\% increase of CPU time and one additional open instance.

The results of the sensitivity analysis of the HGS, in \Cref{tab:analysisHGA}, essentially highlight the importance of the \gls{SP}-based intensification procedure. As visible in Configuration F, the solution quality of HGS significantly decreases (up to 0.82\% average gap on the \textbf{MUN} instances) when this strategy is deactivated. HGS identifies the routes (columns) belonging to the optimal solutions within the available number of iterations, but due to the large number of ships, ship types, and the cargo-ship incompatibility matrix, combining these routes into a complete solution can be a challenging task. In contrast, the \gls{SP} component is perfectly suited for this role, such that the combination of both techniques leads to a particularly effective \emph{matheuristic}.

The impact of other components and parameter settings is less marked: increasing or decreasing the time limit of each SP model (Configurations H and I) does not impact the solution method, due to the fact that most SP models are solved within a few seconds. Moreover, deactivating our ship-dependent neighborhood restrictions (Configuration J) and population-diversity management strategies (Configuration K) leads to a small but significant decrease of solution quality.

\subsection{Experiments on large-scale instances}
\label{sec:perf-larg-inst}

Most shipping companies solve planning problems involving at most a few dozen ships per segment \citep{Wilson2018}. Still, some scenarios may exceptionally require to consider larger instances, for example when evaluating possible merger or coordinated freight operations. Efficient optimization tools are needed to react in such situations. The goal of this section is to evaluate the scalability of our algorithms in such cases. For this analysis, we produced $32$ new instances with the same problem generator as \citet{Hemmati2014}, with up to $89$ ships and $260$ cargoes (i.e., $520$ pickups and deliveries). These instances are accessible at \url{https://w1.cirrelt.ca/~vidalt/en/VRP-resources.html}. We apply the B\&P\textsubscript{2} and the HGS\textsubscript{2} with the same parameters as previously, and increase the CPU-time limits by a factor of four. Therefore, the HGS\textsubscript{2} is run with $T_\textsc{max} = 1\,\text{h}$ and $T^\textsc{sp}_\textsc{max} = 2\,\text{min}$. Similarly, the B\&P\textsubscript{2} is run until a time limit of four hours is exceeded. However, to obtain lower bounds for all instances, we do not interrupt the B\&P\textsubscript{2} until it has at least completed the solution of the root node.

Table~\ref{tab:scalability} reports the results of this experiment using the same conventions as Tables~\ref{tbl:exact}~and~\ref{tab:heuristic}. For these instances, we could not obtain optimal solutions on all instances, such that the column ``Gap'' for the HGS\textsubscript{2} reports the gap relative to the best lower bound found by the B\&P\textsubscript{2}. In addition, detailed solution values for each instance are provided in the electronic companion.

\begin{table}[htbp]
\centering
	\vspace*{-0.5cm}
	\renewcommand{\arraystretch}{1.05}
	\setlength\tabcolsep{6pt}
	\caption{Performance on larger instances}\label{tab:scalability}
	\scalebox{0.65}{
          \begin{tabular}{c@{\hspace*{0.6cm}}l@{\hspace*{0.6cm}}cccccccccccccc}
            \toprule
        & & \multicolumn{10}{c}{Branch-and-Price + Enumeration + SRCs (B\&P\textsubscript{2})} & & \multicolumn{3}{c}{HGS\textsubscript{2}}  \\ \cline{3-12} \cline{14-16}
& Instance                            & T\textsubscript{E}	& R\textsubscript{E}     & Gap\textsubscript{0} & T\textsubscript{0}   & Cuts\textsubscript{0} & R\textsubscript{F}    & Gap\textsubscript{F} & T\textsubscript{F}    & Cuts\textsubscript{F} & N\textsubscript{F} & & Gap  & T     & T*    \\ \midrule
\parbox[t]{2mm}{\multirow{8}{*}{\rotatebox[origin=c]{90}{{\large SS-MUN}}}} & C143-V41 & 5.22 & 385K & 0.13 & 20.39 & 44 & 143K & 0.00 & 24.38 & 209 & 33 &  & 0.04 & 15.85 & 10.69\\
 & C156-V45 & 7.40 & 1M & 0.12 & 35.42 & 58 & 509K & 0.00 & 71.98 & 502 & 129 &  & 0.01 & 28.15 & 18.19\\
 & C169-V48 & 6.83 & 199K & 0.06 & 39.08 & 53 & 88K & 0.00 & 41.71 & 159 & 19 &  & 0.01 & 26.32 & 19.28\\
 & C182-V52 & 13.22 & 762K & 0.10 & 53.72 & 38 & 482K & 0.00 & 88.32 & 406 & 101 &  & 0.03 & 42.22 & 31.34\\
 & C195-V56 & 29.25 & 297K & 0.06 & 133.40 & 71 & 180K & 0.00 & 147.68 & 282 & 47 &  & 0.04 & 50.21 & 39.07\\
 & C208-V59 & 57.49 & 8M & 0.12 & 135.82 & 59 & 2M & 0.02 & 240.00 & 625 & 157 &  & 0.03 & 56.17 & 46.63\\
 & C221-V63$^\ddagger$ & - & - & 0.17 & 495.02 & - & - & 0.17 & 495.02 & - & - &  & 0.20 & 56.32 & 45.02\\
 & C260-V74$^\ddagger$ & - & - & 0.17 & 1196.78 & - & - & 0.17 & 1196.78 & - & - &  & 0.25 & 60.16 & 53.57\\ \midrule
 & Overall & 19.90 & 2M & 0.10 & 69.64 & 53.8 & 559K & 0.04 & 258.23 & 363.8 & 81.0 &  & 0.08 & 41.93 & 32.97\\ \midrule
\parbox[t]{2mm}{\multirow{8}{*}{\rotatebox[origin=c]{90}{{\large SS-FUN}}}} & C110-V52 & 0.01 & 2K & 0.01 & 0.07 & 0 & 2K & 0.00 & 0.14 & 0 & 3 &  & 0.00 & 4.01 & 1.88\\
 & C120-V53 & 0.01 & 3K & 0.01 & 0.10 & 4 & 3K & 0.00 & 0.29 & 7 & 5 &  & 0.00 & 4.40 & 1.87\\
 & C130-V58 & 0.01 & 4K & 0.01 & 0.14 & 10 & 3K & 0.00 & 0.35 & 22 & 5 &  & 0.00 & 5.58 & 2.67\\
 & C140-V62 & 0.01 & 1K & 0.00 & 0.14 & 0 & 1K & 0.00 & 0.14 & 0 & 1 &  & 0.00 & 6.49 & 2.71\\
 & C150-V67 & 0.02 & 4K & 0.00 & 0.20 & 3 & 4K & 0.00 & 0.34 & 3 & 3 &  & 0.00 & 7.66 & 3.33\\
 & C160-V71 & 0.03 & 6K & 0.01 & 0.32 & 5 & 6K & 0.00 & 1.31 & 19 & 11 &  & 0.00 & 9.14 & 3.96\\
 & C170-V76 & 0.03 & 2K & 0.00 & 0.34 & 0 & 2K & 0.00 & 0.34 & 0 & 1 &  & 0.00 & 13.05 & 6.95\\
 & C200-V89 & 0.05 & 4K & 0.00 & 0.78 & 0 & 4K & 0.00 & 0.78 & 0 & 1 &  & 0.01 & 20.74 & 11.52\\ \midrule
 & Overall & 0.02 & 3K & 0.00 & 0.26 & 2.8 & 3K & 0.00 & 0.46 & 6.4 & 3.8 &  & 0.00 & 8.88 & 4.36\\ \midrule
\parbox[t]{2mm}{\multirow{8}{*}{\rotatebox[origin=c]{90}{{\large DS-MUN}}}} & C143-V41 & 5.51 & 16M & 0.39 & 11.63 & 78 & 8M & 0.00 & 33.66 & 315 & 37 &  & 0.04 & 42.36 & 34.89\\
 & C156-V45 & 1.14 & 102K & 0.01 & 7.73 & 78 & 12K & 0.00 & 8.09 & 78 & 3 &  & 0.04 & 43.33 & 34.79\\
 & C169-V48$^\dagger$ & - & - & 0.35 & 240.00 & - & - & 0.35 & 240.00 & - & - &  & 0.40 & 55.06 & 47.54\\
 & C182-V52$^\dagger$ & - & - & 0.51 & 240.00 & - & - & 0.51 & 240.00 & - & - &  & 0.60 & 57.21 & 49.01\\
 & C195-V56$^\dagger$ & - & - & 0.44 & 240.00 & - & - & 0.44 & 240.00 & - & - &  & 0.62 & 57.95 & 49.53\\
 & C208-V59$^\dagger$ & - & - & 1.28 & 240.00 & - & - & 1.28 & 240.00 & - & - &  & 1.50 & 56.32 & 47.13\\
 & C221-V63$^\ddagger$ & - & - & 0.46 & 481.31 & - & - & 0.46 & 481.31 & - & - &  & 0.62 & 59.39 & 50.24\\
 & C260-V74$^\ddagger$ & - & - & 1.04 & 3848.88 & - & - & 1.04 & 3848.88 & - & - &  & 1.32 & 60.00 & 52.28\\ \midrule
 & Overall & 3.33 & 8M & 0.20 & 9.68 & 78.0 & 4M & 0.51 & 666.49 & 196.5 & 20.0 &  & 0.64 & 53.95 & 45.67\\ \midrule
\parbox[t]{2mm}{\multirow{8}{*}{\rotatebox[origin=c]{90}{{\large DS-FUN}}}} & C110-V52 & 0.01 & 2K & 0.00 & 0.04 & 0 & 2K & 0.00 & 0.04 & 0 & 1 &  & 0.00 & 3.72 & 1.40\\
 & C120-V53 & 0.01 & 2K & 0.00 & 0.06 & 5 & 2K & 0.00 & 0.06 & 5 & 1 &  & 0.00 & 4.75 & 2.12\\
 & C130-V58 & 0.02 & 2K & 0.00 & 0.07 & 0 & 2K & 0.00 & 0.07 & 0 & 1 &  & 0.00 & 5.70 & 2.54\\
 & C140-V62 & 0.02 & 5K & 0.00 & 0.12 & 0 & 5K & 0.00 & 0.12 & 0 & 1 &  & 0.00 & 6.87 & 3.16\\
 & C150-V67 & 0.03 & 17K & 0.00 & 0.14 & 2 & 17K & 0.00 & 0.96 & 14 & 9 &  & 0.00 & 8.35 & 3.97\\
 & C160-V71 & 0.03 & 2K & 0.00 & 0.15 & 0 & 2K & 0.00 & 0.15 & 0 & 1 &  & 0.00 & 13.46 & 8.19\\
 & C170-V76 & 0.03 & 12K & 0.00 & 0.18 & 5 & 12K & 0.00 & 0.39 & 5 & 3 &  & 0.00 & 14.74 & 8.32\\
 & C200-V89 & 0.06 & 77K & 0.00 & 0.45 & 9 & 77K & 0.00 & 8.54 & 73 & 37 &  & 0.00 & 17.95 & 9.37\\ \midrule
 & Overall & 0.03 & 15K & 0.00 & 0.15 & 2.6 & 15K & 0.00 & 1.29 & 12.1 & 6.8 &  & 0.00 & 9.44 & 4.88\\ \bottomrule
\multicolumn{16}{l}{$\dagger$ Root-node solution was completed within four hours but enumeration exceeded four hours, triggering termination.}\\
\multicolumn{16}{l}{$\ddagger$ Root-node solution exceeded four hours.}\\
		\end{tabular}
	}
\end{table}

The results on the new large-scale instances are consistent with our previous findings. All the full load instances are solved to optimality within a few minutes. In contrast, the mixed load instances are significantly more difficult since the deliveries can occur in any position after their associated pickups. This leads to a larger search space which effectively requires to arrange up to 520 visits (i.e., 260 pickups and deliveries) in the largest case. Seven out of the $16$ large mixed load instances are solved to optimality. For the remaining $9$ instances, the optimality gap is always smaller than $1.28\%$, but the root-node solution took up to 64 hours in the largest case (DS-MUN-C260-V74).
The solutions of the proposed heuristic are also remarkably accurate.
For the full load instances, the HGS\textsubscript{2} always finds near-optimal solutions (average gap below $0.01\%$) in an average time of $9.16$ minutes.
For the mixed load instances, the HGS\textsubscript{2} complements very well the exact method by producing consistently good solutions (average gap of $0.36\%$) within a controllable CPU time ($47.94$ minutes on average).

\section{Conclusions and Perspectives}
\label{sec:conclusions}

As demonstrated in this paper, the literature on maritime logistics has attained a turning point where state-of-the-art exact algorithms can solve industrial and tramp ship routing optimization problems of a realistic scale. The B\&P algorithm that we designed for this purpose capitalizes on multiple methodological elements to find a good balance between relaxation strength and pricing speed. As demonstrated by our sensitivity analyses, its most critical method components concern the efficiency of the DP pricing and enumeration algorithms: our sophisticated preprocessing techniques and filters, 
the use of DSSR to reintroduce elementarity and the successive completion bounds are essential to solve large \glspl{ITSRSP} instances. These observations are  in line with the works of \cite{Pecin2016a,Sadykov2017} and the general research on \gls{MIP} for \glspl{VRP} which, for a large part, focuses on finding tighter relaxations and faster DP algorithms.

From the heuristic viewpoint, our experiments with a problem-tailored HGS show that the routes of optimal solutions can usually be quickly identified, but that crossover and local search methods are easily tricked into suboptimal route selections. Hybridizing the HGS with an SP solver fixes this issue and allows to attain near-optimal solutions within minutes.

Overall, the algorithms presented in this paper have contributed to push the limits of performance and problem tractability, but multiple avenues of research remain open concerning model accuracy. Indeed, despite its relevance for maritime transportation companies, the \glspl{ITSRSP} remains a mere simplification of reality. As highlighted in \cite{Christiansen2014}, it does not consider load-dependent fuel consumption, possible load splitting and flexible cargo quantities, or the possibility of slow-steaming on selected route segments. Emission control areas and sea conditions (e.g.\ depth and currents) are also largely ignored. Last but not the least, considerable reductions of turnaround time may be achieved by jointly optimizing ship routing and port operations within integrated supply chains. Adapting state-of-the-art exact and heuristic algorithms to handle these complex attributes remain a significant challenge for the future.

\section{Acknowledgements}
\label{sec:ack}

This work was supported by the Conselho Nacional de Desenvolvimento Cient\'ifico e Tecnol\'ogico (CNPq) under Grant numbers 308528/2018-2, 134795/2016-4, 425962/2016-4 and 313521/2017-4, and Funda\c{c}\~ao Carlos Chagas Filho de Amparo \`a Pesquisa do Estado do Rio de Janeiro (FAPERJ) under Grant number E-26/202.790/2019.


\input{ejor2019-R3-ArXiV.bbl}
\input{companion}

\end{document}

%% file: companion.tex
\section*{Appendix -- Detailed Results}

\Crefrange{tab:ss_mun}{tab:large_full} give the detailed results of B\&P\textsubscript{2} and HGS\textsubscript{2} (with the set partitioning component) on each individual instance.
For the B\&P\textsubscript{2}, Columns ``Opt'' and ``T'' represent the optimal value and the total CPU time (in minutes).
For the HGS\textsubscript{2}, Columns ``Best'', ``Avg'' and ``Worst'' represent the best, average and worst cost found over $10$ runs. Column ``T'' gives the total CPU time in minutes, and column ``T*'' gives the \emph{attainment} time (in minutes) needed to reach the final solution.
In \Cref{tab:large_full}, Columns ``LB'' and ``UB'' represent the bounds found by the B\&P\textsubscript{2}.

\begin{table}[htbp]
        \vspace*{-0.7cm}
	\renewcommand{\arraystretch}{1.0}
	\setlength\tabcolsep{7pt}
	\centering
	\caption{Results for short sea mixed load instances}\label{tab:ss_mun}
	\scalebox{0.7}{
		\begin{tabular}{lrrrrrrrrr}
			\toprule
			\multirow{2}{*}{Instance} & \multicolumn{2}{c}{B\&P\textsubscript{2}} && \multicolumn{5}{c}{HGS\textsubscript{2}} \\ \cline{2-3}\cline{5-9}
			& \multicolumn{1}{c}{Opt} & \multicolumn{1}{c}{T} && \multicolumn{1}{c}{Best} & \multicolumn{1}{c}{Avg} & \multicolumn{1}{c}{Worst} & \multicolumn{1}{c}{T} & \multicolumn{1}{c}{T*} \\ \midrule
SHORTSEA\_MUN\_C7\_V3\_HE\_1  & 1476444 & 0.00 &  & 1476444 & 1476444.0 & 1476444 & 0.02 & 0.00\\ 
SHORTSEA\_MUN\_C7\_V3\_HE\_2   & 1134176 & 0.00 &  & 1134176 & 1134176.0 & 1134176 & 0.02 & 0.00\\ 
SHORTSEA\_MUN\_C7\_V3\_HE\_3   & 1196466 & 0.00 &  & 1196466 & 1196466.0 & 1196466 & 0.02 & 0.00\\ 
SHORTSEA\_MUN\_C7\_V3\_HE\_4   & 1256139 & 0.00 &  & 1256139 & 1256139.0 & 1256139 & 0.02 & 0.00\\ 
SHORTSEA\_MUN\_C7\_V3\_HE\_5   & 1160394 & 0.00 &  & 1160394 & 1160394.0 & 1160394 & 0.02 & 0.00\\ 
SHORTSEA\_MUN\_C10\_V3\_HE\_1   & 2083965 & 0.00 &  & 2083965 & 2083965.0 & 2083965 & 0.04 & 0.00\\ 
SHORTSEA\_MUN\_C10\_V3\_HE\_2   & 2012364 & 0.00 &  & 2012364 & 2012364.0 & 2012364 & 0.04 & 0.00\\ 
SHORTSEA\_MUN\_C10\_V3\_HE\_3   & 1986779 & 0.00 &  & 1986779 & 1986779.0 & 1986779 & 0.04 & 0.00\\ 
SHORTSEA\_MUN\_C10\_V3\_HE\_4   & 2125461 & 0.00 &  & 2125461 & 2125461.0 & 2125461 & 0.04 & 0.00\\ 
SHORTSEA\_MUN\_C10\_V3\_HE\_5   & 2162453 & 0.00 &  & 2162453 & 2162453.0 & 2162453 & 0.04 & 0.00\\ 
SHORTSEA\_MUN\_C15\_V4\_HE\_1   & 1959153 & 0.00 &  & 1959153 & 1959153.0 & 1959153 & 0.08 & 0.00\\ 
SHORTSEA\_MUN\_C15\_V4\_HE\_2   & 2560004 & 0.00 &  & 2560004 & 2560004.0 & 2560004 & 0.08 & 0.00\\ 
SHORTSEA\_MUN\_C15\_V4\_HE\_3   & 2582912 & 0.00 &  & 2582912 & 2582912.0 & 2582912 & 0.08 & 0.01\\ 
SHORTSEA\_MUN\_C15\_V4\_HE\_4   & 2265396 & 0.00 &  & 2265396 & 2265396.0 & 2265396 & 0.09 & 0.00\\ 
SHORTSEA\_MUN\_C15\_V4\_HE\_5   & 2230861 & 0.00 &  & 2230861 & 2230861.0 & 2230861 & 0.08 & 0.00\\ 
SHORTSEA\_MUN\_C18\_V5\_HE\_1   & 2374420 & 0.00 &  & 2374420 & 2374420.0 & 2374420 & 0.11 & 0.00\\ 
SHORTSEA\_MUN\_C18\_V5\_HE\_2   & 2987358 & 0.00 &  & 2987358 & 2987358.0 & 2987358 & 0.11 & 0.01\\ 
SHORTSEA\_MUN\_C18\_V5\_HE\_3   & 2301308 & 0.00 &  & 2301308 & 2301308.0 & 2301308 & 0.12 & 0.01\\ 
SHORTSEA\_MUN\_C18\_V5\_HE\_4   & 2400016 & 0.00 &  & 2400016 & 2402999.2 & 2414932 & 0.16 & 0.05\\ 
SHORTSEA\_MUN\_C18\_V5\_HE\_5   & 2813167 & 0.00 &  & 2813167 & 2813167.0 & 2813167 & 0.17 & 0.06\\ 
SHORTSEA\_MUN\_C22\_V6\_HE\_1   & 3928483 & 0.00 &  & 3928483 & 3928483.0 & 3928483 & 0.17 & 0.03\\ 
SHORTSEA\_MUN\_C22\_V6\_HE\_2   & 3683436 & 0.01 &  & 3683436 & 3683436.0 & 3683436 & 0.16 & 0.02\\ 
SHORTSEA\_MUN\_C22\_V6\_HE\_3   & 3264770 & 0.00 &  & 3264770 & 3264770.0 & 3264770 & 0.18 & 0.04\\ 
SHORTSEA\_MUN\_C22\_V6\_HE\_4   & 3228262 & 0.00 &  & 3228262 & 3228262.0 & 3228262 & 0.22 & 0.07\\ 
SHORTSEA\_MUN\_C22\_V6\_HE\_5   & 3770560 & 0.00 &  & 3770560 & 3770560.0 & 3770560 & 0.17 & 0.03\\ 
SHORTSEA\_MUN\_C23\_V13\_HE\_1   & 2276832 & 0.01 &  & 2276832 & 2276832.0 & 2276832 & 0.22 & 0.04\\ 
SHORTSEA\_MUN\_C23\_V13\_HE\_2   & 2255469 & 0.01 &  & 2255469 & 2255469.0 & 2255469 & 0.22 & 0.06\\ 
SHORTSEA\_MUN\_C23\_V13\_HE\_3   & 2362503 & 0.00 &  & 2362503 & 2362503.0 & 2362503 & 0.23 & 0.08\\ 
SHORTSEA\_MUN\_C23\_V13\_HE\_4   & 2250110 & 0.02 &  & 2250110 & 2250110.0 & 2250110 & 0.25 & 0.08\\ 
SHORTSEA\_MUN\_C23\_V13\_HE\_5   & 2325941 & 0.00 &  & 2325941 & 2325941.0 & 2325941 & 0.22 & 0.06\\ 
SHORTSEA\_MUN\_C30\_V6\_HE\_1   & 4958542 & 0.00 &  & 4958542 & 4958542.0 & 4958542 & 0.31 & 0.10\\ 
SHORTSEA\_MUN\_C30\_V6\_HE\_2   & 4549708 & 0.01 &  & 4549708 & 4549708.0 & 4549708 & 0.34 & 0.13\\ 
SHORTSEA\_MUN\_C30\_V6\_HE\_3   & 4098111 & 0.02 &  & 4098111 & 4098111.0 & 4098111 & 0.33 & 0.11\\ 
SHORTSEA\_MUN\_C30\_V6\_HE\_4   & 4449449 & 0.02 &  & 4449449 & 4449485.3 & 4449812 & 0.36 & 0.15\\ 
SHORTSEA\_MUN\_C30\_V6\_HE\_5   & 4528514 & 0.01 &  & 4528514 & 4528514.0 & 4528514 & 0.29 & 0.09\\ 
SHORTSEA\_MUN\_C35\_V7\_HE\_1   & 4893734 & 0.03 &  & 4893734 & 4898167.9 & 4913975 & 0.59 & 0.32\\ 
SHORTSEA\_MUN\_C35\_V7\_HE\_2   & 4533265 & 0.21 &  & 4533265 & 4534290.3 & 4543518 & 0.53 & 0.23\\ 
SHORTSEA\_MUN\_C35\_V7\_HE\_3   & 4433847 & 0.03 &  & 4433847 & 4433847.0 & 4433847 & 0.46 & 0.20\\ 
SHORTSEA\_MUN\_C35\_V7\_HE\_4   & 4580935 & 0.05 &  & 4580935 & 4580935.0 & 4580935 & 0.51 & 0.23\\ 
SHORTSEA\_MUN\_C35\_V7\_HE\_5   & 5511661 & 0.01 &  & 5511661 & 5513388.5 & 5523201 & 0.61 & 0.35\\ 
SHORTSEA\_MUN\_C60\_V13\_HE\_1   & 8133385 & 0.96 &  & 8133385 & 8147210.5 & 8163045 & 2.71 & 1.70\\ 
SHORTSEA\_MUN\_C60\_V13\_HE\_2   & 7971476 & 0.88 &  & 7971476 & 7972935.0 & 7984871 & 1.84 & 1.05\\ 
SHORTSEA\_MUN\_C60\_V13\_HE\_3   & 7604198 & 0.75 &  & 7604198 & 7632301.3 & 7647547 & 2.56 & 1.74\\ 
SHORTSEA\_MUN\_C60\_V13\_HE\_4   & 8505125 & 0.87 &  & 8505125 & 8505971.4 & 8508321 & 2.09 & 1.24\\ 
SHORTSEA\_MUN\_C60\_V13\_HE\_5   & 8921750 & 1.08 &  & 8921750 & 8931617.8 & 8942531 & 2.08 & 1.11\\ 
SHORTSEA\_MUN\_C80\_V20\_HE\_1   & 10289573 & 1.67 &  & 10289573 & 10294261.3 & 10305785 & 5.04 & 3.37\\ 
SHORTSEA\_MUN\_C80\_V20\_HE\_2   & 10240618 & 0.91 &  & 10240618 & 10241641.4 & 10246354 & 3.64 & 2.32\\ 
SHORTSEA\_MUN\_C80\_V20\_HE\_3   & 9606530 & 2.84 &  & 9606530 & 9606573.1 & 9606961 & 2.91 & 1.66\\ 
SHORTSEA\_MUN\_C80\_V20\_HE\_4   & 11302476 & 0.45 &  & 11302476 & 11311027.3 & 11333280 & 4.78 & 3.21\\ 
SHORTSEA\_MUN\_C80\_V20\_HE\_5   & 10862563 & 0.31 &  & 10862563 & 10863032.1 & 10867254 & 2.85 & 1.63\\ 
SHORTSEA\_MUN\_C100\_V30\_HE\_1   & 12626988 & 2.53 &  & 12626988 & 12627601.8 & 12633126 & 4.59 & 2.48\\ 
SHORTSEA\_MUN\_C100\_V30\_HE\_2   & 12774864 & 2.98 &  & 12774864 & 12775622.5 & 12776760 & 5.80 & 3.23\\ 
SHORTSEA\_MUN\_C100\_V30\_HE\_3   & 11935332 & 7.82 &  & 11935332 & 11935349.0 & 11935502 & 5.14 & 2.92\\ 
SHORTSEA\_MUN\_C100\_V30\_HE\_4   & 13605352 & 1.82 &  & 13605352 & 13610213.4 & 13612134 & 5.20 & 2.96\\ 
SHORTSEA\_MUN\_C100\_V30\_HE\_5   & 13240648 & 1.20 &  & 13240648 & 13241314.4 & 13244485 & 5.83 & 3.54\\ 
SHORTSEA\_MUN\_C130\_V40\_HE\_1   & 16316051 & 13.45 &  & 16316388 & 16318378.8 & 16319526 & 13.31 & 8.71\\ 
SHORTSEA\_MUN\_C130\_V40\_HE\_2   & 16260579 & 27.23 &  & 16260579 & 16263866.0 & 16272543 & 12.64 & 8.63\\ 
SHORTSEA\_MUN\_C130\_V40\_HE\_3   & 15537963 & 17.97 &  & 15537963 & 15543747.8 & 15551928 & 12.68 & 9.03\\ 
SHORTSEA\_MUN\_C130\_V40\_HE\_4   & 17011065 & 6.65 &  & 17011065 & 17012795.2 & 17014853 & 11.82 & 7.84\\ 
SHORTSEA\_MUN\_C130\_V40\_HE\_5   & 18273893 & 10.11 &  & 18273893 & 18275423.4 & 18281922 & 11.71 & 7.75\\ \bottomrule
		\end{tabular}
	}
\end{table}

\begin{table}[htbp]
        \vspace*{-0.7cm}
	\renewcommand{\arraystretch}{1.0}
	\setlength\tabcolsep{7pt}
	\centering
	\caption{Results for short sea full load instances}\label{tab:ss_fun}
	\scalebox{0.7}{
		\begin{tabular}{lrrrrrrrrr}
			\toprule
			\multirow{2}{*}{Instance} & \multicolumn{2}{c}{B\&P\textsubscript{2}} && \multicolumn{5}{c}{HGS\textsubscript{2}} \\ \cline{2-3}\cline{5-9}
			& \multicolumn{1}{c}{Opt} & \multicolumn{1}{c}{T} && \multicolumn{1}{c}{Best} & \multicolumn{1}{c}{Avg} & \multicolumn{1}{c}{Worst} & \multicolumn{1}{c}{T} & \multicolumn{1}{c}{T*} \\ \midrule
SHORTSEA\_FUN\_C8\_V3\_HE\_1   & 1391997 & 0.00 &  & 1391997 & 1391997.0 & 1391997 & 0.01 & 0.00\\ 
SHORTSEA\_FUN\_C8\_V3\_HE\_2   & 1246273 & 0.00 &  & 1246273 & 1246273.0 & 1246273 & 0.01 & 0.00\\ 
SHORTSEA\_FUN\_C8\_V3\_HE\_3   & 1698102 & 0.00 &  & 1698102 & 1698102.0 & 1698102 & 0.01 & 0.00\\ 
SHORTSEA\_FUN\_C8\_V3\_HE\_4   & 1777637 & 0.00 &  & 1777637 & 1777637.0 & 1777637 & 0.01 & 0.00\\ 
SHORTSEA\_FUN\_C8\_V3\_HE\_5   & 1636788 & 0.00 &  & 1636788 & 1636788.0 & 1636788 & 0.01 & 0.00\\ 
SHORTSEA\_FUN\_C11\_V4\_HE\_1   & 1052463 & 0.00 &  & 1052463 & 1052463.0 & 1052463 & 0.02 & 0.00\\ 
SHORTSEA\_FUN\_C11\_V4\_HE\_2   & 1067139 & 0.00 &  & 1067139 & 1067139.0 & 1067139 & 0.02 & 0.00\\ 
SHORTSEA\_FUN\_C11\_V4\_HE\_3   & 1212388 & 0.00 &  & 1212388 & 1212388.0 & 1212388 & 0.02 & 0.00\\ 
SHORTSEA\_FUN\_C11\_V4\_HE\_4   & 1185465 & 0.00 &  & 1185465 & 1185465.0 & 1185465 & 0.02 & 0.00\\ 
SHORTSEA\_FUN\_C11\_V4\_HE\_5   & 1310285 & 0.00 &  & 1310285 & 1310285.0 & 1310285 & 0.02 & 0.00\\ 
SHORTSEA\_FUN\_C13\_V5\_HE\_1   & 2034184 & 0.00 &  & 2034184 & 2034184.0 & 2034184 & 0.02 & 0.00\\ 
SHORTSEA\_FUN\_C13\_V5\_HE\_2   & 2043253 & 0.00 &  & 2043253 & 2043253.0 & 2043253 & 0.03 & 0.00\\ 
SHORTSEA\_FUN\_C13\_V5\_HE\_3   & 2378283 & 0.00 &  & 2378283 & 2378283.0 & 2378283 & 0.03 & 0.00\\ 
SHORTSEA\_FUN\_C13\_V5\_HE\_4   & 2707215 & 0.00 &  & 2707215 & 2707215.0 & 2707215 & 0.02 & 0.00\\ 
SHORTSEA\_FUN\_C13\_V5\_HE\_5   & 3011648 & 0.00 &  & 3011648 & 3011648.0 & 3011648 & 0.02 & 0.00\\ 
SHORTSEA\_FUN\_C16\_V6\_HE\_1   & 3577005 & 0.00 &  & 3577005 & 3577005.0 & 3577005 & 0.03 & 0.00\\ 
SHORTSEA\_FUN\_C16\_V6\_HE\_2   & 3560203 & 0.00 &  & 3560203 & 3560203.0 & 3560203 & 0.03 & 0.00\\ 
SHORTSEA\_FUN\_C16\_V6\_HE\_3   & 4081013 & 0.00 &  & 4081013 & 4081013.0 & 4081013 & 0.04 & 0.00\\ 
SHORTSEA\_FUN\_C16\_V6\_HE\_4   & 3667080 & 0.00 &  & 3667080 & 3667080.0 & 3667080 & 0.03 & 0.00\\ 
SHORTSEA\_FUN\_C16\_V6\_HE\_5   & 3438493 & 0.00 &  & 3438493 & 3438493.0 & 3438493 & 0.03 & 0.00\\ 
SHORTSEA\_FUN\_C17\_V13\_HE\_1   & 2265731 & 0.00 &  & 2265731 & 2265731.0 & 2265731 & 0.05 & 0.00\\ 
SHORTSEA\_FUN\_C17\_V13\_HE\_2   & 3154165 & 0.00 &  & 3154165 & 3154165.0 & 3154165 & 0.05 & 0.00\\ 
SHORTSEA\_FUN\_C17\_V13\_HE\_3   & 2699378 & 0.00 &  & 2699378 & 2699378.0 & 2699378 & 0.06 & 0.01\\ 
SHORTSEA\_FUN\_C17\_V13\_HE\_4   & 2806231 & 0.00 &  & 2806231 & 2806231.0 & 2806231 & 0.05 & 0.00\\ 
SHORTSEA\_FUN\_C17\_V13\_HE\_5   & 2910814 & 0.00 &  & 2910814 & 2910814.0 & 2910814 & 0.05 & 0.00\\ 
SHORTSEA\_FUN\_C20\_V6\_HE\_1   & 2973381 & 0.00 &  & 2973381 & 2973381.0 & 2973381 & 0.05 & 0.00\\ 
SHORTSEA\_FUN\_C20\_V6\_HE\_2   & 3206514 & 0.00 &  & 3206514 & 3206514.0 & 3206514 & 0.05 & 0.01\\ 
SHORTSEA\_FUN\_C20\_V6\_HE\_3   & 3197445 & 0.00 &  & 3197445 & 3197445.0 & 3197445 & 0.05 & 0.01\\ 
SHORTSEA\_FUN\_C20\_V6\_HE\_4   & 3342130 & 0.00 &  & 3342130 & 3342130.0 & 3342130 & 0.05 & 0.00\\ 
SHORTSEA\_FUN\_C20\_V6\_HE\_5   & 3156378 & 0.00 &  & 3156378 & 3156378.0 & 3156378 & 0.05 & 0.00\\ 
SHORTSEA\_FUN\_C25\_V7\_HE\_1   & 3833588 & 0.00 &  & 3833588 & 3833588.0 & 3833588 & 0.07 & 0.02\\ 
SHORTSEA\_FUN\_C25\_V7\_HE\_2   & 3673666 & 0.00 &  & 3673666 & 3673666.0 & 3673666 & 0.07 & 0.01\\ 
SHORTSEA\_FUN\_C25\_V7\_HE\_3   & 4238213 & 0.00 &  & 4238213 & 4238213.0 & 4238213 & 0.07 & 0.01\\ 
SHORTSEA\_FUN\_C25\_V7\_HE\_4   & 4260762 & 0.00 &  & 4260762 & 4260762.0 & 4260762 & 0.08 & 0.02\\ 
SHORTSEA\_FUN\_C25\_V7\_HE\_5   & 4069693 & 0.00 &  & 4069693 & 4069693.0 & 4069693 & 0.08 & 0.02\\ 
SHORTSEA\_FUN\_C35\_V13\_HE\_1   & 2986667 & 0.00 &  & 2986667 & 2986667.0 & 2986667 & 0.14 & 0.02\\ 
SHORTSEA\_FUN\_C35\_V13\_HE\_2   & 3002973 & 0.00 &  & 3002973 & 3002973.0 & 3002973 & 0.20 & 0.07\\ 
SHORTSEA\_FUN\_C35\_V13\_HE\_3   & 3084339 & 0.00 &  & 3084339 & 3084339.0 & 3084339 & 0.16 & 0.04\\ 
SHORTSEA\_FUN\_C35\_V13\_HE\_4   & 3952461 & 0.00 &  & 3952461 & 3952461.0 & 3952461 & 0.18 & 0.06\\ 
SHORTSEA\_FUN\_C35\_V13\_HE\_5   & 3293086 & 0.00 &  & 3293086 & 3293086.0 & 3293086 & 0.18 & 0.06\\ 
SHORTSEA\_FUN\_C50\_V20\_HE\_1   & 7258266 & 0.00 &  & 7258266 & 7258266.0 & 7258266 & 0.40 & 0.13\\ 
SHORTSEA\_FUN\_C50\_V20\_HE\_2   & 7452465 & 0.01 &  & 7452465 & 7452465.0 & 7452465 & 0.44 & 0.17\\ 
SHORTSEA\_FUN\_C50\_V20\_HE\_3   & 6922293 & 0.01 &  & 6922293 & 6922293.0 & 6922293 & 0.44 & 0.15\\ 
SHORTSEA\_FUN\_C50\_V20\_HE\_4   & 8933846 & 0.01 &  & 8933846 & 8933846.5 & 8933848 & 0.46 & 0.17\\ 
SHORTSEA\_FUN\_C50\_V20\_HE\_5   & 7322307 & 0.01 &  & 7322307 & 7322307.0 & 7322307 & 0.40 & 0.13\\ 
SHORTSEA\_FUN\_C70\_V30\_HE\_1   & 10051856 & 0.02 &  & 10051856 & 10051856.0 & 10051856 & 0.98 & 0.39\\ 
SHORTSEA\_FUN\_C70\_V30\_HE\_2   & 10455468 & 0.01 &  & 10455468 & 10455468.0 & 10455468 & 0.99 & 0.38\\ 
SHORTSEA\_FUN\_C70\_V30\_HE\_3   & 10172541 & 0.01 &  & 10172541 & 10172541.0 & 10172541 & 1.27 & 0.62\\ 
SHORTSEA\_FUN\_C70\_V30\_HE\_4   & 10854036 & 0.06 &  & 10854036 & 10854036.0 & 10854036 & 1.02 & 0.38\\ 
SHORTSEA\_FUN\_C70\_V30\_HE\_5   & 10886838 & 0.10 &  & 10886838 & 10886838.0 & 10886838 & 1.16 & 0.35\\ 
SHORTSEA\_FUN\_C90\_V40\_HE\_1   & 13361947 & 0.03 &  & 13361947 & 13362943.0 & 13371155 & 1.94 & 0.82\\ 
SHORTSEA\_FUN\_C90\_V40\_HE\_2   & 13828112 & 0.02 &  & 13828112 & 13828112.0 & 13828112 & 1.88 & 0.71\\ 
SHORTSEA\_FUN\_C90\_V40\_HE\_3   & 12627125 & 0.08 &  & 12627125 & 12627476.1 & 12628003 & 1.77 & 0.63\\ 
SHORTSEA\_FUN\_C90\_V40\_HE\_4   & 14406428 & 0.03 &  & 14406428 & 14406689.8 & 14409031 & 1.82 & 0.64\\ 
SHORTSEA\_FUN\_C90\_V40\_HE\_5   & 13560830 & 0.06 &  & 13560830 & 13560835.3 & 13560853 & 2.23 & 1.01\\ 
SHORTSEA\_FUN\_C100\_V50\_HE\_1   & 13800823 & 0.03 &  & 13800823 & 13800823.0 & 13800823 & 2.94 & 1.35\\ 
SHORTSEA\_FUN\_C100\_V50\_HE\_2   & 14644836 & 0.19 &  & 14644836 & 14645381.1 & 14647299 & 3.02 & 1.30\\ 
SHORTSEA\_FUN\_C100\_V50\_HE\_3   & 13135505 & 0.05 &  & 13135505 & 13135756.6 & 13136396 & 2.65 & 1.02\\ 
SHORTSEA\_FUN\_C100\_V50\_HE\_4   & 14841840 & 0.14 &  & 14841840 & 14841840.4 & 14841841 & 3.06 & 1.41\\ 
SHORTSEA\_FUN\_C100\_V50\_HE\_5   & 14009874 & 0.07 &  & 14009874 & 14009971.3 & 14010827 & 2.48 & 0.93\\ \bottomrule
		\end{tabular}
	}
\end{table}

\begin{table}[htbp]
        \vspace*{-0.7cm}
	\renewcommand{\arraystretch}{1.0}
	\setlength\tabcolsep{7pt}
	\centering
	\caption{Results for deep sea mixed load instances}\label{tab:ds_mun}
	\scalebox{0.7}{
		\begin{tabular}{lrrrrrrrrr}
			\toprule
			\multirow{2}{*}{Instance} & \multicolumn{2}{c}{B\&P\textsubscript{2}} && \multicolumn{5}{c}{HGS\textsubscript{2}} \\ \cline{2-3}\cline{5-9}
			& \multicolumn{1}{c}{Opt} & \multicolumn{1}{c}{T} && \multicolumn{1}{c}{Best} & \multicolumn{1}{c}{Avg} & \multicolumn{1}{c}{Worst} & \multicolumn{1}{c}{T} & \multicolumn{1}{c}{T*} \\ \midrule
DEEPSEA\_MUN\_C7\_V3\_HE\_1   & 5233464 & 0.00 &  & 5233464 & 5233464.0 & 5233464 & 0.02 & 0.00\\ 
DEEPSEA\_MUN\_C7\_V3\_HE\_2   & 6053699 & 0.00 &  & 6053699 & 6053699.0 & 6053699 & 0.02 & 0.00\\ 
DEEPSEA\_MUN\_C7\_V3\_HE\_3   & 5888949 & 0.00 &  & 5888949 & 5888949.0 & 5888949 & 0.02 & 0.00\\ 
DEEPSEA\_MUN\_C7\_V3\_HE\_4   & 6510656 & 0.00 &  & 6510656 & 6510656.0 & 6510656 & 0.02 & 0.00\\ 
DEEPSEA\_MUN\_C7\_V3\_HE\_5   & 7220458 & 0.00 &  & 7220458 & 7220458.0 & 7220458 & 0.02 & 0.00\\ 
DEEPSEA\_MUN\_C10\_V3\_HE\_1   & 7986248 & 0.00 &  & 7986248 & 7986248.0 & 7986248 & 0.04 & 0.00\\ 
DEEPSEA\_MUN\_C10\_V3\_HE\_2   & 7754484 & 0.00 &  & 7754484 & 7754484.0 & 7754484 & 0.04 & 0.00\\ 
DEEPSEA\_MUN\_C10\_V3\_HE\_3   & 9499357 & 0.00 &  & 9499357 & 9499357.0 & 9499357 & 0.04 & 0.00\\ 
DEEPSEA\_MUN\_C10\_V3\_HE\_4   & 8617192 & 0.00 &  & 8617192 & 8617192.0 & 8617192 & 0.04 & 0.00\\ 
DEEPSEA\_MUN\_C10\_V3\_HE\_5   & 8653992 & 0.00 &  & 8653992 & 8653992.0 & 8653992 & 0.04 & 0.00\\ 
DEEPSEA\_MUN\_C15\_V4\_HE\_1   & 13467090 & 0.00 &  & 13467090 & 13467090.0 & 13467090 & 0.07 & 0.00\\ 
DEEPSEA\_MUN\_C15\_V4\_HE\_2   & 12457251 & 0.00 &  & 12457251 & 12457251.0 & 12457251 & 0.10 & 0.02\\ 
DEEPSEA\_MUN\_C15\_V4\_HE\_3   & 12567396 & 0.00 &  & 12567396 & 12567396.0 & 12567396 & 0.08 & 0.00\\ 
DEEPSEA\_MUN\_C15\_V4\_HE\_4   & 11764241 & 0.00 &  & 11764241 & 11764241.0 & 11764241 & 0.08 & 0.01\\ 
DEEPSEA\_MUN\_C15\_V4\_HE\_5   & 10833640 & 0.00 &  & 10833640 & 10833640.0 & 10833640 & 0.08 & 0.00\\ 
DEEPSEA\_MUN\_C18\_V5\_HE\_1   & 43054055 & 0.00 &  & 43054055 & 43054055.0 & 43054055 & 0.14 & 0.02\\ 
DEEPSEA\_MUN\_C18\_V5\_HE\_2   & 25068287 & 0.00 &  & 25068287 & 25068287.0 & 25068287 & 0.13 & 0.02\\ 
DEEPSEA\_MUN\_C18\_V5\_HE\_3   & 29211238 & 0.00 &  & 29211238 & 29211238.0 & 29211238 & 0.12 & 0.00\\ 
DEEPSEA\_MUN\_C18\_V5\_HE\_4   & 32281904 & 0.00 &  & 32281904 & 32281904.0 & 32281904 & 0.11 & 0.00\\ 
DEEPSEA\_MUN\_C18\_V5\_HE\_5   & 40718028 & 0.00 &  & 40718028 & 40718028.0 & 40718028 & 0.12 & 0.02\\ 
DEEPSEA\_MUN\_C22\_V6\_HE\_1   & 41176718 & 0.00 &  & 41176718 & 41176718.0 & 41176718 & 0.16 & 0.03\\ 
DEEPSEA\_MUN\_C22\_V6\_HE\_2   & 37236363 & 0.00 &  & 37236363 & 37236363.0 & 37236363 & 0.17 & 0.05\\ 
DEEPSEA\_MUN\_C22\_V6\_HE\_3   & 38215238 & 0.00 &  & 38215238 & 38215238.0 & 38215238 & 0.17 & 0.03\\ 
DEEPSEA\_MUN\_C22\_V6\_HE\_4   & 34129809 & 0.00 &  & 34129809 & 34129809.0 & 34129809 & 0.22 & 0.08\\ 
DEEPSEA\_MUN\_C22\_V6\_HE\_5   & 46379332 & 0.00 &  & 46379332 & 46379332.0 & 46379332 & 0.17 & 0.04\\ 
DEEPSEA\_MUN\_C23\_V13\_HE\_1   & 41002992 & 0.00 &  & 41002992 & 41002992.0 & 41002992 & 0.19 & 0.04\\ 
DEEPSEA\_MUN\_C23\_V13\_HE\_2   & 28014147 & 0.00 &  & 28014147 & 28014147.0 & 28014147 & 0.19 & 0.02\\ 
DEEPSEA\_MUN\_C23\_V13\_HE\_3   & 29090422 & 0.00 &  & 29090422 & 29090422.0 & 29090422 & 0.17 & 0.00\\ 
DEEPSEA\_MUN\_C23\_V13\_HE\_4   & 33685274 & 0.00 &  & 33685274 & 33685274.0 & 33685274 & 0.23 & 0.07\\ 
DEEPSEA\_MUN\_C23\_V13\_HE\_5   & 38664843 & 0.00 &  & 38664843 & 38664843.0 & 38664843 & 0.20 & 0.06\\ 
DEEPSEA\_MUN\_C30\_V6\_HE\_1   & 19227093 & 0.00 &  & 19227093 & 19227093.0 & 19227093 & 0.31 & 0.12\\ 
DEEPSEA\_MUN\_C30\_V6\_HE\_2   & 16784810 & 0.02 &  & 16784810 & 16784810.0 & 16784810 & 0.35 & 0.11\\ 
DEEPSEA\_MUN\_C30\_V6\_HE\_3   & 21183928 & 0.01 &  & 21183928 & 21213100.9 & 21298546 & 0.35 & 0.16\\ 
DEEPSEA\_MUN\_C30\_V6\_HE\_4   & 21076728 & 0.00 &  & 21076728 & 21076728.0 & 21076728 & 0.28 & 0.09\\ 
DEEPSEA\_MUN\_C30\_V6\_HE\_5   & 24490671 & 0.01 &  & 24490671 & 24490671.0 & 24490671 & 0.33 & 0.12\\ 
DEEPSEA\_MUN\_C35\_V7\_HE\_1   & 65082675 & 0.01 &  & 65082675 & 65086315.3 & 65119078 & 0.50 & 0.21\\ 
DEEPSEA\_MUN\_C35\_V7\_HE\_2   & 54810586 & 0.03 &  & 54810586 & 54810586.0 & 54810586 & 0.45 & 0.18\\ 
DEEPSEA\_MUN\_C35\_V7\_HE\_3   & 56182502 & 0.00 &  & 56182502 & 56182502.0 & 56182502 & 0.42 & 0.14\\ 
DEEPSEA\_MUN\_C35\_V7\_HE\_4   & 61354812 & 0.02 &  & 61354812 & 61354812.0 & 61354812 & 0.46 & 0.16\\ 
DEEPSEA\_MUN\_C35\_V7\_HE\_5   & 63904705 & 0.00 &  & 63904705 & 63904705.0 & 63904705 & 0.39 & 0.15\\ 
DEEPSEA\_MUN\_C60\_V13\_HE\_1   & 80649895 & 3.25 &  & 80649895 & 80696507.0 & 80708160 & 2.13 & 1.14\\ 
DEEPSEA\_MUN\_C60\_V13\_HE\_2   & 74881109 & 0.10 &  & 74881109 & 74881109.4 & 74881110 & 1.86 & 1.21\\ 
DEEPSEA\_MUN\_C60\_V13\_HE\_3   & 91766747 & 0.03 &  & 91766747 & 91768529.1 & 91782830 & 1.47 & 0.78\\ 
DEEPSEA\_MUN\_C60\_V13\_HE\_4   & 89702352 & 0.05 &  & 89702352 & 89763856.0 & 89863541 & 1.90 & 1.17\\ 
DEEPSEA\_MUN\_C60\_V13\_HE\_5   & 88486544 & 0.08 &  & 88486544 & 88498544.6 & 88606550 & 1.71 & 1.00\\ 
DEEPSEA\_MUN\_C80\_V20\_HE\_1   & 70718084 & 0.33 &  & 70718084 & 70799785.6 & 70922338 & 3.20 & 1.78\\ 
DEEPSEA\_MUN\_C80\_V20\_HE\_2   & 73558165 & 1.95 &  & 73558165 & 73589043.3 & 73603212 & 5.07 & 3.47\\ 
DEEPSEA\_MUN\_C80\_V20\_HE\_3   & 78250612 & 0.12 &  & 78250612 & 78251002.0 & 78254512 & 2.56 & 1.43\\ 
DEEPSEA\_MUN\_C80\_V20\_HE\_4   & 75962439 & 0.32 &  & 75962439 & 75994306.3 & 76061556 & 3.67 & 2.34\\ 
DEEPSEA\_MUN\_C80\_V20\_HE\_5   & 74162521 & 0.22 &  & 74162521 & 74169783.2 & 74207930 & 3.29 & 1.97\\ 
DEEPSEA\_MUN\_C100\_V30\_HE\_1   & 150481912 & 1.25 &  & 150481912 & 150508649.2 & 150525751 & 10.03 & 7.17\\ 
DEEPSEA\_MUN\_C100\_V30\_HE\_2   & 150826322 & 0.52 &  & 150826322 & 150834992.9 & 150866157 & 12.08 & 9.54\\ 
DEEPSEA\_MUN\_C100\_V30\_HE\_3   & 151027805 & 1.83 &  & 151027805 & 151036770.9 & 151039246 & 8.58 & 4.48\\ 
DEEPSEA\_MUN\_C100\_V30\_HE\_4   & 151193009 & 6.33 &  & 151193009 & 151218563.7 & 151331951 & 10.65 & 7.34\\ 
DEEPSEA\_MUN\_C100\_V30\_HE\_5   & 159789021 & 0.75 &  & 159789021 & 159799748.9 & 159828402 & 8.02 & 5.67\\ 
DEEPSEA\_MUN\_C130\_V40\_HE\_1   & 232582224 & 5.30 &  & 232625726 & 232672618.6 & 232835502 & 14.74 & 12.29\\ 
DEEPSEA\_MUN\_C130\_V40\_HE\_2   & 228036360 & 265.06 &  & 228205988 & 228285255.3 & 228365518 & 15.00 & 13.56\\ 
DEEPSEA\_MUN\_C130\_V40\_HE\_3   & 235657072 & 9.49 &  & 235666249 & 235772248.0 & 236033885 & 15.00 & 13.66\\ 
DEEPSEA\_MUN\_C130\_V40\_HE\_4   & 220357686 & 6.40 &  & 220360711 & 220564140.8 & 220783879 & 14.31 & 11.66\\ 
DEEPSEA\_MUN\_C130\_V40\_HE\_5   & 235381937 & 7.03 &  & 235487434 & 235739525.3 & 235849521 & 15.01 & 13.23\\ \bottomrule
		\end{tabular}
	}
\end{table}

\begin{table}[htbp]
        \vspace*{-0.7cm}
	\renewcommand{\arraystretch}{1.0}
	\setlength\tabcolsep{7pt}
	\centering
	\caption{Results for deep sea full load instances}\label{tab:ds_fun}
	\scalebox{0.7}{
		\begin{tabular}{lrrrrrrrrr}
			\toprule
			\multirow{2}{*}{Instance} & \multicolumn{2}{c}{B\&P\textsubscript{2}} && \multicolumn{5}{c}{HGS\textsubscript{2}} \\ \cline{2-3}\cline{5-9}
			& \multicolumn{1}{c}{Opt} & \multicolumn{1}{c}{T} && \multicolumn{1}{c}{Best} & \multicolumn{1}{c}{Avg} & \multicolumn{1}{c}{Worst} & \multicolumn{1}{c}{T} & \multicolumn{1}{c}{T*} \\ \midrule
DEEPSEA\_FUN\_C8\_V3\_HE\_1   & 9584863 & 0.00 &  & 9584863 & 9584863.0 & 9584863 & 0.01 & 0.00\\ 
DEEPSEA\_FUN\_C8\_V3\_HE\_2   & 9369654 & 0.00 &  & 9369654 & 9369654.0 & 9369654 & 0.01 & 0.00\\ 
DEEPSEA\_FUN\_C8\_V3\_HE\_3   & 4596681 & 0.00 &  & 4596681 & 4596681.0 & 4596681 & 0.01 & 0.00\\ 
DEEPSEA\_FUN\_C8\_V3\_HE\_4   & 6899730 & 0.00 &  & 6899730 & 6899730.0 & 6899730 & 0.01 & 0.00\\ 
DEEPSEA\_FUN\_C8\_V3\_HE\_5   & 6815253 & 0.00 &  & 6815253 & 6815253.0 & 6815253 & 0.01 & 0.00\\ 
DEEPSEA\_FUN\_C11\_V4\_HE\_1   & 34854819 & 0.00 &  & 34854819 & 34854819.0 & 34854819 & 0.02 & 0.00\\ 
DEEPSEA\_FUN\_C11\_V4\_HE\_2   & 25454434 & 0.00 &  & 25454434 & 25454434.0 & 25454434 & 0.02 & 0.00\\ 
DEEPSEA\_FUN\_C11\_V4\_HE\_3   & 29627143 & 0.00 &  & 29627143 & 29627143.0 & 29627143 & 0.02 & 0.00\\ 
DEEPSEA\_FUN\_C11\_V4\_HE\_4   & 33111680 & 0.00 &  & 33111680 & 33111680.0 & 33111680 & 0.02 & 0.00\\ 
DEEPSEA\_FUN\_C11\_V4\_HE\_5   & 28175914 & 0.00 &  & 28175914 & 28175914.0 & 28175914 & 0.03 & 0.00\\ 
DEEPSEA\_FUN\_C13\_V5\_HE\_1   & 11629005 & 0.00 &  & 11629005 & 11629005.0 & 11629005 & 0.04 & 0.00\\ 
DEEPSEA\_FUN\_C13\_V5\_HE\_2   & 11820655 & 0.00 &  & 11820655 & 11820655.0 & 11820655 & 0.04 & 0.00\\ 
DEEPSEA\_FUN\_C13\_V5\_HE\_3   & 9992593 & 0.00 &  & 9992593 & 9992593.0 & 9992593 & 0.04 & 0.00\\ 
DEEPSEA\_FUN\_C13\_V5\_HE\_4   & 12819619 & 0.00 &  & 12819619 & 12819619.0 & 12819619 & 0.03 & 0.00\\ 
DEEPSEA\_FUN\_C13\_V5\_HE\_5   & 10534892 & 0.00 &  & 10534892 & 10534892.0 & 10534892 & 0.03 & 0.00\\ 
DEEPSEA\_FUN\_C16\_V6\_HE\_1   & 51127590 & 0.00 &  & 51127590 & 51127590.0 & 51127590 & 0.04 & 0.00\\ 
DEEPSEA\_FUN\_C16\_V6\_HE\_2   & 44342796 & 0.00 &  & 44342796 & 44342796.0 & 44342796 & 0.03 & 0.00\\ 
DEEPSEA\_FUN\_C16\_V6\_HE\_3   & 45391842 & 0.00 &  & 45391842 & 45391842.0 & 45391842 & 0.03 & 0.00\\ 
DEEPSEA\_FUN\_C16\_V6\_HE\_4   & 39687114 & 0.00 &  & 39687114 & 39687114.0 & 39687114 & 0.04 & 0.00\\ 
DEEPSEA\_FUN\_C16\_V6\_HE\_5   & 42855603 & 0.00 &  & 42855603 & 42855603.0 & 42855603 & 0.04 & 0.00\\ 
DEEPSEA\_FUN\_C17\_V13\_HE\_1   & 17316720 & 0.00 &  & 17316720 & 17316720.0 & 17316720 & 0.04 & 0.00\\ 
DEEPSEA\_FUN\_C17\_V13\_HE\_2   & 12194861 & 0.00 &  & 12194861 & 12194861.0 & 12194861 & 0.05 & 0.00\\ 
DEEPSEA\_FUN\_C17\_V13\_HE\_3   & 12091554 & 0.00 &  & 12091554 & 12091554.0 & 12091554 & 0.05 & 0.00\\ 
DEEPSEA\_FUN\_C17\_V13\_HE\_4   & 12847653 & 0.00 &  & 12847653 & 12847653.0 & 12847653 & 0.05 & 0.01\\ 
DEEPSEA\_FUN\_C17\_V13\_HE\_5   & 13213406 & 0.00 &  & 13213406 & 13213406.0 & 13213406 & 0.05 & 0.00\\ 
DEEPSEA\_FUN\_C20\_V6\_HE\_1   & 16406738 & 0.00 &  & 16406738 & 16406738.0 & 16406738 & 0.05 & 0.00\\ 
DEEPSEA\_FUN\_C20\_V6\_HE\_2   & 16079401 & 0.00 &  & 16079401 & 16079401.0 & 16079401 & 0.05 & 0.00\\ 
DEEPSEA\_FUN\_C20\_V6\_HE\_3   & 17342200 & 0.00 &  & 17342200 & 17342200.0 & 17342200 & 0.04 & 0.00\\ 
DEEPSEA\_FUN\_C20\_V6\_HE\_4   & 16529748 & 0.00 &  & 16529748 & 16529748.0 & 16529748 & 0.05 & 0.01\\ 
DEEPSEA\_FUN\_C20\_V6\_HE\_5   & 17449378 & 0.00 &  & 17449378 & 17449378.0 & 17449378 & 0.05 & 0.00\\ 
DEEPSEA\_FUN\_C25\_V7\_HE\_1   & 22773158 & 0.00 &  & 22773158 & 22773158.0 & 22773158 & 0.07 & 0.01\\ 
DEEPSEA\_FUN\_C25\_V7\_HE\_2   & 20206329 & 0.00 &  & 20206329 & 20206329.0 & 20206329 & 0.08 & 0.01\\ 
DEEPSEA\_FUN\_C25\_V7\_HE\_3   & 19108952 & 0.00 &  & 19108952 & 19108952.0 & 19108952 & 0.07 & 0.01\\ 
DEEPSEA\_FUN\_C25\_V7\_HE\_4   & 22668675 & 0.00 &  & 22668675 & 22668675.0 & 22668675 & 0.07 & 0.01\\ 
DEEPSEA\_FUN\_C25\_V7\_HE\_5   & 23036603 & 0.00 &  & 23036603 & 23036603.0 & 23036603 & 0.08 & 0.02\\ 
DEEPSEA\_FUN\_C35\_V13\_HE\_1   & 86951609 & 0.00 &  & 86951609 & 86951609.0 & 86951609 & 0.22 & 0.09\\ 
DEEPSEA\_FUN\_C35\_V13\_HE\_2   & 83422071 & 0.00 &  & 83422071 & 83422071.0 & 83422071 & 0.19 & 0.07\\ 
DEEPSEA\_FUN\_C35\_V13\_HE\_3   & 83898591 & 0.00 &  & 83898591 & 83898591.0 & 83898591 & 0.21 & 0.08\\ 
DEEPSEA\_FUN\_C35\_V13\_HE\_4   & 91970481 & 0.00 &  & 91970481 & 91970481.0 & 91970481 & 0.23 & 0.09\\ 
DEEPSEA\_FUN\_C35\_V13\_HE\_5   & 91123040 & 0.00 &  & 91123040 & 91123040.0 & 91123040 & 0.20 & 0.07\\ 
DEEPSEA\_FUN\_C50\_V20\_HE\_1   & 41310946 & 0.00 &  & 41310946 & 41310946.0 & 41310946 & 0.46 & 0.18\\ 
DEEPSEA\_FUN\_C50\_V20\_HE\_2   & 37784994 & 0.00 &  & 37784994 & 37784994.0 & 37784994 & 0.46 & 0.18\\ 
DEEPSEA\_FUN\_C50\_V20\_HE\_3   & 39841724 & 0.00 &  & 39841724 & 39841724.0 & 39841724 & 0.41 & 0.13\\ 
DEEPSEA\_FUN\_C50\_V20\_HE\_4   & 43941098 & 0.00 &  & 43941098 & 43941098.0 & 43941098 & 0.47 & 0.20\\ 
DEEPSEA\_FUN\_C50\_V20\_HE\_5   & 41947437 & 0.00 &  & 41947437 & 41947437.0 & 41947437 & 0.46 & 0.18\\ 
DEEPSEA\_FUN\_C70\_V30\_HE\_1   & 142679953 & 0.01 &  & 142679953 & 142679953.0 & 142679953 & 1.03 & 0.38\\ 
DEEPSEA\_FUN\_C70\_V30\_HE\_2   & 135031988 & 0.02 &  & 135031988 & 135031988.0 & 135031988 & 1.06 & 0.36\\ 
DEEPSEA\_FUN\_C70\_V30\_HE\_3   & 162759203 & 0.01 &  & 162759203 & 162759203.0 & 162759203 & 1.02 & 0.40\\ 
DEEPSEA\_FUN\_C70\_V30\_HE\_4   & 155855123 & 0.01 &  & 155855123 & 155855123.0 & 155855123 & 1.06 & 0.44\\ 
DEEPSEA\_FUN\_C70\_V30\_HE\_5   & 156557723 & 0.01 &  & 156557723 & 156557723.0 & 156557723 & 0.95 & 0.35\\ 
DEEPSEA\_FUN\_C90\_V40\_HE\_1   & 190627186 & 0.06 &  & 190627186 & 190630992.5 & 190641592 & 2.08 & 0.83\\ 
DEEPSEA\_FUN\_C90\_V40\_HE\_2   & 189770977 & 0.02 &  & 189770977 & 189771678.3 & 189777990 & 2.63 & 1.45\\ 
DEEPSEA\_FUN\_C90\_V40\_HE\_3   & 211038412 & 0.02 &  & 211038412 & 211038684.4 & 211041136 & 2.15 & 0.94\\ 
DEEPSEA\_FUN\_C90\_V40\_HE\_4   & 210449287 & 0.02 &  & 210449287 & 210449654.6 & 210451528 & 2.02 & 0.93\\ 
DEEPSEA\_FUN\_C90\_V40\_HE\_5   & 197804917 & 0.05 &  & 197804917 & 197805398.0 & 197809727 & 2.11 & 0.87\\ 
DEEPSEA\_FUN\_C100\_V50\_HE\_1   & 205826535 & 0.08 &  & 205826535 & 205831890.5 & 205844919 & 2.96 & 1.24\\ 
DEEPSEA\_FUN\_C100\_V50\_HE\_2   & 207809147 & 0.03 &  & 207809147 & 207813969.7 & 207833395 & 4.33 & 2.62\\ 
DEEPSEA\_FUN\_C100\_V50\_HE\_3   & 217000928 & 0.02 &  & 217000928 & 217000928.0 & 217000928 & 3.08 & 1.41\\ 
DEEPSEA\_FUN\_C100\_V50\_HE\_4   & 220879632 & 0.03 &  & 220879632 & 220879794.0 & 220880172 & 2.49 & 0.83\\ 
DEEPSEA\_FUN\_C100\_V50\_HE\_5   & 223265017 & 0.02 &  & 223265017 & 223265583.6 & 223270683 & 2.79 & 1.10\\ \bottomrule
		\end{tabular}
	}
      \end{table}

\begin{table}[htbp]
\centering
  \vspace*{-0.7cm}
  \renewcommand{\arraystretch}{1.0}
  \setlength\tabcolsep{7pt}
  \caption{Results for new large scale instances}\label{tab:large_full}
  \scalebox{0.7}{
    \begin{tabular}{lrrrrrrrrr}
      \toprule
      \multirow{2}{*}{Instance} & \multicolumn{3}{c}{B\&P\textsubscript{2}} && \multicolumn{5}{c}{HGS\textsubscript{2}} \\ \cline{2-4}\cline{6-10}
                                & \multicolumn{1}{c}{LB} & \multicolumn{1}{c}{UB} & \multicolumn{1}{c}{T} && \multicolumn{1}{c}{Best} & \multicolumn{1}{c}{Avg} & \multicolumn{1}{c}{Worst} & \multicolumn{1}{c}{T} & \multicolumn{1}{c}{T*} \\ \midrule
SHORTSEA\_MUN\_C143\_V41	&17799119.0	&17799119	&24.38	&	&17804172	&17806545.5	&17810359	&15.85	&10.69\\
SHORTSEA\_MUN\_C156\_V45	&19342942.0	&19342942	&71.98	&	&19343517	&19344605.7	&19346784	&28.15	&18.19\\
SHORTSEA\_MUN\_C169\_V48	&21268109.0	&21268109	&41.71	&	&21268109	&21270048.5	&21278516	&26.32	&19.28\\
SHORTSEA\_MUN\_C182\_V52	&22980177.0	&22980177	&88.32	&	&22982740	&22986270.9	&22991134	&42.22	&31.34\\
SHORTSEA\_MUN\_C195\_V56	&24440859.0	&24440859	&147.68	&	&24440859	&24450058.2	&24468278	&50.21	&39.07\\
SHORTSEA\_MUN\_C208\_V59	&25708244.1	&25712849	&240.00	&	&25712849	&25716650.8	&25723526	&56.17	&46.62\\
SHORTSEA\_MUN\_C221\_V63	&26989861.5	&27034941	&495.02	&	&27034941	&27042556.8	&27058195	&56.32	&45.02\\
SHORTSEA\_MUN\_C260\_V74	&32407978.0	&32464488	&1196.78	&	&32464488	&32488515.1	&32538633	&60.16	&53.57\\
SHORTSEA\_FUN\_C110\_V52	&15133771.0	&15133771	&0.14	&	&15133771	&15133926.9	&15134691	&4.01	&1.88\\
SHORTSEA\_FUN\_C120\_V53	&16558958.0	&16558958	&0.29	&	&16558993	&16559104.9	&16559264	&4.40	&1.87\\
SHORTSEA\_FUN\_C130\_V58	&17649436.0	&17649436	&0.35	&	&17649436	&17649659.7	&17650377	&5.58	&2.67\\
SHORTSEA\_FUN\_C140\_V62	&19179026.0	&19179026	&0.14	&	&19179026	&19179026.2	&19179027	&6.49	&2.71\\
SHORTSEA\_FUN\_C150\_V67	&20281407.0	&20281407	&0.34	&	&20281407	&20281794.0	&20283296	&7.65	&3.33\\
SHORTSEA\_FUN\_C160\_V71	&21695647.0	&21695647	&1.31	&	&21696335	&21696659.8	&21696858	&9.14	&3.96\\
SHORTSEA\_FUN\_C170\_V76	&23110481.0	&23110481	&0.34	&	&23110481	&23111291.2	&23114048	&13.06	&6.95\\
SHORTSEA\_FUN\_C200\_V89	&27690324.0	&27690324	&0.78	&	&27690324	&27693029.9	&27703323	&20.74	&11.52\\
DEEPSEA\_MUN\_C143\_V41	&254254590.0	&254254590	&33.66	&	&254254591	&254349635.7	&254608621	&42.36	&34.89\\
DEEPSEA\_MUN\_C156\_V45	&270058045.0	&270058045	&8.09	&	&270085318	&270176938.6	&270330689	&43.32	&34.79\\
DEEPSEA\_MUN\_C169\_V48	&289887680.2	&290916269	&240.00	&	&290916269	&291038620.0	&291257150	&55.06	&47.54\\
DEEPSEA\_MUN\_C182\_V52	&300137926.8	&301663875	&240.00	&	&301663875	&301925296.5	&302379995	&57.21	&49.01\\
DEEPSEA\_MUN\_C195\_V56	&310191451.9	&311546377	&240.00	&	&311546377	&312129552.3	&313791486	&57.95	&49.53\\
DEEPSEA\_MUN\_C208\_V59	&341832426.9	&346223133	&240.00	&	&346223133	&346961721.4	&347822705	&56.32	&47.13\\
DEEPSEA\_MUN\_C221\_V63	&350836918.3	&352462189	&481.31	&	&352462189	&352995236.0	&354581124	&59.39	&50.24\\
DEEPSEA\_MUN\_C260\_V74	&404166031.4	&408352976	&3848.88	&	&408352976	&409513515.0	&411652613	&60.00	&52.28\\
DEEPSEA\_FUN\_C110\_V52	&240011111.0	&240011111	&0.05	&	&240011111	&240014104.1	&240026076	&3.72	&1.40\\
DEEPSEA\_FUN\_C120\_V53	&248614953.0	&248614953	&0.06	&	&248614953	&248618591.1	&248633514	&4.75	&2.12\\
DEEPSEA\_FUN\_C130\_V58	&288771846.0	&288771846	&0.08	&	&288771846	&288771846.0	&288771846	&5.70	&2.54\\
DEEPSEA\_FUN\_C140\_V62	&303231470.0	&303231470	&0.12	&	&303231470	&303231470.0	&303231470	&6.87	&3.16\\
DEEPSEA\_FUN\_C150\_V67	&323442552.0	&323442552	&0.96	&	&323442822	&323455751.7	&323466563	&8.35	&3.97\\
DEEPSEA\_FUN\_C160\_V71	&370429540.0	&370429540	&0.16	&	&370429540	&370439735.8	&370471324	&13.46	&8.19\\
DEEPSEA\_FUN\_C170\_V76	&395641818.0	&395641818	&0.39	&	&395641818	&395648175.4	&395663480	&14.74	&8.32\\
DEEPSEA\_FUN\_C200\_V89	&430966915.0	&430966915	&8.54	&	&430970779	&430984012.9	&431008839	&17.95	&9.37\\ \bottomrule
    \end{tabular}
  }
\end{table}

\clearpage
\section*{Appendix -- Comparison with \citet{Hemmati2016}}

\citet{Hemmati2016} have presented detailed results of six ALNS variants on a subset of the \citet{Hemmati2014} instances. \Cref{tab:vs2016} compares these results with those of HGS\textsubscript{2}. Each line represents a group of instances, and the best results are highlighted in boldface.  For each method, column ``Gap'' reports the percentage gap relative to the \emph{optimal} solutions, and column ``T'' gives the total CPU time in minutes.

\begin{table}[htbp]
	\renewcommand{\arraystretch}{1.1}
	\setlength\tabcolsep{5pt}
	\centering
	\caption{Performance comparison with~\citet{Hemmati2016}}\label{tab:vs2016}
	\scalebox{0.85}{
	\begin{tabular}{clrrrrrrrrrrrrrr}
		\toprule
		                                                                              &         & \multicolumn{2}{c}{ALNS\textsubscript{1}} & \multicolumn{2}{c}{ALNS\textsubscript{2}} & \multicolumn{2}{c}{ALNS\textsubscript{3}} & \multicolumn{2}{c}{ALNS\textsubscript{4}} & \multicolumn{2}{c}{ALNS\textsubscript{5}} & \multicolumn{2}{c}{ALNS\textsubscript{6}} & \multicolumn{2}{c}{HGS\textsubscript{2}} \\
		                                                                              &         &           Gap &                         T &           Gap &                         T &  Gap &                                  T &  Gap &                                  T &           Gap &                         T &           Gap &                         T &           Gap &                        T \\ \midrule
		\parbox[t]{2mm}{\multirow{5}{*}{\rotatebox[origin=c]{90}{{SHORTSEA}}}} & C22-V6  &          0.29 &                      0.18 &          0.29 &                      0.20 & 0.23 &                               0.19 & 0.17 &                               0.19 &          0.35 &                      0.20 &          0.53 &                      0.18 & \textbf{0.00} &                     0.18 \\
		                                                                              & C23-V13 &          0.82 &                      0.25 &          0.34 &                      0.25 & 0.35 &                               0.26 & 0.28 &                               0.26 &          0.24 &                      0.25 &          0.46 &                      0.25 & \textbf{0.00} &                     0.23 \\
		                                                                              & C30-V6  &          1.32 &                      0.35 &          0.66 &                      0.40 & 1.14 &                               0.41 & 0.90 &                               0.42 &          1.04 &                      0.39 &          1.95 &                      0.38 & \textbf{0.00} &                     0.33 \\
		                                                                              & C35-V7  &          1.40 &                      0.51 &          1.51 &                      0.58 & 1.69 &                               0.58 & 1.21 &                               0.58 &          1.03 &                      0.55 &          1.97 &                      0.53 & \textbf{0.01} &                     0.53 \\
		                                                                              & C60-V13 &          2.89 &                      2.05 &          2.29 &                      2.18 & 2.83 &                               2.29 & 2.30 &                               2.42 &          2.63 &                      2.06 &          1.81 &                      1.95 & \textbf{0.13} &                     2.14 \\ \midrule
		                                                                              & Overall &          1.34 &                      0.67 &          1.02 &                      0.72 & 1.25 &                               0.75 & 0.97 &                               0.77 &          1.06 &                      0.69 &          1.34 &                      0.66 & \textbf{0.03} &                     0.68 \\ \midrule
		\parbox[t]{2mm}{\multirow{5}{*}{\rotatebox[origin=c]{90}{{DEEPSEA}}}}  & C22-V6  &          0.35 &                      0.18 &          0.31 &                      0.19 & 0.15 &                               0.20 & 0.15 &                               0.20 &          0.19 &                      0.19 &          1.66 &                      0.18 & \textbf{0.00} &                     0.18 \\
		                                                                              & C23-V13 & \textbf{0.00} &                      0.26 & \textbf{0.00} &                      0.26 & 0.02 &                               0.28 & 0.01 &                               0.28 & \textbf{0.00} &                      0.25 & \textbf{0.00} &                      0.24 & \textbf{0.00} &                     0.20 \\
		                                                                              & C30-V6  &          0.67 &                      0.36 &          0.34 &                      0.42 & 0.36 &                               0.42 & 0.34 &                               0.42 &          0.42 &                      0.39 &          0.58 &                      0.37 & \textbf{0.03} &                     0.33 \\
		                                                                              & C35-V7  &          0.61 &                      0.52 &          0.51 &                      0.60 & 0.50 &                               0.60 & 0.58 &                               0.63 &          0.37 &                      0.54 &          0.69 &                      0.51 & \textbf{0.00} &                     0.43 \\
		                                                                              & C60-V13 &          4.37 &                      2.18 &          3.50 &                      2.41 & 4.11 &                               2.39 & 4.12 &                               2.39 &          3.23 &                      1.99 &          3.41 &                      1.85 & \textbf{0.02} &                     1.73 \\ \midrule
		                                                                              & Overall &          1.20 &                      0.70 &          0.93 &                      0.78 & 1.03 &                               0.78 & 1.04 &                               0.78 &          0.84 &                      0.67 &          1.27 &                      0.63 & \textbf{0.01} &                     0.57 \\ \bottomrule
	\end{tabular}
	}
\end{table}

%% file: ejor2019-R3-ArXiV.bbl
\begin{thebibliography}{49}
\expandafter\ifx\csname natexlab\endcsname\relax\def\natexlab#1{#1}\fi
\providecommand{\url}[1]{\texttt{#1}}
\providecommand{\href}[2]{#2}
\providecommand{\path}[1]{#1}
\providecommand{\DOIprefix}{doi:}
\providecommand{\ArXivprefix}{arXiv:}
\providecommand{\URLprefix}{URL: }
\providecommand{\Pubmedprefix}{pmid:}
\providecommand{\doi}[1]{\href{http://dx.doi.org/#1}{\path{#1}}}
\providecommand{\Pubmed}[1]{\href{pmid:#1}{\path{#1}}}
\providecommand{\bibinfo}[2]{#2}
\ifx\xfnm\relax \def\xfnm[#1]{\unskip,\space#1}\fi
\bibitem[{Andersson et~al.(2011)Andersson, Christiansen and
  Fagerholt}]{Andersson2011}
\bibinfo{author}{Andersson, H.}, \bibinfo{author}{Christiansen, M.},
  \bibinfo{author}{Fagerholt, K.}, \bibinfo{year}{2011}.
\newblock \bibinfo{title}{The maritime pickup and delivery problem with time
  windows and split loads}.
\newblock \bibinfo{journal}{{INFOR}: Information Systems and Operational
  Research} \bibinfo{volume}{49}, \bibinfo{pages}{79--91}.
\bibitem[{Baldacci et~al.(2011)Baldacci, Mingozzi and Roberti}]{Baldacci2011}
\bibinfo{author}{Baldacci, R.}, \bibinfo{author}{Mingozzi, A.},
  \bibinfo{author}{Roberti, R.}, \bibinfo{year}{2011}.
\newblock \bibinfo{title}{New route relaxation and pricing strategies for the
  vehicle routing problem}.
\newblock \bibinfo{journal}{Operations Research} \bibinfo{volume}{59},
  \bibinfo{pages}{1269--1283}.
\bibitem[{Bertsimas et~al.(2019)Bertsimas, Jaillet and Martin}]{Bertsimas2018}
\bibinfo{author}{Bertsimas, D.}, \bibinfo{author}{Jaillet, P.},
  \bibinfo{author}{Martin, S.}, \bibinfo{year}{2019}.
\newblock \bibinfo{title}{{Online vehicle routing: The edge of optimization in
  large-scale applications}}.
\newblock \bibinfo{journal}{Operations Research} \bibinfo{volume}{67},
  \bibinfo{pages}{143--162}.
\bibitem[{Borthen et~al.(2018)Borthen, Loennechen, Wang, Fagerholt and
  Vidal}]{Borthen2017}
\bibinfo{author}{Borthen, T.}, \bibinfo{author}{Loennechen, H.},
  \bibinfo{author}{Wang, X.}, \bibinfo{author}{Fagerholt, K.},
  \bibinfo{author}{Vidal, T.}, \bibinfo{year}{2018}.
\newblock \bibinfo{title}{A genetic search-based heuristic for a fleet size and
  periodic routing problem with application to offshore supply planning}.
\newblock \bibinfo{journal}{{EURO} Journal on Transportation and Logistics}
  \bibinfo{volume}{7}, \bibinfo{pages}{121--150}.
\bibitem[{Br{\o}nmo et~al.(2007)Br{\o}nmo, Christiansen, Fagerholt and
  Nygreen}]{Bronmo2007}
\bibinfo{author}{Br{\o}nmo, G.}, \bibinfo{author}{Christiansen, M.},
  \bibinfo{author}{Fagerholt, K.}, \bibinfo{author}{Nygreen, B.},
  \bibinfo{year}{2007}.
\newblock \bibinfo{title}{A multi-start local search heuristic for ship
  scheduling{\textemdash}a computational study}.
\newblock \bibinfo{journal}{Computers {\&} Operations Research}
  \bibinfo{volume}{34}, \bibinfo{pages}{900--917}.
\bibitem[{Brown et~al.(1987)Brown, Graves and Ronen}]{Brown1987}
\bibinfo{author}{Brown, G.G.}, \bibinfo{author}{Graves, G.W.},
  \bibinfo{author}{Ronen, D.}, \bibinfo{year}{1987}.
\newblock \bibinfo{title}{Scheduling ocean transportation of crude oil}.
\newblock \bibinfo{journal}{Management Science} \bibinfo{volume}{33},
  \bibinfo{pages}{335--346}.
\bibitem[{Bulh{\~{o}}es et~al.(2018)Bulh{\~{o}}es, H{\`{a}}, Martinelli and
  Vidal}]{Bulhoes2017}
\bibinfo{author}{Bulh{\~{o}}es, T.}, \bibinfo{author}{H{\`{a}}, M.},
  \bibinfo{author}{Martinelli, R.}, \bibinfo{author}{Vidal, T.},
  \bibinfo{year}{2018}.
\newblock \bibinfo{title}{{The vehicle routing problem with service level
  constraints}}.
\newblock \bibinfo{journal}{European Journal of Operational Research}
  \bibinfo{volume}{265}, \bibinfo{pages}{544--558}.
\bibitem[{Christiansen and Fagerholt(2014)}]{Christiansen2014}
\bibinfo{author}{Christiansen, M.}, \bibinfo{author}{Fagerholt, K.},
  \bibinfo{year}{2014}.
\newblock \bibinfo{title}{Ship routing and scheduling in industrial and tramp
  shipping}, in: \bibinfo{editor}{Toth, P.}, \bibinfo{editor}{Vigo, D.} (Eds.),
  \bibinfo{booktitle}{Vehicle Routing}. \bibinfo{publisher}{Society for
  Industrial and Applied Mathematics}, pp. \bibinfo{pages}{381--408}.
\bibitem[{Christiansen et~al.(2007)Christiansen, Fagerholt, Nygreen and
  Ronen}]{Christiansen2007}
\bibinfo{author}{Christiansen, M.}, \bibinfo{author}{Fagerholt, K.},
  \bibinfo{author}{Nygreen, B.}, \bibinfo{author}{Ronen, D.},
  \bibinfo{year}{2007}.
\newblock \bibinfo{title}{Maritime transportation}, in:
  \bibinfo{editor}{Barnhart, C.}, \bibinfo{editor}{Laporte, G.} (Eds.),
  \bibinfo{booktitle}{Transportation}. \bibinfo{publisher}{Elsevier}, pp.
  \bibinfo{pages}{189--284}.
\bibitem[{Christiansen et~al.(2013)Christiansen, Fagerholt, Nygreen and
  Ronen}]{Christiansen2013}
\bibinfo{author}{Christiansen, M.}, \bibinfo{author}{Fagerholt, K.},
  \bibinfo{author}{Nygreen, B.}, \bibinfo{author}{Ronen, D.},
  \bibinfo{year}{2013}.
\newblock \bibinfo{title}{Ship routing and scheduling in the new millennium}.
\newblock \bibinfo{journal}{European Journal of Operational Research}
  \bibinfo{volume}{228}, \bibinfo{pages}{467--483}.
\bibitem[{Christofides et~al.(1981)Christofides, Mingozzi and
  Toth}]{Christofides1981}
\bibinfo{author}{Christofides, N.}, \bibinfo{author}{Mingozzi, A.},
  \bibinfo{author}{Toth, P.}, \bibinfo{year}{1981}.
\newblock \bibinfo{title}{{Exact algorithms for the vehicle routing problem,
  based on spanning tree and shortest path relaxations}}.
\newblock \bibinfo{journal}{Mathematical Programming} \bibinfo{volume}{20},
  \bibinfo{pages}{255--282}.
\bibitem[{Contardo and Martinelli(2014)}]{Contardo2014}
\bibinfo{author}{Contardo, C.}, \bibinfo{author}{Martinelli, R.},
  \bibinfo{year}{2014}.
\newblock \bibinfo{title}{A new exact algorithm for the multi-depot vehicle
  routing problem under capacity and route length constraints}.
\newblock \bibinfo{journal}{Discrete Optimization} \bibinfo{volume}{12},
  \bibinfo{pages}{129--146}.
\bibitem[{Desaulniers et~al.(2008)Desaulniers, Lessard and
  Hadjar}]{Desaulniers2008}
\bibinfo{author}{Desaulniers, G.}, \bibinfo{author}{Lessard, F.},
  \bibinfo{author}{Hadjar, A.}, \bibinfo{year}{2008}.
\newblock \bibinfo{title}{Tabu search, partial elementarity, and generalized
  k-path inequalities for the vehicle routing problem with time windows}.
\newblock \bibinfo{journal}{Transportation Science} \bibinfo{volume}{42},
  \bibinfo{pages}{387--404}.
\bibitem[{Duhamel et~al.(2011)Duhamel, Lacomme and Prodhon}]{Duhamel2011a}
\bibinfo{author}{Duhamel, C.}, \bibinfo{author}{Lacomme, P.},
  \bibinfo{author}{Prodhon, C.}, \bibinfo{year}{2011}.
\newblock \bibinfo{title}{Efficient frameworks for greedy split and new depth
  first search split procedures for routing problems}.
\newblock \bibinfo{journal}{Computers {\&} Operations Research}
  \bibinfo{volume}{38}, \bibinfo{pages}{723--739}.
\bibitem[{Dumas et~al.(1991)Dumas, Desrosiers and Soumis}]{Dumas1991}
\bibinfo{author}{Dumas, Y.}, \bibinfo{author}{Desrosiers, J.},
  \bibinfo{author}{Soumis, F.}, \bibinfo{year}{1991}.
\newblock \bibinfo{title}{The pickup and delivery problem with time windows}.
\newblock \bibinfo{journal}{European Journal of Operational Research}
  \bibinfo{volume}{54}, \bibinfo{pages}{7--22}.
\bibitem[{Fagerholt(2004)}]{Fagerholt2004}
\bibinfo{author}{Fagerholt, K.}, \bibinfo{year}{2004}.
\newblock \bibinfo{title}{{A computer-based decision support system for vessel
  fleet scheduling - Experience and future research}}.
\newblock \bibinfo{journal}{Decision Support Systems} \bibinfo{volume}{37},
  \bibinfo{pages}{35--47}.
\bibitem[{Fagerholt and Christiansen(2000a)}]{Fagerholt2000b}
\bibinfo{author}{Fagerholt, K.}, \bibinfo{author}{Christiansen, M.},
  \bibinfo{year}{2000}a.
\newblock \bibinfo{title}{A combined ship scheduling and allocation problem}.
\newblock \bibinfo{journal}{Journal of the Operational Research Society}
  \bibinfo{volume}{51}, \bibinfo{pages}{834--842}.
\bibitem[{Fagerholt and Christiansen(2000b)}]{Fagerholt2000}
\bibinfo{author}{Fagerholt, K.}, \bibinfo{author}{Christiansen, M.},
  \bibinfo{year}{2000}b.
\newblock \bibinfo{title}{A travelling salesman problem with allocation, time
  window and precedence constraints — an application to ship scheduling}.
\newblock \bibinfo{journal}{International Transactions in Operational Research}
  \bibinfo{volume}{7}, \bibinfo{pages}{231--244}.
\bibitem[{Glover and Hao(2009)}]{Glover2009}
\bibinfo{author}{Glover, F.}, \bibinfo{author}{Hao, J.K.},
  \bibinfo{year}{2009}.
\newblock \bibinfo{title}{The case for strategic oscillation}.
\newblock \bibinfo{journal}{Annals of Operations Research}
  \bibinfo{volume}{183}, \bibinfo{pages}{163--173}.
\bibitem[{Gschwind et~al.(2018)Gschwind, Irnich, Rothenbächer and
  Tilk}]{Gschwind2018}
\bibinfo{author}{Gschwind, T.}, \bibinfo{author}{Irnich, S.},
  \bibinfo{author}{Rothenbächer, A.K.}, \bibinfo{author}{Tilk, C.},
  \bibinfo{year}{2018}.
\newblock \bibinfo{title}{Bidirectional labeling in column-generation
  algorithms for pickup-and-delivery problems}.
\newblock \bibinfo{journal}{European Journal of Operational Research}
  \bibinfo{volume}{266}, \bibinfo{pages}{521--530}.
\bibitem[{Hemmati and Hvattum(2016)}]{Hemmati2016}
\bibinfo{author}{Hemmati, A.}, \bibinfo{author}{Hvattum, L.M.},
  \bibinfo{year}{2016}.
\newblock \bibinfo{title}{Evaluating the importance of randomization in
  adaptive large neighborhood search}.
\newblock \bibinfo{journal}{International Transactions in Operational Research}
  \bibinfo{volume}{24}, \bibinfo{pages}{929--942}.
\bibitem[{Hemmati et~al.(2014)Hemmati, Hvattum, Fagerholt and
  Norstad}]{Hemmati2014}
\bibinfo{author}{Hemmati, A.}, \bibinfo{author}{Hvattum, L.M.},
  \bibinfo{author}{Fagerholt, K.}, \bibinfo{author}{Norstad, I.},
  \bibinfo{year}{2014}.
\newblock \bibinfo{title}{Benchmark suite for industrial and tramp ship routing
  and scheduling problems}.
\newblock \bibinfo{journal}{{INFOR}: Information Systems and Operational
  Research} \bibinfo{volume}{52}, \bibinfo{pages}{28--38}.
\bibitem[{Jepsen et~al.(2008)Jepsen, Petersen, Spoorendonk and
  Pisinger}]{Jepsen2008}
\bibinfo{author}{Jepsen, M.}, \bibinfo{author}{Petersen, B.},
  \bibinfo{author}{Spoorendonk, S.}, \bibinfo{author}{Pisinger, D.},
  \bibinfo{year}{2008}.
\newblock \bibinfo{title}{Subset-row inequalities applied to the
  vehicle-routing problem with time windows}.
\newblock \bibinfo{journal}{Operations Research} \bibinfo{volume}{56},
  \bibinfo{pages}{497--511}.
\bibitem[{Korsvik et~al.(2009)Korsvik, Fagerholt and Laporte}]{Korsvik2009}
\bibinfo{author}{Korsvik, J.E.}, \bibinfo{author}{Fagerholt, K.},
  \bibinfo{author}{Laporte, G.}, \bibinfo{year}{2009}.
\newblock \bibinfo{title}{A tabu search heuristic for ship routing and
  scheduling}.
\newblock \bibinfo{journal}{Journal of the Operational Research Society}
  \bibinfo{volume}{61}, \bibinfo{pages}{594--603}.
\bibitem[{Korsvik et~al.(2011)Korsvik, Fagerholt and Laporte}]{Korsvik2011}
\bibinfo{author}{Korsvik, J.E.}, \bibinfo{author}{Fagerholt, K.},
  \bibinfo{author}{Laporte, G.}, \bibinfo{year}{2011}.
\newblock \bibinfo{title}{A large neighbourhood search heuristic for ship
  routing and scheduling with split loads}.
\newblock \bibinfo{journal}{Computers {\&} Operations Research}
  \bibinfo{volume}{38}, \bibinfo{pages}{474--483}.
\bibitem[{Martinelli et~al.(2014)Martinelli, Pecin and Poggi}]{Martinelli2014}
\bibinfo{author}{Martinelli, R.}, \bibinfo{author}{Pecin, D.},
  \bibinfo{author}{Poggi, M.}, \bibinfo{year}{2014}.
\newblock \bibinfo{title}{Efficient elementary and restricted non-elementary
  route pricing}.
\newblock \bibinfo{journal}{European Journal of Operational Research}
  \bibinfo{volume}{239}, \bibinfo{pages}{102--111}.
\bibitem[{Muter et~al.(2010)Muter, Birbil and Sahin}]{Muter2010}
\bibinfo{author}{Muter, I.}, \bibinfo{author}{Birbil, S.},
  \bibinfo{author}{Sahin, G.}, \bibinfo{year}{2010}.
\newblock \bibinfo{title}{{Combination of metaheuristic and exact algorithms
  for solving set covering-type optimization problems}}.
\newblock \bibinfo{journal}{INFORMS Journal on Computing} \bibinfo{volume}{22},
  \bibinfo{pages}{603--619}.
\bibitem[{Nagata et~al.(2010)Nagata, Br\"{a}ysy and Dullaert}]{Nagata2010b}
\bibinfo{author}{Nagata, Y.}, \bibinfo{author}{Br\"{a}ysy, O.},
  \bibinfo{author}{Dullaert, W.}, \bibinfo{year}{2010}.
\newblock \bibinfo{title}{A penalty-based edge assembly memetic algorithm for
  the vehicle routing problem with time windows}.
\newblock \bibinfo{journal}{Computers {\&} Operations Research}
  \bibinfo{volume}{37}, \bibinfo{pages}{724--737}.
\bibitem[{Pecin et~al.(2017)Pecin, Contardo, Desaulniers and
  Uchoa}]{Pecin2016a}
\bibinfo{author}{Pecin, D.}, \bibinfo{author}{Contardo, C.},
  \bibinfo{author}{Desaulniers, G.}, \bibinfo{author}{Uchoa, E.},
  \bibinfo{year}{2017}.
\newblock \bibinfo{title}{{New enhancements for the exact solution of the
  vehicle routing problem with time windows}}.
\newblock \bibinfo{journal}{INFORMS Journal on Computing} \bibinfo{volume}{29},
  \bibinfo{pages}{489--502}.
\bibitem[{Prins(2004)}]{Prins2004}
\bibinfo{author}{Prins, C.}, \bibinfo{year}{2004}.
\newblock \bibinfo{title}{A simple and effective evolutionary algorithm for the
  vehicle routing problem}.
\newblock \bibinfo{journal}{Computers \& Operations Research}
  \bibinfo{volume}{31}, \bibinfo{pages}{1985--2002}.
\bibitem[{Prins et~al.(2009)Prins, Labadi and Reghioui}]{Prins2009a}
\bibinfo{author}{Prins, C.}, \bibinfo{author}{Labadi, N.},
  \bibinfo{author}{Reghioui, M.}, \bibinfo{year}{2009}.
\newblock \bibinfo{title}{{Tour splitting algorithms for vehicle routing
  problems}}.
\newblock \bibinfo{journal}{International Journal of Production Research}
  \bibinfo{volume}{47}, \bibinfo{pages}{507--535}.
\bibitem[{Righini and Salani(2008)}]{Righini2008}
\bibinfo{author}{Righini, G.}, \bibinfo{author}{Salani, M.},
  \bibinfo{year}{2008}.
\newblock \bibinfo{title}{New dynamic programming algorithms for the resource
  constrained elementary shortest path problem}.
\newblock \bibinfo{journal}{Networks} \bibinfo{volume}{51},
  \bibinfo{pages}{155--170}.
\bibitem[{Ronen(1983)}]{Ronen1983}
\bibinfo{author}{Ronen, D.}, \bibinfo{year}{1983}.
\newblock \bibinfo{title}{Cargo ships routing and scheduling: survey of models
  and problems}.
\newblock \bibinfo{journal}{European Journal of Operational Research}
  \bibinfo{volume}{12}, \bibinfo{pages}{119--126}.
\bibitem[{Ropke and Cordeau(2009)}]{Ropke2009}
\bibinfo{author}{Ropke, S.}, \bibinfo{author}{Cordeau, J.F.},
  \bibinfo{year}{2009}.
\newblock \bibinfo{title}{Branch and cut and price for the pickup and delivery
  problem with time windows}.
\newblock \bibinfo{journal}{Transportation Science} \bibinfo{volume}{43},
  \bibinfo{pages}{267--286}.
\bibitem[{Sadykov et~al.(2017)Sadykov, Uchoa and Pessoa}]{Sadykov2017}
\bibinfo{author}{Sadykov, R.}, \bibinfo{author}{Uchoa, E.},
  \bibinfo{author}{Pessoa, A.}, \bibinfo{year}{2017}.
\newblock \bibinfo{title}{A bucket graph based labeling algorithm with
  application to vehicle routing}.
\newblock \bibinfo{type}{Technical Report} \bibinfo{number}{L-2017-7}. Cadernos
  do LOGIS-UFF. \bibinfo{address}{Niter{\'o}i, Brazil}.
\bibitem[{Sigurd et~al.(2005)Sigurd, Ulstein, Nygreen and Ryan}]{Sigurd2005}
\bibinfo{author}{Sigurd, M.M.}, \bibinfo{author}{Ulstein, N.L.},
  \bibinfo{author}{Nygreen, B.}, \bibinfo{author}{Ryan, D.M.},
  \bibinfo{year}{2005}.
\newblock \bibinfo{title}{Ship scheduling with recurring visits and visit
  separation requirements}, in: \bibinfo{editor}{Desaulniers, G.},
  \bibinfo{editor}{Desrosiers, J.}, \bibinfo{editor}{Solomon, M.M.} (Eds.),
  \bibinfo{booktitle}{Column Generation}. \bibinfo{publisher}{Springer US},
  \bibinfo{address}{Boston, MA}, pp. \bibinfo{pages}{225--245}.
\bibitem[{St{\aa}lhane et~al.(2012)St{\aa}lhane, Andersson, Christiansen,
  Cordeau and Desaulniers}]{Stlhane2012}
\bibinfo{author}{St{\aa}lhane, M.}, \bibinfo{author}{Andersson, H.},
  \bibinfo{author}{Christiansen, M.}, \bibinfo{author}{Cordeau, J.F.},
  \bibinfo{author}{Desaulniers, G.}, \bibinfo{year}{2012}.
\newblock \bibinfo{title}{A branch-price-and-cut method for a ship routing and
  scheduling problem with split loads}.
\newblock \bibinfo{journal}{Computers {\&} Operations Research}
  \bibinfo{volume}{39}, \bibinfo{pages}{3361--3375}.
\bibitem[{Subramanian et~al.(2013)Subramanian, Uchoa and
  Ochi}]{Subramanian2013}
\bibinfo{author}{Subramanian, A.}, \bibinfo{author}{Uchoa, E.},
  \bibinfo{author}{Ochi, L.S.}, \bibinfo{year}{2013}.
\newblock \bibinfo{title}{A hybrid algorithm for a class of vehicle routing
  problems}.
\newblock \bibinfo{journal}{Computers {\&} Operations Research}
  \bibinfo{volume}{40}, \bibinfo{pages}{2519--2531}.
\bibitem[{UNCTAD(2017)}]{UNCTAD2017}
\bibinfo{author}{UNCTAD}, \bibinfo{year}{2017}.
\newblock \bibinfo{title}{Review of maritime transport} \URLprefix
  \url{https://unctad.org/en/PublicationsLibrary/rmt2017_en.pdf}.
  \bibinfo{note}{accessed: 2018-09-06}.
\bibitem[{Velasco et~al.(2009)Velasco, Castagliola, Dejax, Gu{\'e}ret and
  Prins}]{Velasco2009}
\bibinfo{author}{Velasco, N.}, \bibinfo{author}{Castagliola, P.},
  \bibinfo{author}{Dejax, P.}, \bibinfo{author}{Gu{\'e}ret, C.},
  \bibinfo{author}{Prins, C.}, \bibinfo{year}{2009}.
\newblock \bibinfo{title}{A memetic algorithm for a pick-up and delivery
  problem by helicopter}, in: \bibinfo{editor}{Pereira, F.B.},
  \bibinfo{editor}{Tavares, J.} (Eds.), \bibinfo{booktitle}{Bio-inspired
  Algorithms for the Vehicle Routing Problem}. \bibinfo{publisher}{Springer
  Berlin Heidelberg}, pp. \bibinfo{pages}{173--190}.
\bibitem[{Vidal(2016)}]{Vidal2016}
\bibinfo{author}{Vidal, T.}, \bibinfo{year}{2016}.
\newblock \bibinfo{title}{Technical note: Split algorithm in {O(n)} for the
  capacitated vehicle routing problem}.
\newblock \bibinfo{journal}{Computers {\&} Operations Research}
  \bibinfo{volume}{69}, \bibinfo{pages}{40--47}.
\bibitem[{Vidal et~al.(2012)Vidal, Crainic, Gendreau, Lahrichi and
  Rei}]{Vidal2012}
\bibinfo{author}{Vidal, T.}, \bibinfo{author}{Crainic, T.G.},
  \bibinfo{author}{Gendreau, M.}, \bibinfo{author}{Lahrichi, N.},
  \bibinfo{author}{Rei, W.}, \bibinfo{year}{2012}.
\newblock \bibinfo{title}{A hybrid genetic algorithm for multidepot and
  periodic vehicle routing problems}.
\newblock \bibinfo{journal}{Operations Research} \bibinfo{volume}{60},
  \bibinfo{pages}{611--624}.
\bibitem[{Vidal et~al.(2013)Vidal, Crainic, Gendreau and Prins}]{Vidal2013}
\bibinfo{author}{Vidal, T.}, \bibinfo{author}{Crainic, T.G.},
  \bibinfo{author}{Gendreau, M.}, \bibinfo{author}{Prins, C.},
  \bibinfo{year}{2013}.
\newblock \bibinfo{title}{A hybrid genetic algorithm with adaptive diversity
  management for a large class of vehicle routing problems with time-windows}.
\newblock \bibinfo{journal}{Computers \& Operations Research}
  \bibinfo{volume}{40}, \bibinfo{pages}{475--489}.
\bibitem[{Vidal et~al.(2014)Vidal, Crainic, Gendreau and Prins}]{Vidal2014}
\bibinfo{author}{Vidal, T.}, \bibinfo{author}{Crainic, T.G.},
  \bibinfo{author}{Gendreau, M.}, \bibinfo{author}{Prins, C.},
  \bibinfo{year}{2014}.
\newblock \bibinfo{title}{A unified solution framework for multi-attribute
  vehicle routing problems}.
\newblock \bibinfo{journal}{European Journal of Operational Research}
  \bibinfo{volume}{234}, \bibinfo{pages}{658--673}.
\bibitem[{Vidal et~al.(2015a)Vidal, Crainic, Gendreau and Prins}]{Vidal2015}
\bibinfo{author}{Vidal, T.}, \bibinfo{author}{Crainic, T.G.},
  \bibinfo{author}{Gendreau, M.}, \bibinfo{author}{Prins, C.},
  \bibinfo{year}{2015}a.
\newblock \bibinfo{title}{Time-window relaxations in vehicle routing
  heuristics}.
\newblock \bibinfo{journal}{Journal of Heuristics} \bibinfo{volume}{21},
  \bibinfo{pages}{329--358}.
\bibitem[{Vidal et~al.(2015b)Vidal, Crainic, Gendreau and Prins}]{Vidal2015b}
\bibinfo{author}{Vidal, T.}, \bibinfo{author}{Crainic, T.G.},
  \bibinfo{author}{Gendreau, M.}, \bibinfo{author}{Prins, C.},
  \bibinfo{year}{2015}b.
\newblock \bibinfo{title}{Timing problems and algorithms: time decisions for
  sequences of activities}.
\newblock \bibinfo{journal}{Networks} \bibinfo{volume}{65},
  \bibinfo{pages}{102--128}.
\bibitem[{Vidal et~al.(2016)Vidal, Maculan, Ochi and Penna}]{Vidal2014bb}
\bibinfo{author}{Vidal, T.}, \bibinfo{author}{Maculan, N.},
  \bibinfo{author}{Ochi, L.}, \bibinfo{author}{Penna, P.},
  \bibinfo{year}{2016}.
\newblock \bibinfo{title}{{Large neighborhoods with implicit customer selection
  for vehicle routing problems with profits}}.
\newblock \bibinfo{journal}{Transportation Science} \bibinfo{volume}{50},
  \bibinfo{pages}{720--734}.
\bibitem[{Wilson(2018)}]{Wilson2018}
\bibinfo{author}{Wilson}, \bibinfo{year}{2018}.
\newblock \URLprefix \url{https://www.wilsonship.no/en}.
  \bibinfo{note}{accessed: 2018-09-06}.
\bibitem[{Zachariadis and Kiranoudis(2010)}]{Zachariadis2010b}
\bibinfo{author}{Zachariadis, E.}, \bibinfo{author}{Kiranoudis, C.},
  \bibinfo{year}{2010}.
\newblock \bibinfo{title}{{A strategy for reducing the computational complexity
  of local search-based methods for the vehicle routing problem}}.
\newblock \bibinfo{journal}{Computers {\&} Operations Research}
  \bibinfo{volume}{37}, \bibinfo{pages}{2089--2105}.

\end{thebibliography}
